\documentclass[twocolumn, amssymb, amsmath, nobibnotes, aps, pra, reprint]{revtex4-2}
\usepackage{dcolumn}
\usepackage{bm}
\usepackage{flushend}
\usepackage{balance}
\usepackage{lipsum}
\usepackage{mathtools}
\usepackage{cuted}

\usepackage[caption=false]{subfig}
\usepackage{float}
\usepackage{silence}
\newcommand{\be}{\begin{equation}}

\newcommand{\ee}{\end{equation}}
\newcommand{\ben}{\begin{eqnarray}}
\newcommand{\een}{\end{eqnarray}}
\newcommand{\bes}{\begin{subequations}}
\newcommand{\ees}{\end{subequations}}
\newcommand{\bF}{\begin{figure}}
\newcommand{\eF}{\end{figure}}

\def\tr{ {\rm{Tr }}\,}

\newcommand{\ket}[1]{\left|#1\right\rangle}
\newcommand{\bra}[1]{\left\langle#1\right|}

\newcommand{\supket}[1]{\left|#1\right)}
\newcommand{\supbra}[1]{\left(#1\right|}
\newcommand{\supbraket}[2]{\left(#1| #2\right) }

\newcommand{\mbf}[1]{\mathbf{#1}}
\usepackage{lineno}
\usepackage[colorlinks,citecolor=blue,urlcolor=blue,bookmarks=false,hypertexnames=true]{hyperref} 

\modulolinenumbers[5]

\begin{document}

\title{Quantifying operator spreading and chaos in Krylov subspaces with quantum state reconstruction}

	\author{Abinash Sahu}
\email{ommabinash@physics.iitm.ac.in}
\affiliation{Department of Physics, Indian Institute of Technology Madras, Chennai, India, 600036}
\affiliation{Center for Quantum Information, Communication and Computing,
	Indian Institute of Technology Madras, Chennai, India 600036}
\author{Naga Dileep Varikuti}
%\email{vndileep@physics.iitm.ac.in}
\affiliation{Department of Physics, Indian Institute of Technology Madras, Chennai, India, 600036}
\affiliation{Center for Quantum Information, Communication and Computing,
	Indian Institute of Technology Madras, Chennai, India 600036}
 \author{Bishal Kumar Das}
%\email{dasbishalkumar4@gmail.com dasbishalkumar4@gmail.com }
\affiliation{Department of Physics, Indian Institute of Technology Madras, Chennai, India, 600036}
\affiliation{Center for Quantum Information, Communication and Computing,
	Indian Institute of Technology Madras, Chennai, India 600036}
\author{Vaibhav Madhok}
\email{madhok@physics.iitm.ac.in}
\affiliation{Department of Physics, Indian Institute of Technology Madras, Chennai, India, 600036}
\affiliation{Center for Quantum Information, Communication and Computing,
	Indian Institute of Technology Madras, Chennai, India 600036}

%%%%%%%%%%%%%%%%%%%%%%%%%%%%%%%%%%%%%%%%%%%%%%%%%%%%%%%%%%%%%%%%%%%%%%%%%%%%%%%%%%%%%%%%%%

\begin{abstract}
We study operator spreading in many-body quantum systems by its potential to generate an informationally complete measurement record in quantum tomography. We adopt continuous weak measurement tomography for this purpose. We generate the measurement record as a series of expectation values of an observable evolving under the desired dynamics, which can show a transition from integrability to complete chaos. We find that the amount of operator spreading, as quantified by the fidelity in quantum tomography, increases with the degree of chaos in the system. We also observe a remarkable increase in information gain when the dynamics transitions from integrable to nonintegrable. We find our approach in quantifying operator spreading is a more consistent indicator of quantum chaos than Krylov complexity as the latter may correlate/anti-correlate or show no explicit behavior with the level of chaos in the dynamics. We support our argument through various metrics of information gain for two models: the Ising spin chain with a tilted magnetic field and the Heisenberg XXZ spin chain with an integrability-breaking field. Our paper gives an operational interpretation for operator spreading in quantum chaos.
\end{abstract}

\maketitle

\section{Introduction}

Operator spreading characterizes a process in which the Heisenberg evolution of a local operator under the dynamics of a many-body Hamiltonian extends over the entire system \cite{von2018operator}. The operator spreading also serves as a probe for scrambling of quantum information that is inaccessible to local measurements. {Once the information is scrambled, the information is now delocalized over the entire operator space in complex observables.} Thus, operator spreading is also connected to the understanding of the questions of chaos, nonintegrability, and thermalization in many-body quantum systems \cite{deutsch1991quantum, srednicki1994chaos, tasaki1998quantum, rigol2008thermalization, rigol2010quantum, torres2013effects}. Intense research has been directed towards the study of operator spreading in various fields such as black hole physics \cite{hayden2007black, sekino2008fast, hosur2016chaos, shenker2014black, mcginley2022quantifying}, holography \cite{bhattacharyya2022quantum}, integrable systems \cite{
xu2020does, rozenbaum2020early, pilatowsky2020positive}, random unitary circuits \cite{nahum2018dynamics, nahum2018operator, khemani2018operator, rakovszky2018diffusive}, quantum field theories \cite{roberts2015diagnosing, stanford2016many, chowdhury2017onset, patel2017quantum}, and chaotic spin-chains \cite{luitz2017information, heyl2018detecting, lin2018out, geller2022quantum}. 

Quantum chaos is the study of signatures of classical chaos in quantum systems whose classical counterpart is chaotic. For many quantum systems, operator spreading is a reliable indicator of chaos in the dynamics \cite{moudgalya2019operator, omanakuttan2023scrambling}. One can quantify the spreading of the operator through out-of-time-ordered correlators (OTOCs) \cite{maldacena2016bound, maldacena2016bound, swingle2018unscrambling, seshadri2018tripartite, prakash2020scrambling, xu2020accessing, sreeram2021out, varikuti2022out}, operator entanglement \cite{nie2019signature, wang2019barrier, alba2019operator, styliaris2021information}, memory matrix formalism \cite{mcculloch2022operator} or Krylov complexity \cite{parker2019universal, yates2021strong, rabinovici2021operator,  noh2021operator, dymarsky2021krylov, caputa2022geometry, rabinovici2022krylov, avdoshkin2022krylov, rabinovici2022k, bhattacharya2022operator, bhattacharya2023krylov, suchsland2023krylov}. OTOCs, which measure the incompatibility between a stationary operator and another operator evolving with time in the Heisenberg picture, have been studied extensively to witness operator growth. However, measuring OTOCs in the laboratory is challenging even with state-of-the-art experimental techniques, which require backward evolution in time, that is, the ability to completely reverse the Hamiltonian \cite{li2017measuring, green2022experimental, garttner2017measuring, zhu2016measurement, swingle2016measuring, yao2016interferometric, halpern2017jarzynski, bohrdt2017scrambling, tsuji2017exact, nie2020experimental, dressel2018strengthening, joshi2020quantum, asban2021interferometric}. {Recently, certain protocols have been discussed where one can avoid the problem of backward time evolution \cite{vermersch2019probing, blocher2022measuring, sundar2022proposal}.}
{Furthermore, the Krylov complexity, another quantifier of operator spreading, is computed when the operator at a given time is expressed in an orthonormal sequence of operators generated from the Lanczos algorithm. The Liouvillian superoperator of a time-independent Hamiltonian is repeatedly applied on the initial operator to construct the Krylov basis. In the Appendices, we have detailed the procedure for obtaining Krylov subspace in the Lanczos algorithm and quantifying Krylov complexity. Nevertheless, the saturation value of the Krylov complexity depends on the choice of initial observable. Also, the initial growth of Krylov complexity has been observed to be exponential for certain non-chaotic dynamics \cite{dymarsky2021krylov, avdoshkin2022krylov}. Thus, the Krylov complexity does not serve as an unambiguous indicator of chaos.} 

%%%%%%%%%%%%%%%%%%%%%%%%% Graphical Abstract %%%%%%%%%%%%%%%%%%%%%%%%%
%%%%%%%%%%%%%%%%%%%%%%%%%%%%%%%%%%%%%%%%%%%%%%%%%%%%%%%%%%%%%%%%%%%%%%%

 \begin{figure*}[htbp]
  	%\centering   
  	\centering
    \includegraphics[scale=0.65]{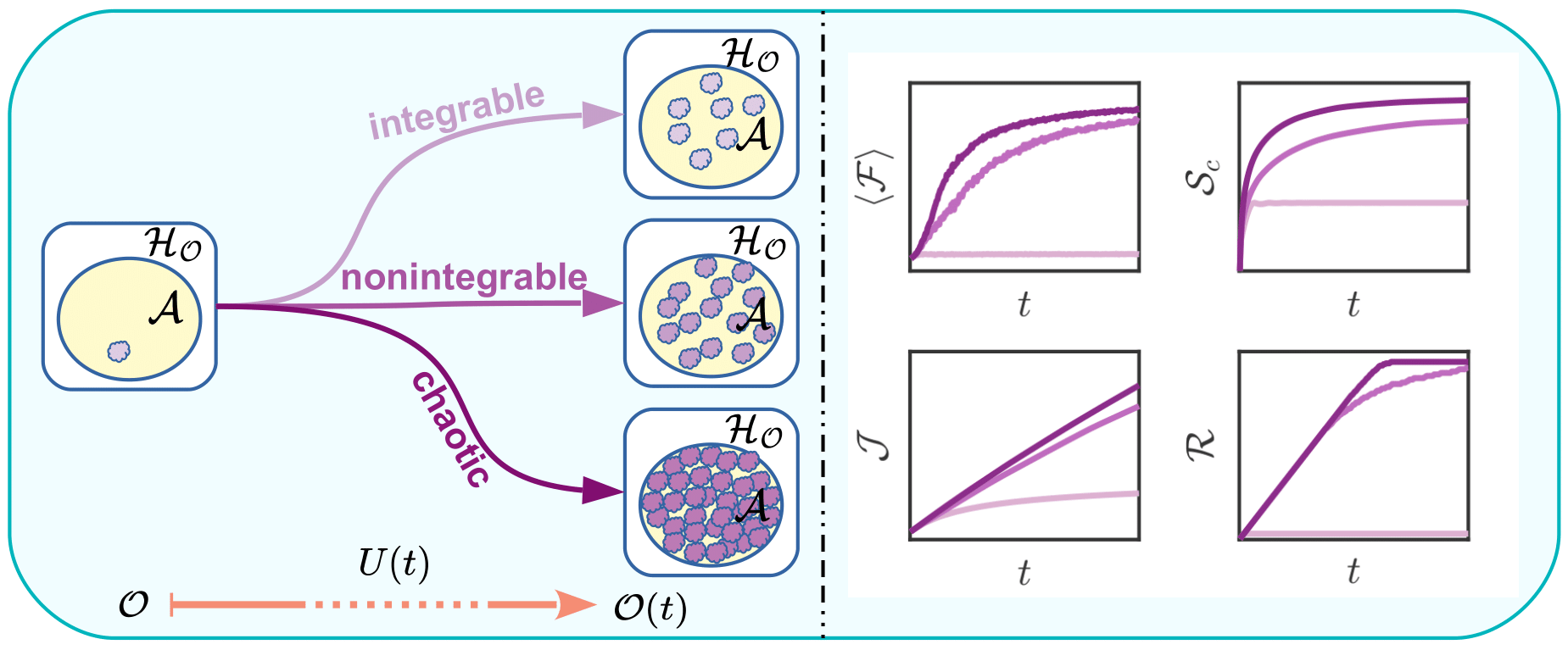}
  	
  	\caption {An illustration of operator spreading in the Hilbert space of operators $\mathcal{H}_{\mathcal{O}}$. {The initial operator $\mathcal{O}$ evolves under the system dynamics and generates a set of operators that spans the subspace $\mathcal{A}$.} The dimension of the subspace is more as the dynamics becomes nonintegrable and finally chaotic. However, fully chaotic dynamics helps to span the largest subspace possible with $\mathrm{dim}(\mathcal{A})=d^2-d+1$. In this paper, we quantify the operator spreading through the rate of information gain in quantum tomography with certain information-theoretic quantifiers like average reconstruction fidelity $\langle\mathcal{F}\rangle$, Shannon entropy $\mathcal{S}_c$, Fisher information $\mathcal{J}$, and rank of covariance matrix $\mathcal{R}$. }
  	\label{fig:graphical}
  \end{figure*}
  %%%%%%%%%%%%%%%%%%%%%%%%%%%%%%%%%%%%%%%%%%%%%%%%%%%%%%%%%%%%%
  %%%%%%%%%%%%%%%%%%%%%%%%%%%%%%%%%%%%%%%%%%%%%%%%%%%%%%%%%%%

In this paper, we take an alternate route and quantify operator spreading through the performance of a concrete quantum information processing task quantum tomography.
How does the system dynamics drive operator complexity, which affects the information gain in quantum tomography?
We answer the above question by quantifying operator spreading in integrable, nonintegrable, and chaotic many-body systems via their ability to generate an optimal measurement record for quantum tomography. Intuitively, an evolution of a fiducial operator with a single random unitary will lead to maximal operator spreading over the entire operator space \cite{merkel2010random, sreeram2021quantum}. 
{Krylov subspace for operators is generated by repeated application of a map to an initial operator. Thus, such a random unitary evolution will also saturate the maximum possible dimension of the Krylov subspace. In our paper, the operators of Krylov subspace obtained from an initial operator upon time evolution have a simple interpretation. These are elements in the operator Hilbert space that will be measured in tomography.} How many of these directions and with what signal-to-noise ratio they are measured as the dynamics becomes nonintegrable and increasingly chaotic will give an operational and physically motivated way to quantify operator spreading as illustrated in Fig. \ref{fig:graphical}. 

The Lyapunov exponents quantify the rapid divergence rate of neighboring trajectories in a classically chaotic system. The quantum counterpart of these diverging trajectories is the growth of incompatibility of operators as quantified by the OTOCs which give the ``quantum Lyapunov exponents" of the dynamics. Therefore, we unify the connections between information gain, scrambling, and chaos through an actual physical process. 

 A connection between tomography which is about information gain of an unknown state and chaos, which represents unpredictability, seems to be at odds with each other.
However, a deeper analysis reveals fundamental connections. 
Classically chaotic systems show exponential sensitivity to perturbations of the initial conditions, as measured by the Kolmogorov-Sinai (KS) entropy \cite{pesin1977characteristic} which is equal to the sum of positive Lyapunov exponents of the system. On the flip side, KS entropy also measures the rate at which successive measurements on a classically chaotic system provide information about the initial condition. The missing information in deterministic chaos is the initial condition. A time history of a trajectory at discrete times is an archive of information about the initial conditions given perfect knowledge about the dynamics. Moreover, if the dynamics is chaotic the rate at which we learn information increases is given by the KS entropy. This is precisely analogous to quantum tomography where the missing information is the unknown initial state of the quantum system.

% Previous studies have shown how the information gain is connected to the degree of chaos in the dynamics \cite{madhok2014information, madhok2016characterizing, sahu2022effect}. \textcolor{red}{cite more of us} 

In this paper, we consider the dynamics of the 1D Ising model with a tilted magnetic field \cite{prosen2007chaos, pineda2007universal, kukuljan2017weak, karthik2007entanglement, prosen2007efficiency}, and the 1D anisotropic Heisenberg XXZ model with an integrability-breaking field \cite{santos2004integrability, santos2004entanglement, barivsic2009incoherent, rigol2010quantum, santos2011domain, santos2012chaos, gubin2012quantum, brenes2018high, brenes2020low, pandey2020adiabatic} to study the growth of operator spreading and its connection to chaos in quantum tomography. They manifest a range of behavior from integrable to fully chaotic. The Hamiltonian of the Ising model with either a time-independent tilted field or a time-dependent delta-kicked tilted field shows integrable to chaos transition, and we explore both models for our tomography process.

This paper is organized as follows. Sec. \ref{Background} briefly reviews the continuous measurement tomography protocol. We then describe both the models we have considered in this paper: the Ising model with a tilted magnetic field and the Heisenberg XXZ model with an integrability-breaking field. In Sec. \ref{results}, the heart of the manuscript, we quantify the operator spreading using various information-theoretic quantifiers by connecting it to the rate of information gain in tomography for both models. We also show that our quantifiers for operator spreading are independent of the choice of the initial operator and act as an unambiguous way of measuring chaos, unlike Krylov complexity. We conclude in Sec. \ref{conclusion} with a summary and some final remarks.

%%%%%%%%%%%%%%%%%%%%%%%%%%%%%%%%%%%%%%%%%%%%%%%%%%%%%%%%%%%%%%%%%%%%%%%%%%%%%%%%%%%%%%%%%%%%%%%%%%%%%%%%%%%%%%%%%%%%%%%%%%%%%%%%
\section{BACKGROUND}
\label{Background}

\subsection{Continuous weak measurement tomography} 
%In this subsection we outline the protocol for continuous weak measurement tomography~\cite{silberfarb2005quantum, smith2006efficient, riofrio2011quantum, smith2013quantum, merkel2010random, madhok2014information, madhok2016characterizing, sreeram2021quantum, sahu2022effect, sahu2022quantum}. 
Quantum state tomography uses the statistics of measurement records on an ensemble of identical systems to best estimate an unknown quantum state $\rho_0$ \cite{paris2004quantum, d2003quantum}. {Strong projective measurements 
 of an informationally complete set of observables have traditionally been used to extract information for state reconstruction. It is a time-consuming and tedious procedure when applied to systems of large dimensions. Also, one needs to reprepare the ensemble and reconfigure the measurement apparatus after each measurement in certain experimental settings \cite{smith2006efficient}. On the other hand, weak measurements help in reducing the amount of resources required for the reconstruction process as they cause
minimal disturbance to the state.  In this paper, we are interested in continuous weak measurement tomography~\cite{silberfarb2005quantum, smith2006efficient, riofrio2011quantum, smith2013quantum, merkel2010random, madhok2014information, madhok2016characterizing, sreeram2021quantum, sahu2022effect, sahu2022quantum}. Over a period of time, one can generate an informationally complete set of measurement records by continuously probing an ensemble of identically prepared, collectively and coherently evolved systems.}

A time series of operators is generated by evolving a physical observable under the dynamics of a many-body system in the Heisenberg picture. We exploit this choice of dynamics for time-evolution and explore operator spreading across the entire system or in the Hilbert space of operators. {An ensemble of $N_{s}$ identical systems undergo separable time evolution by a unitary $U(t)$ where all the systems of the ensemble evolve independently under the chosen dynamics. In our paper, we consider an ensemble of systems consisting of $N_s$ number of identical tilted field Ising spin chains or Heisenberg XXZ models. 
The collective observable $\sum_{j}^{N_s}\mathcal{O}^{(j)}$ is a sum of single system operators, which is being evolved under the collective dynamics and probed. We use $\mathcal{O}$, a single system operator, for all our calculations since all the systems evolve independently under the chosen dynamics. We generate the measurement record from weak continuous measurement of an observable $\mathcal{O}$ through a probe coupled to the ensemble of identical systems.} For sufficiently weak coupling, the randomness of the measurement outcomes is dominated by the quantum noise (shot noise) in the probe rather than the quantum fluctuations of measurement outcomes intrinsic to the state (known as projection noise). In such a case, the backaction on the quantum state is negligible throughout the measurement, and the state of the ensemble remains approximately separable. We can write the measurement record as
\begin{equation}
 M(t)=X(t)/{N_{s}}=Tr[\mathcal{O}(t)\rho_0]+W(t),
 \label{tom_noise}
\end{equation}
where $W(t)$ is a Gaussian white noise with spread $\sigma/N_{s}$, and $\mathcal{O}(t)=U^{\dag}(t)\mathcal{O}U(t)$ is the time evolved operator in Heisenberg picture.

A generalized Bloch vector $\bf r$ describes any arbitrary density matrix of Hilbert-space dimension $\mathrm{dim}(\mathcal{H})=d$, when expressed in an orthonormal basis of $d^2-1$ traceless Hermitian operators $\{E_\alpha\}$ as $\rho_0=I/d+\Sigma^{d^2-1}_{\alpha=1}\ r_\alpha E_\alpha$.  We consider the measurement record at discrete times as $M_n=M(t_n)=Tr(\mathcal{O}_n\rho_0)+W_n$, that allows one to express the measurement history as
  \begin{equation}
  {\bf M}=\tilde{\mathcal{O}}{\bf r}+{\bf W},
  \label{ms}
  \end{equation}
  where $\tilde{\mathcal{O}}_{n\alpha}=\tr(\mathcal{O}_{n}E_\alpha)$. {In the negligible backaction limit, the probability distribution associated with measurement history $\bf M$ for a given state vector $\bf r$ is Gaussian with spread $\sigma$ \cite{silberfarb2005quantum,smith2006efficient}
\begin{equation} 
\begin{split}
p({\bf M|r}) & \varpropto \mathrm{exp}\ \Big\{-\frac {N_{s}^2}{2\sigma^2}\sum_{i}[M_i-\sum_{\alpha}\tilde{\mathcal{O}}_{i\alpha}r_\alpha]^2\Big\}
\\
& \varpropto \mathrm{exp}\ \Big\{-{\frac {N_{s}^2}{2\sigma^2}\sum_{\alpha,\beta}({\bf r}-{\bf r}_\mathrm{ML})_\alpha\ C^{-1}_{\alpha\beta}\ ({\bf r}-{\bf r}_\mathrm{ML})_\beta\Big\}},
\label{likelihood}
\end{split}
\end{equation}
where ${\bf C^{-1}=(\tilde{\mathcal{O}}^T\tilde{\mathcal{O}}})$ is the inverse of the covariance matrix and the inverse is Moore-Penrose pseudo inverse~\cite{ben2003generalized} in general. The peak of the distribution is the maximum likelihood estimate ${\bf r}_\mathrm{ML}$ of the unknown Bloch vector parameters $\{r_{\alpha}\}$ which is equal to the least-square fit as given by
  \begin{equation}
  {\bf r}_\mathrm{ML}=\bf C\tilde{\mathcal{O}}^{T}M.
  \label{rml1}
  \end{equation}}

 In the presence of measurement noise, or when the measurement record is incomplete, the estimated Bloch vector ${\bf r}_\mathrm{ML}$ may represent an unphysical density matrix $\rho_{ML}$ with negative eigenvalues. Therefore, we impose the positivity constraint on the reconstructed density matrix and obtain the physical state closest to the maximum likelihood estimate, which is the most consistent with our measured data. We employ a convex optimization procedure~\cite{vandenberghe1996semidefinite}, a semidefinite program to obtain the final estimate of the Bloch vector $\bf \bar{r}$ by minimizing the argument
  \begin{equation}
  \|{\bf r}_\mathrm{ML}-{\bf \bar{r}}\|^2=({\bf r}_\mathrm{ML}-{\bf \bar{r}})^T{\bf C}^{-1}({\bf r}_\mathrm{ML}-\bf \bar{r})
  \end{equation}
  subject to the constraint $$I/d+\Sigma^{d^2-1}_{\alpha=1}\ \bar{r}_\alpha E_\alpha\geq0.$$
  The performance of the quantum state tomography protocol is quantified by the fidelity of the reconstructed state $\bar{\rho}$ relative to the actual state $\ket{\psi_0},$ $\mathcal{F}=\bra{\psi_0}\bar{\rho}\ket{\psi_0}$ as a function of time. The reconstruction fidelity $\mathcal{F}$ depends on the informational completeness of the measurement record \cite{merkel2010random, madhok2014information, sreeram2021quantum}, the choice of observables and quantum states \cite{sahu2022effect} as well as the presence of noise in the measurement outcomes \cite{sahu2022quantum}.
 %%%%%%%%%%%%%%%%%%%%%%%%%%%%%%%%%%%%%%%%%%%%%%%%%%%%%%%%%%%%%%%%

\subsection{Models}
\label{model}
%%%%%%%%%%%%%%%%%%%%%% figure %%%%%%%%%%%%%%%%%%%%%
\begin{figure}[htbp]
\centering
\includegraphics[scale=0.34]{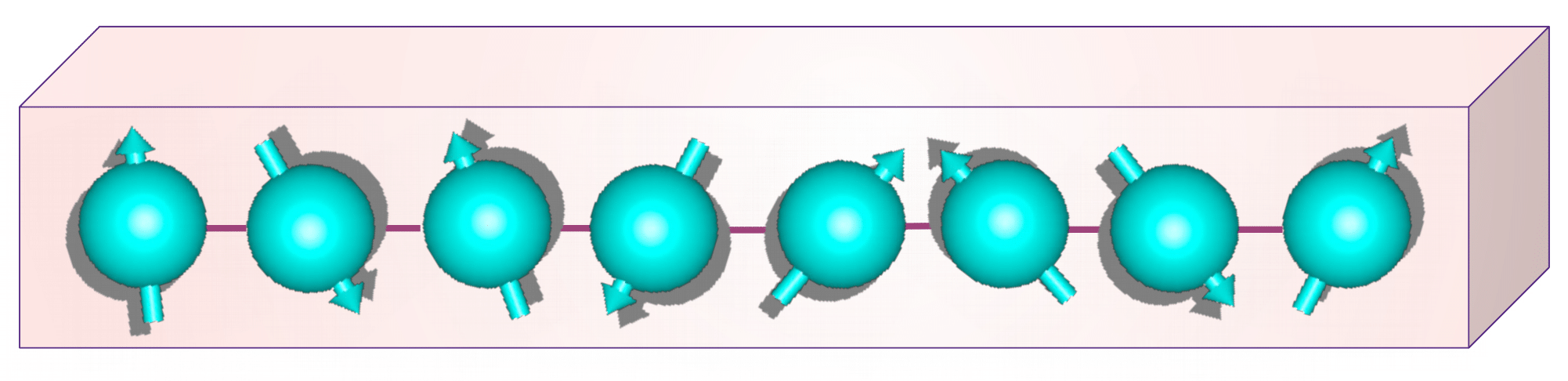}
\caption{Many-body system of spins with nearest-neighbor interaction. One can drive the system from integrable to fully chaotic by tailoring the strength of integrability-breaking fields applied at suitable sites in certain directions.}
\label{spinchain}
\end{figure}
%%%%%%%%%%%%%%%%%%%%%%%%%%%%%%%%%%%%%%%%%%%%%%%%%%%%
We consider two different many-body quantum systems as spin chains as shown in Fig. \ref{spinchain}. The models are Ising spin chain with a tiled magnetic field and Heisenberg XXZ spin chain with an integrability-breaking field as we describe below.
\subsubsection{\bf Ising spin chain with a tilted magnetic field}
The Hamiltonian of the tilted field Kicked Ising model consists of the nearest neighbor Ising interaction term, and the system is periodically kicked with a spatially homogenous but arbitrarily oriented magnetic field \cite{prosen2007chaos, pineda2007universal, kukuljan2017weak}. The Hamiltonian for tilted field kicked Ising model for $L$ spins  is given by
\begin{equation}
 H_{TKI}=\sum_{j=1}^L\big\{J\sigma_{j}^{z}\sigma_{j+1}^{z}+
\big(h_z\sigma_{j}^{z}+h_x\sigma_{j}^{x}\big)\sum_{n}\delta(t-n)\big\},
\end{equation}
where $\sigma_j^{\alpha}$ are the Pauli spin matrices with $\alpha=x, y, z$. This Hamiltonian has three parameters: the Ising coupling $J$, the transverse magnetic field strength $h_x$, and the longitudinal magnetic field strength $h_z$. The Floquet map for the tilted field kicked Ising model for a time period of $\tau=1$ is
\begin{equation}
 U_{TKI}=e^{-iJ\sum_{j}\sigma_{j}^{z}\sigma_{j+1}^{z}}\ e^{-i\sum_{j}
(h_z\sigma_{j}^{z}+h_x\sigma_{j}^{x})}.
\label{tkising}
\end{equation}

We consider the free boundary condition for the model. {The model is integrable when either $h_x$ or $h_z$ is zero.} The Hamiltonian $H_{TKI}$ is integrable for $h_z=0$ because of the Jordan-Wigner transformation. There is another non-trivial completely integrable regime found in the tilted field kicked Ising model when the magnitude of the magnetic field is an integer multiple of $\pi/2$, i.e. $h=\sqrt{h_x^2+h_z^2}=n\pi/2, n\in\mathbb{Z}$ \cite{prosen2007chaos}. Nevertheless, the model is nonintegrable in a general case of a tilted magnetic field when both the components $h_x$ and $h_z$ are non-vanishing, and $2h/\pi$ is non-integer. We fix the Ising coupling strength $J=1$, the transverse magnetic field strength $h_x=1.4$, and vary the longitudinal magnetic field strength $h_z$ to tune the nonintegrability of the dynamics. As we increase the value of $h_z$, the system becomes nonintegrable. Thus, for $h_z=0.4$, the system is weakly nonintegrable, and it will become strongly nonintegrable for $h_z=1.4$, i.e., when the strengths of longitudinal and transverse fields become comparable. 

Interestingly, time dependence is not necessary for making the dynamics nonintegrable. {The tilted field Ising model is nonintegrable for non-zero values of $h_z$ even though there are no delta kicks \cite{prosen2007efficiency}. Here, the system is strongly nonintegrable for a small value of $h_z$ even when $h_z$ and $h_x$ are not of comparable strength \cite{karthik2007entanglement}. Thus, we choose $h_z=0.1$ for the system to be weakly nonintegrable, and for a higher value of $h_z$, it will obey the random matrix theory predictions.} The Hamiltonian for the tilted field Ising model without delta kicks can be expressed as
\begin{equation}
 H_{TI}=\sum_{j=1}^L\big\{J\sigma_{j}^{z}\sigma_{j+1}^{z}+
h_z\sigma_{j}^{z}+h_x\sigma_{j}^{x}\big\}.
\end{equation}
$H_{TI}$ is a time-independent Hamiltonian, so the time evolution unitary operator for this model for time $t$ is given by
\begin{equation}
 U_{TI}(t)=e^{-it\sum_j\{J\sigma_{j}^{z}\sigma_{j+1}^{z}+
h_z\sigma_{j}^{z}+h_x\sigma_{j}^{x}\}}.
\label{tising}
\end{equation}
{For our current work, we explore time-dependent and time-independent models to relate the information gain in tomography to the operator spreading and compare them with random matrix theory.}

\subsubsection{\bf Heisenberg XXZ spin chain with an integrability-breaking field}

The 1D anisotropic Heisenberg XXZ spin chain is an integrable model with nearest-neighbor interaction, which can be proved by Bethe ansatz \cite{shastry1990twisted, cazalilla2011one}. The Hamiltonian for the Heisenberg XXZ spin chain is 
\begin{equation}
    H_{XXZ}=\sum_{j=1}^L \frac{J_{xy}}{4} \big\{ \sigma_{j}^{x}\sigma_{j+1}^{x} + \sigma_{j}^{y}\sigma_{j+1}^{y} \big\} + \frac{J_{zz}}{4} \sigma_{j}^{z}\sigma_{j+1}^{z},
\end{equation}
where $s_j^{\alpha}=1/2 \sigma_j^{\alpha}$.
There are various ways in which we can make the XXZ model nonintegrable. One can introduce a single magnetic impurity at one of the sites \cite{santos2004integrability, santos2011domain, barivsic2009incoherent, brenes2020low, pandey2020adiabatic, rabinovici2022k}, a global staggered transverse field \cite{brenes2018high} or next-to-nearest-neighbour interaction \cite{santos2012chaos, gubin2012quantum, rabinovici2022k} to make the dynamics nonintegrable. In this paper, we consider the single magnetic impurity at one of the sites and explore the operator spreading with an increase in the integrability-breaking field strength. The Hamiltonian for this nonintegrable Heisenberg model is 
\begin{equation}
    H_{HNI}=H_{XXZ}+\frac{g}{2} H_{si},
\end{equation}
where the integrability-breaking field with strength $g$ is $H_{si}=\sigma_l^z$, for site $j=l$. For our analysis, we consider $J_{xy}=1$, $J_{zz}=1.1$, and vary the strength of the integrability-breaking field $g$ as the chaoticity parameter. While changing the value of $g$ from 0 to 1, the fully integrable XXZ model becomes chaotic, which is clear from the level statistics and other properties \cite{santos2004integrability, rabinovici2022k}. The time evolution unitary for time $t$ for this nonintegrable time-independent Hamiltonian is 
\begin{equation}
    U_{HNI}=e^{-it(H_{XXZ}+{g/2}H_{si})}.
\end{equation}

{The XXZ model with periodic boundary condition respects many symmetries, including translation symmetry in the space because of conservation of linear momentum \cite{joel2013introduction}, and we can find many degenerate states \cite{santos2004integrability}.} Thus, we choose the free boundary condition for the XXZ model. {For a spin chain with a very large number of spins, the boundary conditions have no effects, but for numerical calculations, we have to take a finite number of spins. However, even in the deep quantum regime,  we can still witness integrability to chaos transition.} The Hamiltonian has a reflection symmetry about the center of the chain if a single impurity is placed at the center of the chain. {The Hamiltonian $H_{HNI}$ commutes with the total spin along $z$ direction $S_z=\frac{1}{2}\sum_{j=1}^L\sigma_j^{z}$ which makes the system invariant under rotation around the $z-$ axis \cite{joel2013introduction}.}

%%%%%%%%%%%%%%%%%%%%%%%%%%%%%%%%%%%%%%%%%%%%%%%%%%%%%%%%%%%%%%%%%%%%%%%%%%%%%%%%%%%%%%%%%%%%%%%%%%%%%%%%%%%%%%%%%%%%%%%%%%%%%%%%%%%%%%%%%%%%%%%%%%%%%%%%%%%%%%%%%%%%%%%%%%%%%%

\section{Information gain as a paradigm: operator complexity, nonintegrability, and chaos}
\label{results}
In this section, we come to the central question we ask. What are the consequences of operator spreading in quantum information theory? We use continuous weak measurement tomography as a paradigm to study the operator spreading. The measurement record is acquired as expectation values of operators generated by the Heisenberg evolution of a chosen dynamics. We exploit the freedom of choosing the dynamics to explore and explain the effect of chaos in the operator spreading. For our analysis, we consider both time-dependent delta kicked and time-independent 1D tilted field Ising model dynamics and the dynamics of 1D anisotropic Heisenberg XXZ spin chain with an integrability-breaking field to investigate the operator complexity through various information-theoretic metrics. We also relate our information-theoretic way of quantifying operator complexity to the Krylov complexity.

{Krylov subspace is generated by repeated application of a map $\mathcal{M}_K$ on an initial operator $\mathcal{O}$ as $\mathcal{A}=\mathrm{span}\{\mathcal{O},\ \mathcal{M}_K(\mathcal{O}),\ \mathcal{M}^2_K(\mathcal{O}),\ \mathcal{M}^3_K(\mathcal{O}), ...\}$. Here, we are interested in studying the Krylov subspace of linear operators for the Hilbert space $\mathcal{H}$. We have outlined the Lanczos algorithm for constructing Krylov subspace in the Appendix \ref{krylovsubspace}. The dimension of the operator Hilbert space is $\mathrm{dim}(\mathcal{H}_{\mathcal{O}})=d^2-1$.} {However, the maximum dimension of Krylov subspace is $d^2-d+1$, which leaves out a subspace of dimension at least $d-2$ from $\mathcal{H}_{\mathcal{O}}$  [see Appendix \ref{krylov_dim} for proof as given in Ref. \cite{merkel2010random} also Ref. \cite{rabinovici2021operator, rabinovici2022krylov} for an alternate proof]. In this paper, we use a unitary map for certain dynamical systems to generate the Krylov subspace that helps to quantify operator spreading due to the desired dynamics.} We apply a single parameter unitary map $U$ repeatedly to get the time evolved operator after $n$ time steps
\begin{equation}
 \mathcal{O}_n=U^{\dagger n}\mathcal{O}U^n.
 \label{unievol}
\end{equation}
In the superoperator picture, one can write the operator $\mathcal{O}_n$ as
\begin{equation}
 \supket{\mathcal{O}_n}=\mathcal{U}^n_K\supket{\mathcal{O}},
 \label{supOP}
\end{equation}
where $\mathcal{U}_K=U^{\dagger}\otimes U^T$. Thus, we get the Krylov subspace, which is $\mathrm{span}\{\supket{\mathcal{O}_n}\}$, and quantify the operator spreading through various metrics.

We choose $L$ qubit random states with Hilbert space dimension $d=2^L$ for state reconstruction. We evolve an initial local operator $\mathcal{O}$ and get the archive of operators that help in state reconstruction. In the beginning, the observable $\mathcal{O}$ can be a local observable with access to the spin at site $j$; hence, it does not gain any information about other sites. However, we need an informationally complete set of global observables to reconstruct any arbitrary random pure states. We notice that the reconstruction fidelity increases with time, which implies the growth or spread of the initial local operator across the spin chain as a complex operator. Thus, the average reconstruction fidelity serves as a quantifier for the operator complexity. The average is taken over an ensemble of $80$ random pure states sampled from the Haar measure on SU$(d)$, where the Hilbert space dimension $d=2^L$. 

To further quantify the operator complexity, we study certain information-theoretic metrics. The covariance matrix of the joint probability distribution Eq. (\ref{likelihood}) determines the information gain in the continuous measurement tomography protocol. We have the inverse of the covariance matrix as ${\bf C^{-1}}=\tilde{\mathcal{O}}^T\tilde{\mathcal{O}}$. Thus, in the superoperator picture, we can write
\begin{equation}
 {\bf C^{-1}}=\sum_{n=1}^N{\supket{\mathcal{O}_n}\supbra{\mathcal{O}_n}},
\end{equation}
where $\supket{\mathcal{O}_n}$ are produced by repeated application of the Floquet map as given in Eq. (\ref{supOP}) or by applying the time evolution unitary of a time-independent Hamiltonian for time $t=n$. Each eigenvector of $\bf C^{-1}$ represents an orthogonal direction in the operator space we have measured until the final time $t=N$. The eigenvalues of $\bf C^{-1}$ give us the signal-to-noise ratio in that orthogonal direction. {Given a finite time, the operator dynamics needs to be unbiased to get equal information in all the directions of the operator space, which requires the eigenvalues of $\bf C^{-1}$ to be equal. Thus, the information gain in tomography for random states is maximum when all the eigenvalues are equal in magnitude \cite{madhok2014information} [refer to Appendix \ref{MI_Shannon} for details].} One can normalize the eigenvalues to get a probability distribution from the eigenvalue spectrum. Shannon entropy quantifies the bias of this distribution as
\begin{equation}
 \mathcal{S}_c=-\sum_i{\lambda_i}\ln\lambda_i,
 \label{shn_ent}
\end{equation}
where $\{\lambda_i\}$ is the normalized eigenvalue spectrum of $\bf C^{-1}$ \cite{madhok2014information,sreeram2021quantum}. The Shannon entropy is maximum when the eigenvalues are uniformly distributed. The Shannon entropy increases with time and saturates at a higher value. The saturation value of Shannon entropy is higher when the dynamics is fully nonintegrable, and the saturation value increases with an increase in the degree of chaos. As the dynamics becomes chaotic, the operators spread uniformly in the operator space because of the ergodicity. Thus, Shannon entropy $\mathcal{S}_c$ of the spectrum $\{\lambda_i\}$ can be used as a quantifier for operator complexity. 
Krylov entropy $\mathcal{S}_K$ as quantified in Eq. (\ref{kent}) [see Appendix \ref{krylovsubspace} for the details], measures the complexity of the time-evolved operator in the Krylov basis generated from the Lanczos algorithm. In contrast, Shannon entropy $\mathcal{S}_c$ determines the spreading of the operators along the orthogonal directions in the operator space measured till time $t=N$. Shannon entropy $\mathcal{S}_c$ is a very good quantifier of operator complexity that comes naturally while doing a quantum information processing task, the quantum tomography. 

During the reconstruction process, the average Hilbert-Schmidt distance between the true and estimated state in quantum tomography is equal to the  total uncertainty in the Bloch vector components \cite{hradil02}
\begin{equation}
 \mathcal{D}_{HS} =\langle \tr[(\rho_0 - \bar{\rho})^2]\rangle = \sum_\alpha\langle (\Delta r_\alpha)^2 \rangle,
\end{equation}
where $\Delta r_\alpha=r_\alpha- \bar{r}_{\alpha}$.
The Cramer-Rao inequlaity, $\langle (\Delta r_\alpha)^2 \rangle \ge \left[ \mbf{F}^{-1} \right]_{\alpha \alpha}$, relates these uncertainties to the   the Fisher information matrix, $\mbf{F}$, associated with the conditional probability distribution, Eq. (\ref{likelihood}), and thus $\mathcal{D}_{HS} \ge \tr[\mbf{F}^{-1}]$.  Our probability distribution is a multivariate Gaussian regardless of the state, which helps the bound to saturate in the limit of negligible backaction. In that case, the Fisher information matrix equals the inverse of the covariance matrix, $\mbf{F} = \mbf{C}^{-1}$, in units of $N_{s}^2/\sigma^2, $  {  where  $\mbf{C}^{-1}=\mathcal{\tilde{O}}^T \mathcal{\tilde{O}},$ and $\mathcal{\tilde{O}}_{n\alpha}= \tr [\mathcal{O}_n E_\alpha]$} \cite{madhok2014information}. Thus, a metric for the total information gained in tomography is the inverse of this uncertainty,
\begin{equation}
\mathcal{J} = \frac{1}{\tr[\mbf{C}]}  
\end{equation}
which measures the total Fisher information. The inverse covariance matrix is never full rank in this protocol. We regularize $\mbf{C}^{-1}$ by adding to it a small fraction of the identity matrix (see, e.g., \cite{boyd2004convex}). For pure states,  the average Hilbert-Schmidt distance $\mathcal{D}_{HS} =1/\mathcal{J}= 1-\langle \tr\bar{\rho}^2\rangle -2 \langle \mathcal{F} \rangle$ \cite{hradil02}. Fisher information is related to the average reconstruction fidelity; hence, it can be used as a quantifier for the efficiency of the tomography protocol. 

%%%%%%%%%%%%%%%%%%%%%%% figure %%%%%%%%%%%%%%%%%%%%%%%%%
\begin{figure*}[htbp]
	\centering
	\includegraphics[scale=0.523]{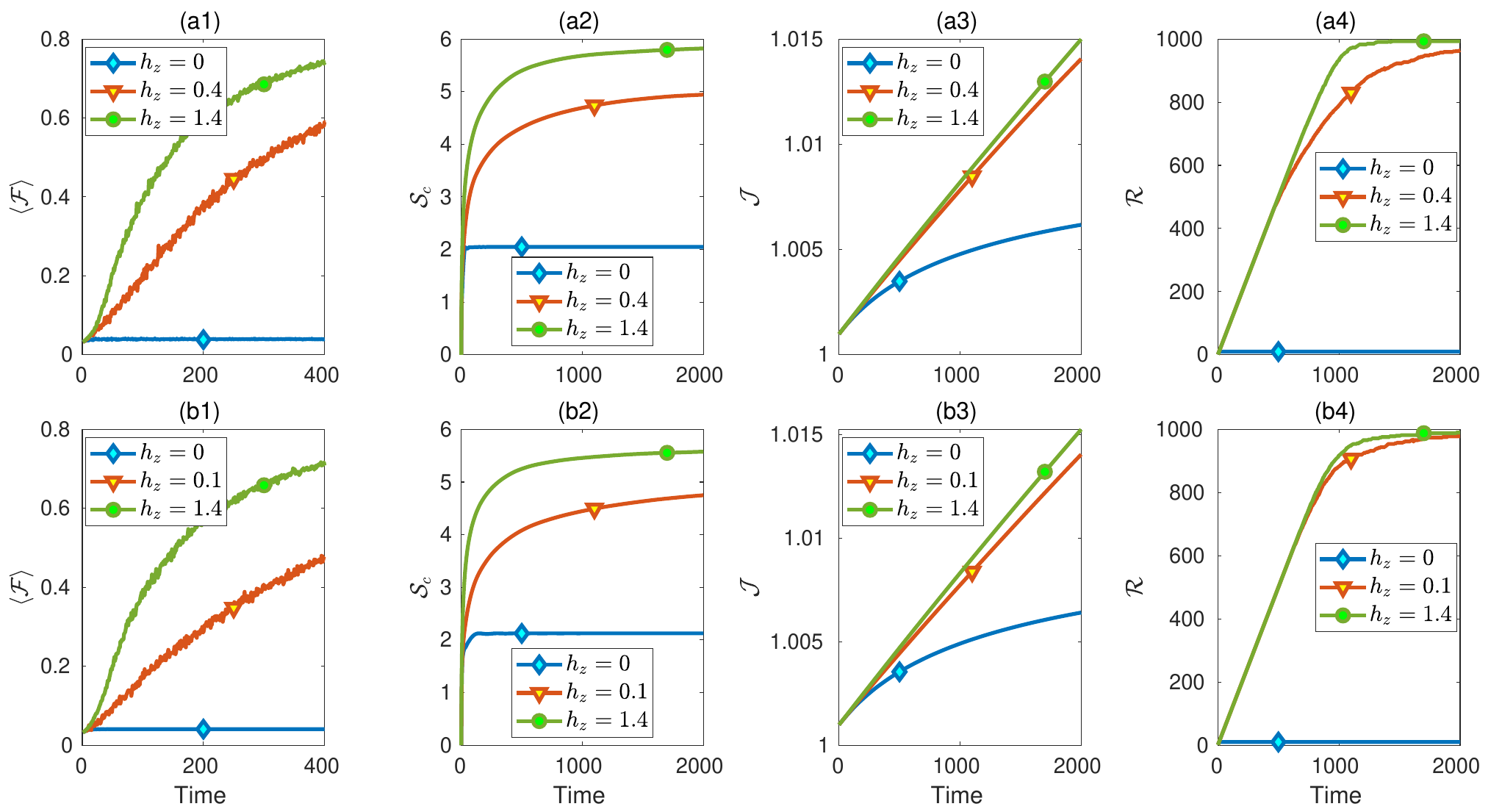}
	\caption{Quantifying operator spreading through various information-theoretic metrics as a function of time with an increase in the extent of chaos. The time series of operators are generated by repeated application of the Floquet map of the time-dependent tilted field kicked Ising model  $U_{TKI}$ as shown in Eq. (\ref{tkising}) for plots (a1)-(a4). The unitary for time-independent tilted field Ising model Eq. (\ref{tising}) generates the time evolved operators for plots (b1)-(b4). All numerical simulations are carried out for the Ising model of $L=5$ spins with $J=1$, $h_x=1.4$, and the initial observable $s_1^{y}$. (a1 and b1) Average reconstruction fidelity $\langle\mathcal{F}\rangle$ as a function of time. (a2 and b2) The Shannon entropy $\mathcal{S}_c$ of the normalized eigenvalues of the inverse of the covariance matrix of the likelihood function. (a3 and b3) The Fisher information $\mathcal{J}$ for parameter (Bloch vector components) estimation. (a4 and b4) Rank $\mathcal{R}$ of the covariance matrix. In all cases, the values of the quantifiers are higher for higher nonintegrability parameter $h_z$. }
	\label{IsingTomo5}
\end{figure*}
%%%%%%%%%%%%%%%%%%%%%%% figure %%%%%%%%%%%%%%%%%%%%%%%%

%%%%%%%%%%%% figure %%%%%%%%%%%%%%%%%%%%%%%
\begin{figure*}[htbp]
\centering
 \includegraphics[scale=0.58]{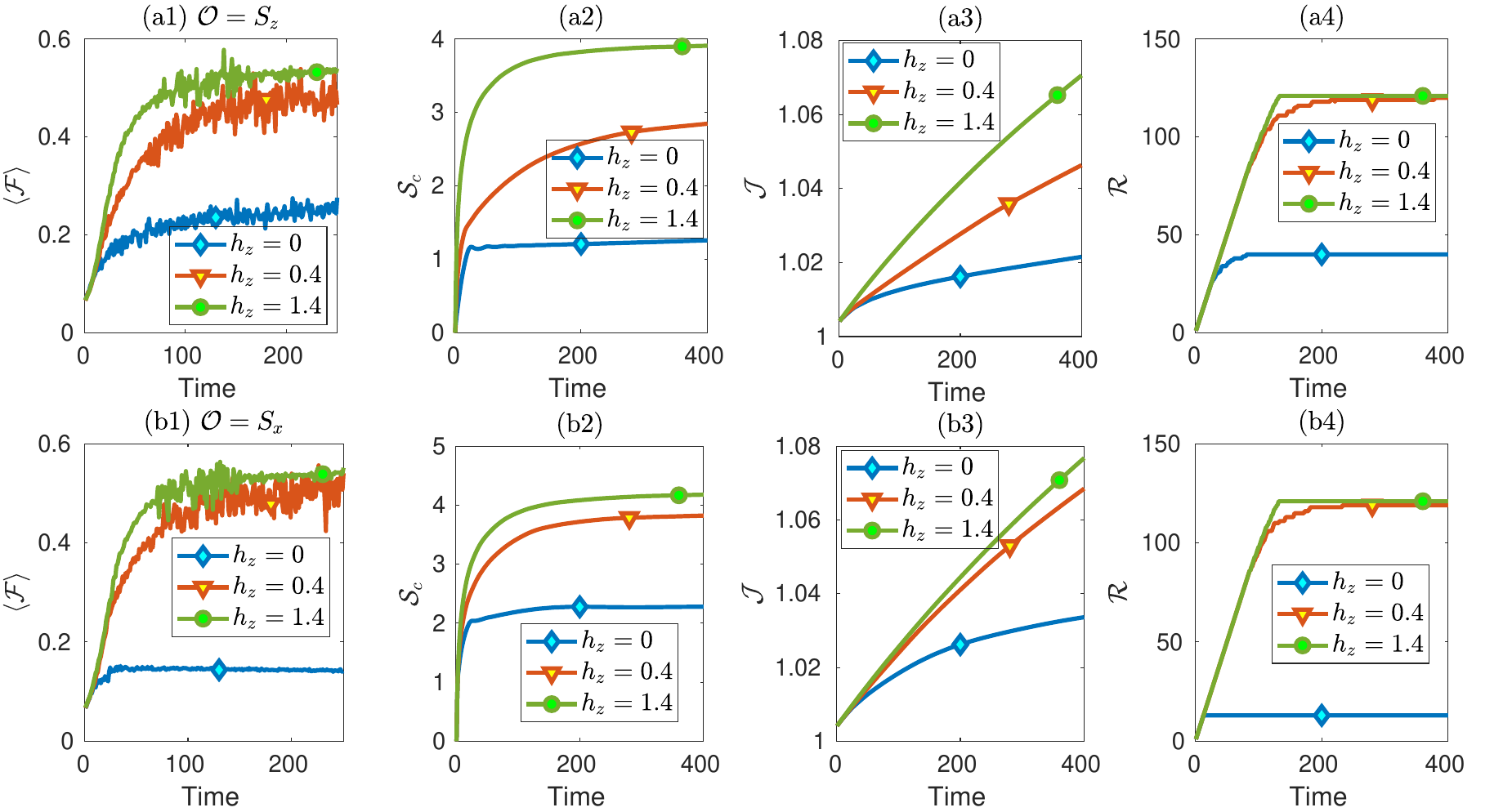}
 \caption{Quantifying operator spreading through various information-theoretic metrics as a function of time with an increase in the strength of the nonintegrability field. The time series of operators are generated by repeated application of the Floquet map of the time-dependent tilted field kicked Ising model  $U_{TKI}$ as visible in Eq. (\ref{tkising}) for all plots. All numerical simulations are carried out for the Ising model of $L=4$ spins with $J=1$, $h_x=1.4$. Plots (a1 - a4) are for the initial observable $S_z$, and plots (b1 - b4) are for the initial observable $S_x$. (a1 and b1) Average reconstruction fidelity $\langle\mathcal{F}\rangle$ as a function of time. (a2 and b2) The Shannon entropy $\mathcal{S}_c$ of the normalized eigenvalues of the inverse of the covariance matrix of the likelihood function. (a3 and b3) The Fisher information $\mathcal{J}$ for parameter (Bloch vector components) estimation. (a4 and b4) Rank $\mathcal{R}$ of the covariance matrix. In all cases, the values of the quantifiers are higher for higher nonintegrability parameter $h_z$. }
 \label{plotIsing4xz}
\end{figure*}
%%%%%%%%%%%%% figure %%%%%%%%%%%%%%%%%%%%

To further elucidate the operator spreading, we calculate the rank of the covariance matrix $\mathcal{R}$. The rank of the covariance matrix determines the dimension of the operator space spanned under the evolution of the system dynamics. Repeated application of a single parameter unitary can generate $K\le d^2-d+1$ number of linearly independent operators \cite{merkel2010random}. Therefore, the rank of the covariance matrix is $\mathcal{R}\le d^2-d+1$, and the maximum rank of the covariance matrix increases with an increase in the extent of chaos, and we can adopt it as a measure of operator spreading. {Our quantifiers work very well for various models irrespective of the choice of initial observables. Here, we have shown the results for certain generic observables. However, these results are valid for other observables as well. In Fig. \ref{plotRMT_ising} of Appendix \ref{app_RMT_Ising}, we have illustrated our findings even for a local random initial observable. We have examined various local and global observables; however, they are not presented in this paper.}

%%%%%%%%%%%%%%%%%%% figure %%%%%%%%%%%%%%%%%
\begin{figure*}[htbp]
	\centering
	\includegraphics[scale=0.58]{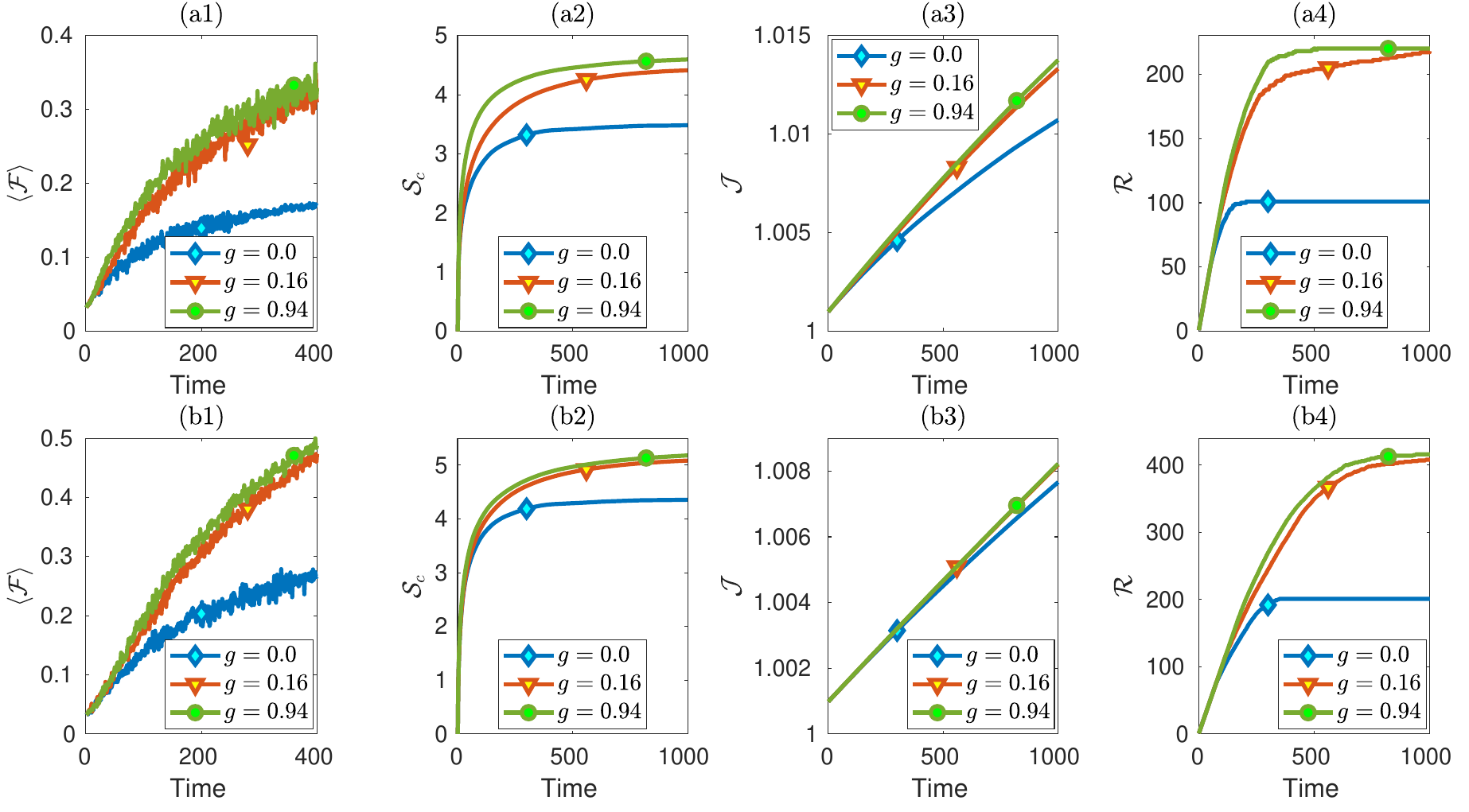}
	\caption{Operator spreading through various information-theoretic metrics as a function of time with an increase in the degree of chaos for Heisenberg XXZ spin chain with an integrability-breaking field $H_{si}=s^y_3$. All numerical simulations are carried out for the Hamiltonian $H_{HNI}$ of $L=5$ spins with $J_{xy}=1$, $J_{zz}=1.1$. Plots (a1 - a4) are for the initial local observable $\mathcal{O}=s^{y}_2+s^y_4$, and plots (b1 - b4) are for the initial observable $\mathcal{O}=s^y_2$. (a1 and b1) Average reconstruction fidelity $\langle\mathcal{F}\rangle$ as a function of time. (a2 and b2) The Shannon entropy $\mathcal{S}_c$ of the normalized eigenvalues of the inverse of the covariance matrix of the likelihood function. (a3 and b3) The Fisher information $\mathcal{J}$ for parameter (Bloch vector components) estimation. (a4 and b4) Rank $\mathcal{R}$ of the covariance matrix. In all cases, the values of the quantifiers are higher for higher integrability-breaking parameter $g$ value.}
	\label{plotHeisenXXZ}
\end{figure*}
%%%%%%%%%%%%%%%%%% figure %%%%%%%%%%%%%%%%%%

\subsection{Results for Ising spin chain with a tilted magnetic field}

We have considered an ensemble of systems consisting of $N_s$ number of identical tilted field Ising spin chains with $L$ number of spins. Thus, $\mathcal{O}$ acts on the site $j$ of each of the systems for collective evolution and measurement. We evolve an initial local operator $\mathcal{O}=s_1^y$ to generate the measurement record. In the beginning, the observable $\mathcal{O}=s_1^y$ has access to the spin at site $j=1$; hence, it does not gain any information about other sites. We notice that the reconstruction fidelity increases with time, as it is apparent in Fig. \ref{IsingTomo5} (a1) for the time-dependent tilted field kicked Ising model and Fig. \ref{IsingTomo5} (b1) for time-independent tilted field Ising model. Thus, the average reconstruction fidelity serves as a quantifier for the operator spreading. {For the kicked Ising model we use the parameters $h_z=\{0.0, 0.4, 1.4\}$, and for the time-independent Ising model we consider the parameter set $h_z=\{0.0, 0.1, 1.4\}$ since the later becomes strongly nonintegrable for a small value of $h_z$ even when $h_z$ and $h_x$ are not of comparable strength \cite{karthik2007entanglement}.} It is evident from Fig. \ref{IsingTomo5} (a2) and (b2) that the saturation value of Shannon entropy is more when the dynamics is fully nonintegrable, and the saturation value increases with an increase in the degree of chaos. Figures \ref{IsingTomo5} (a3) and \ref{IsingTomo5} (b3) display the Fisher information for random states as a function of time with an increase in the level of chaos. We can observe how the rise in Fisher information is correlated with the chaos in the dynamics, making it fit as a quantifier for operator complexity. It is pretty clear from Fig. \ref{IsingTomo5} (a4) and (b4) that the rank is more when the dynamics is chaotic as opposed to when it is integrable. The dimension of the Krylov subspace $K$ also matches with the maximal $\mathcal{R}$ when the dynamics is fully chaotic [see Fig. \ref{Lanczos} of Appendix \ref{FOalgorithm}, and Fig. \ref{IsingTomo234} of Appendix \ref{appnd_Ising} for this]. Thus, rank $R$ represents a natural measure for operator spreading. {In Fig. \ref{plotRMT_ising} of Appendix \ref{app_RMT_Ising}, we show the values of our information-theoretic quantifiers in the fully chaotic regime are consistent with random matrix theory.}

Operator spreading measured from the saturation value of Krylov complexity $\mathcal{C}_K$ as quantified in Eq. (\ref{kcomplexity}) depends on the choice of initial observable (see ref \cite{espanol2023assessing}). Previously it is demonstrated that the late-time saturation value of Krylov complexity correlates with the level of chaos for some operators like the collective spin operator $S_x=\frac{1}{2}\sum_{j=1}^L\sigma_j^{x}$, and anti-correlates with some other operators like $S_z=\frac{1}{2}\sum_{j=1}^L\sigma_j^{z}$ [see Fig. \ref{complexity_plot} in Appendix \ref{krylovsubspace}]. Also, there are operators for which the Krylov complexity does not exhibit any systematic behavior with the level of chaos. Remarkably, the information-theoretic measures we use here give us unambiguous signatures of chaos, as is evident in Fig. \ref{plotIsing4xz}. We have shown the behavior of the information-theoretic quantifiers for the collective spin operators $S_x$ and $S_z$. The rate of information gain in tomography for random states increases with an increase in the level of chaos in the dynamics, which is valid for any initial physical observable. We have not desymmetrized the Hamiltonian in this paper or considered any symmetric subspace for our calculations. { Here, the collective spin operators respect the reflection symmetry of the tilted field Ising model. The set of operators from the time evolution of $S_x$ or $S_z$ will not generate the informationally complete set of operators as the operators will be restricted to respective symmetric subspaces. Thus, one can not achieve informationally completeness of the measurement record, which leads to a lower value of reconstruction fidelity, and the saturation value of $\mathcal{R}$ is less than the maximum value.} One has to study the effects of symmetry on the reconstruction closely to have a better understanding. Nevertheless, the Shannon entropy $\mathcal{S}_c$, the Fisher information $\mathcal{J}$, and the rank of the covariance matrix $\mathcal{R}$ serve as natural measures for operator complexity through a concrete physical task.

\subsection{Results for Heisenberg XXZ spin chain with an integrability-breaking field}
Here, we illustrate the operator spreading for the Heisenberg XXZ model for $L=5$ number of spins with an integrability-breaking field. The single impurity $H_{si}=s^z_3$ is placed at the center of the spin chain, and the strength of the field $g$ is varied to drive the dynamics from integrable to chaotic. {We choose the parameter set $g=\{0.0, 0.16, 0.94\}$, where the dynamics is fully chaotic and level statistics follow random matrix predictions for $g=0.94$ \cite{rabinovici2022krylov}.} We have considered two initial observables $\mathcal{O}=s^y_2+s^y_4$, and $\mathcal{O}=s^y_2$. The observable $\mathcal{O}=s^y_2+s^y_4$ respects the reflection symmetry about the center of the spin chain, whereas the operator $\mathcal{O}=s^y_2$ does not respect the reflection symmetry. However, both the initial observables do not have support over the full spin chain. In Fig. \ref{plotHeisenXXZ}, we have shown the average reconstruction fidelity $\langle\mathcal{F}\rangle$, the Shannon entropy $\mathcal{S}_c$, the Fisher information $\mathcal{J}$ and the rank of covariance matrix $\mathcal{R}$ for both the initial observables. We notice that the operator spreading is more when the dynamics becomes more chaotic irrespective of the choice of initial observable. {We can see the effects of symmetries in the saturation value of all the quantities other than the Fisher information, which does not saturate. Fisher information, which is highly sensitive to vanishingly small eigenvalues associated with the covariance matrix, is not very sensitive to the chaoticity parameter, unlike the Shannon entropy. While we observe that the Fisher information computed for the fully chaotic and weakly nonintegrable dynamics can be close to each other, it certainly does preserve the correlation with the degree of chaos.}

\section{Discussion}
\label{conclusion}

Characterizing chaos in the quantum world and its manifestations in quantum information processing is currently being vigorously pursued. In this paper, we explore quantum chaos and its connections to operator spreading in many-body quantum systems. We connect the operator spreading to the rate of information gain in quantum tomography - a protocol at the heart of quantum information processing. Our work gives an operational interpretation of operator spreading and relates/contrasts to the Krylov complexity in the study of quantum chaos.

The operator, which is initially localized, will evolve under the chaotic dynamics and spread in the operator space as a more complex operator. %This work shows an operational interpretation of operator spreading through continuous weak measurement quantum tomography. 
Interestingly, various information-theoretic measures like Shannon entropy, Fisher information, and rank of the covariance matrix not only quantify the information gain but also support us in assessing the operator complexity. We show an unambiguous way of measuring operator complexity and operator scrambling as the dynamics becomes chaotic for the 1D Ising model with a tilted magnetic field and the 1D anisotropic Heisenberg XXZ spin chain with an integrability-breaking field. The rate of operator scrambling is positively correlated with the rate of information gain for random states, which increases with an increase in the level of chaos in the dynamics. Our information-theoretic quantifiers are suitable for both time-independent as well as time-dependent Hamiltonians.

The idea of operator spreading in the operator space can be compared to the exploration of the classical trajectory in the classical phase space. Kolmogorov-Sinai (KS) entropy is known to quantify the rate of exploration in the classical phase space, which is equal to the sum of positive Lyapunov exponents. Here, we can easily relate the rate of the operator spreading to the degree of chaos in the dynamics through our information-theoretic metrics. Thus, the information gain in quantum tomography is connected to the operator spreading in the operator space and the KS entropy.

Simulating quantum chaos on a quantum computer \cite{lysne2020small, munoz2020simulating, krithika2023nmr, maurya2022control} and exploring its information-theoretic signatures like operator scrambling is an exciting avenue being vigorously pursued. 
Our work, connecting continuous measurement tomography and information scrambling, paves the way to realize and interpret such experiments in the laboratory \cite{smith2006efficient, chaudhury2009quantum}.

%It is challenging to build a universal quantum simulator that can imitate any quantum process, including the physical world, as proposed by Feynmann. \textcolor{blue}{However, if one expects the quantum simulator to simulate certain interesting physical systems that classical computers can not achieve, it would be easier to construct \cite{lysne2020small, munoz2020simulating, krithika2023nmr, maurya2022control}.} Operator scrambling is such a quantum phenomenon that can be simulated via continuous measurement tomography with state-of-the-art quantum simulators \cite{smith2006efficient, chaudhury2009quantum}. Our work paves the way to realize operator spreading like various other exciting quantum phenomena through certain concrete physical tasks \cite{sahu2022quantum, sahu2022effect}.
%%%%%%%%%%%%%%%%%%%%%%%%%%%%%%%%%%%%%%%%%%%%%%%%%%%%%%%%%%%%%%%%%%%%%%%%%%%%%%%%%%%%%%%%%%%%%%%%%%%%%%%%%%%%%%%%%%%%%%%%%%%%%%%%%%
\section*{ACKNOWLEDGEMENTS}
We are grateful to Arul Lakshminarayan for useful discussions. We thank Sreeram PG for the helpful discussions. We acknowledge the anonymous referees for their valuable feedback. We thank HPCE, IIT Madras, for providing the computational facility for numerical simulations. This work was supported in part by Grant No.  DST/ICPS/QusT/Theme-3/2019/Q69 and a new faculty Seed Grant from IIT Madras. The authors were supported, in part, by a grant from Mphasis to the Centre for Quantum Information, Communication, and Computing (CQuICC) at IIT Madras.

%\newpage
\appendix

\section{Krylov subspace from Lanczos algorithm and operator complexity}
\label{krylovsubspace}
Generally, Krylov subspace for operators is generated by repeated application of a map to an initial operator, and its span helps to quantify operator spreading due to the dynamics.
Here, we outline the method to generate the Krylov basis through the Lanczos algorithm. In the Lanczos scheme, the generator of the Krylov basis is a Hermitian operator (a Hamiltonian $H$) \cite{viswanath2008recursion}.  %In the Arnoldi iteration method \cite{arnoldi1951principle}, a unitary operator $U$ generates the Krylov subspace.%$\mathcal{A}_A$.
The time evolution of an operator $\mathcal{O}$ is given by 
\begin{equation}
 \mathcal{O}(t)=e^{iHt}\mathcal{O}e^{-iHt}=e^{i\mathcal{L}t}\mathcal{O},
 \label{Hevol}
\end{equation}
where the Liouvillian operator is defined as $\mathcal{L}=[H,.].$ In the superoperator notation the operator $\mathcal{O}$ is written as $\supket{\mathcal{O}}$, which allows us to write the Eq. (\ref{Hevol}) as
\begin{equation}
 \supket{\mathcal{O}(t)}= \sum_{k=0}^{\infty} \frac{(it)^k}{k!}\mathcal{L}^k\supket{\mathcal{O}}.
\end{equation}
Now, one can orthonormalize the operators $\{\mathcal{L}^k\supket{\mathcal{O}}\}_{k=0}^{\infty}$ using Gram-Schmidt orthogonalization procedure to get the Krylov basis $\mathcal{A}_L=\{\supket{\mathcal{Q}_k}\}_{k=0}^{K-1}$, where $K$ is the dimension of Krylov subspace and $K\le d^2-d+1$ \cite{rabinovici2021operator,rabinovici2022krylov}. This is an iterative process to get the orthonormal basis known as Lanczos algorithm for a well-defined Hilbert-Schmidt inner product $\supbraket{\mathcal{Q}_k}{\mathcal{Q}_l}=\tr(\mathcal{Q}_k^{\dagger}\mathcal{Q}_l)=\delta_{kl}$. In this process the first two basis operators are $\supket{\mathcal{Q}_0}=\supket{\mathcal{O}}/\supbraket{\mathcal{O}}{\mathcal{O}}^{1/2}$ and $\supket{\mathcal{Q}_1}=b_1^{-1}\mathcal{L}\supket{\mathcal{Q}_0}$, where $b_1=\sqrt{|\mathcal{L}\supket{\mathcal{Q}_0}|^2}$. Following the Lanczos algorithm, we can get the other basis operators $\supket{\mathcal{Q}_k}$ in a recursive method as
\begin{equation}
\supket{\mathcal{Q}_k}=b_k^{-1}\big(\mathcal{L}\supket{\mathcal{Q}_{k-1}}-b_{k-1}\supket{\mathcal{Q}_{k-2}}\big),
 \label{Liouvillian}
\end{equation}
where $b_k=\sqrt{|\mathcal{L}\supket{\mathcal{Q}_{k-1}}-b_{k-1}\supket{\mathcal{Q}_{k-2}}|^2}$ are called the Lanczos coefficients. 

\begin{figure}[htbp]
\centering
 \includegraphics[scale=0.56]{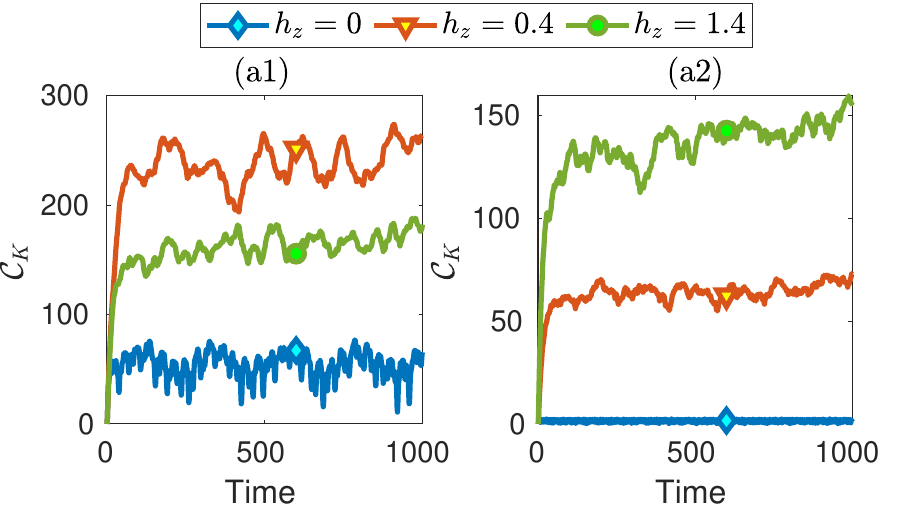}
 \caption{Krylov complexity $\mathcal{C}_K$ as a function of time for the dynamics of 1D Ising model with a tilted magnetic field for an increase in the field strength $h_z$. We have considered $L=5$ spins with $J=1$ and $h_x=1.4$ for the numerical simulations. Two different initial observables (a1) $\mathcal{O}=S_z$ and (a2) $\mathcal{O}=S_x$. For generating this plot, we have not desymmetrized the Hamiltonian.}
 \label{complexity_plot}
\end{figure}

The Liouvillian operator is a tridiagonal matrix in the Krylov basis, as it is apparent from Eq. (\ref{Liouvillian}). Thus, we can express the time-evolved operator in the Krylov basis as
\begin{equation}
 \supket{\mathcal{O}(t)}=\sum_k^{K-1}{i^k\varphi_k(t)\supket{\mathcal{Q}_k}},
\end{equation}
where $\varphi_k=i^{-k}\supbraket{\mathcal{Q}_k}{\mathcal{O}(t)}$ are the time-dependent real probability amplitudes that describe the distribution of the time-evolved operator over the Krylov basis. Recently, specific features of this probability distribution have been explored to quantify nonintegrability and chaos in the dynamics \cite{parker2019universal, rabinovici2021operator,  noh2021operator, caputa2022geometry, rabinovici2022krylov, rabinovici2022k, bhattacharya2022operator, bhattacharya2023krylov, espanol2023assessing}. Krylov complexity is a measure of the average position of the operator distribution on the ordered Krylov basis, which is defined as
\begin{equation}
 \mathcal{C}_K(t)=\sum_{k=0}^{K-1}k|\varphi_k(t)|^2.
 \label{kcomplexity}
\end{equation}
In the thermodynamic limit, the Krylov complexity $\mathcal{C}_K$ grows exponentially with time initially, and the Lanczos coefficients grow linearly as $b_k\propto k$ when the dynamics is chaotic\cite{parker2019universal,caputa2022geometry}. For finite-dimensional systems, the long-time saturation value of the Krylov complexity is higher when the dynamics is chaotic, which is identified as a signature of chaos \cite{rabinovici2022k}. However, these signatures depend highly on the choice of initial observable \cite{espanol2023assessing} as shown in Fig. \ref{complexity_plot}. Another measure known as Krylov entropy quantifies how evenly the operator is distributed over the Krylov subspace 
\begin{equation}
 \mathcal{S}_K(t)=-\sum_{k=0}^{K-1}|\varphi_k(t)|^2 \log |\varphi_k(t)|^2.
 \label{kent}
\end{equation}
%%%%%%%%%%%%%%%%%%%%% figure %%%%%%%%%%%%%%%%%%%%%%%%%
%%%%%%%%%%%%%%%%%%%% figure %%%%%%%%%%%%%%%%%%%%%%%%%%

Lanczos algorithm is suitable for analytical calculations of the orthogonal operators $\supket{\mathcal{Q}_k}$ and the Lanczos coefficients $b_k$. Unfortunately, it is numerically not feasible to generate the Krylov basis and the coefficients because of the unavoidable errors accumulated from floating-point rounding in the Hilbert-Schmidt inner products. Thus, it is required to use alternative methods to address this issue. Full orthogonalization method \cite{parlett1998thesym} is such a method that performs a brute-force re-orthogonalization of the newly constructed Krylov element with respect to the previous ones at every iteration of the Lanczos algorithm. In this paper, we also present the full orthogonalization method to compute the Lanczos coefficients numerically [see Appendix \ref{FOalgorithm} for the detailed algorithm].

%We now discuss the Arnoldi method of construction for Krylov subspace that is generated from the repeated application of a unitary $U$ on the initial operator $\mathcal{O}$ \cite{arnoldi1951principle,yates2021strong}. For a unitary map $U$, the time evolved operator after $n$ time steps is 
%\begin{equation}
 %\mathcal{O}_n=U^{\dagger n}\mathcal{O}U^n.
 %\label{Arnoldi}
%\end{equation}
%In the superoperator picture, we can write the operator $\mathcal{O}_n$ as
%\begin{equation}
% \supket{\mathcal{O}_n}=\mathcal{U}^n\supket{\mathcal{O}},
 %\label{arnoldi}
%\end{equation}
%where $\mathcal{U}=U^{\dagger}\otimes U^T$.
%The Krylov basis can be obtained by orthonormalizing the operators $\{\supket{\mathcal{O}_n}\}_{n=0}^{K-1}$ using Gram-Schmidt orthogonalization procedure. In this method also, one can have maximum $K\le d^2-d+1$ linearly independent basis operators as we see in the Lanczos algorithm [see Appendix \ref{krylov_dim} for the proof as shown in Ref. \cite{merkel2010random}]. The operator $\mathcal{U}$ is a unitary, unlike the Liouvillian $\mathcal{L}$, which is a Hermitian tridiagonal matrix. In the literature, there are no quantifiers for operator complexity using the operator distribution over the Krylov basis obtained from the Arnoldi iteration method. The next section will give an information-theoretic explanation for operator complexity. The operators are generated from repeated application of a unitary operator as shown in Eq. (\ref{Arnoldi}).

\section{Full orthogonalization algorithm}
\label{FOalgorithm}
Lanczos algorithm, which makes use of the two previous operators in the construction of each Krylov element, encounters numerical instabilities because of the accumulation of errors from the finite precision arithmetic, and the orthogonality of the Krylov basis is lost in a few steps. Residual overlaps between the Krylov elements increase with the number of iterations $k$, giving rise to unreliable Lanczos coefficients $b_k$. The full orthogonalization method \cite{parlett1998thesym} performs a brute-force re-orthogonalization of the newly constructed Krylov element with respect to the previous ones at every iteration of the Lanczos algorithm that ensures the orthonormality of the Krylov basis up to machine precision $\epsilon$. 
Full orthogonalization performs Gram-Schmidt orthogonalization at every iteration in the Lanczos algorithm to ensure orthogonality up to the machine precision $\epsilon$. For optimality, it is better to adopt Gram-Schmidt twice every time. The algorithm reads as follows:
%%%%%%%%%%%%%%%%%%%%%%%%%%%%%%%%%%%%%
%%%%%%%%%%%%%%%%%%%%% figure %%%%%%%%%%%%%%%%%%%%%
\begin{figure}[htbp]
\centering
 \includegraphics[scale=0.53]{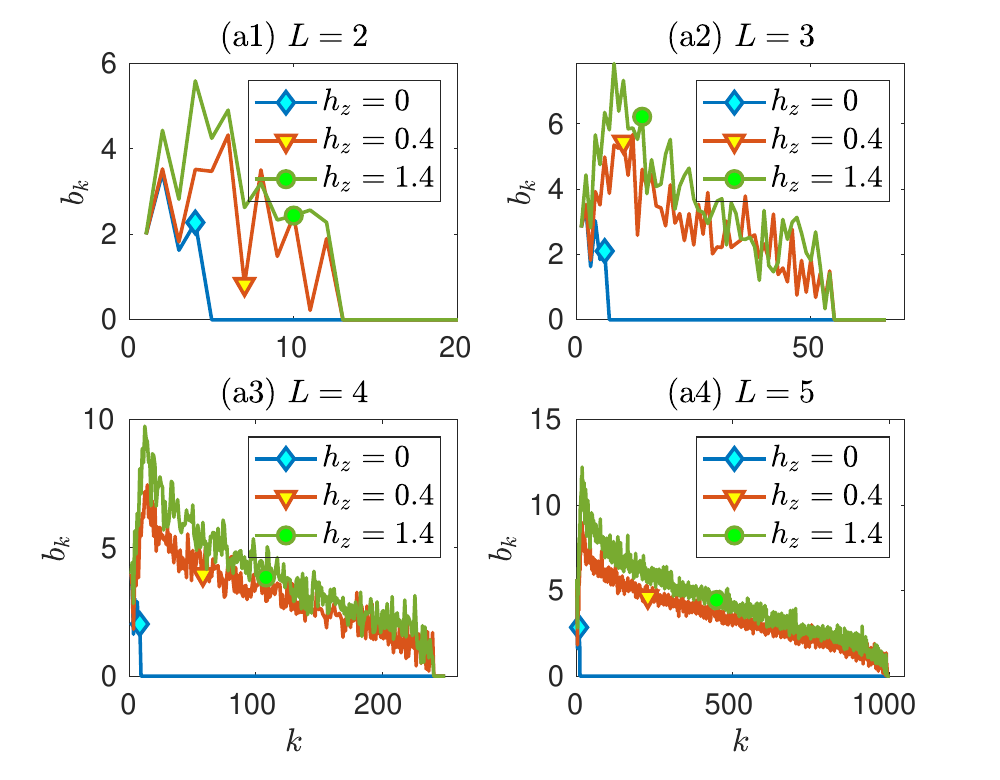}
 \caption{Lanczos coefficients sequence as a function of $k$. The Krylov subspace is obtained for initial observable $s_1^{y}$ using tilted field Ising model Hamiltonian Eq. (\ref{tising}) with $J=1$, and $h_z=1.4$. We vary $h_x$ as the nonintegrability parameter. The Lanczos sequence is plotted for different numbers of spins $L$. $K$ is the dimension of the Krylov subspace when the dynamics is fully nonintegrable. (a1) $L=2$, $K=13$. (a2) $L=3$, $K=57$. (a3) $L=4$, $K=241$. (a4) $L=5$, $K=993$.}
 \label{Lanczos}
\end{figure}
%%%%%%%%%%%%%%%%%%%% figure %%%%%%%%%%%%%%%%%%%%%%
%%%%%%%%%%%%%%%%%%%%%%%%%%%%%%%%%%%%
\begin{itemize}
 \item[1.] $\supket{\mathcal{Q}_0}=\supket{\mathcal{O}}/\supbraket{\mathcal{O}}{\mathcal{O}}^{1/2}.$

 \item[2.] For $k\ge1$: compute $\supket{\mathcal{B}_k}=\mathcal{L}\supket{\mathcal{Q}_{k-1}}$.
 \item[3.] Re-orthogonalize $\supket{\mathcal{B}_k}$ explicitly with respect to all previous Krylov elements: $$\supket{\mathcal{B}_k}\longmapsto \supket{\mathcal{B}_k}-\sum_{m=0}^{k-1}\supket{\mathcal{Q}_m}\supbraket{\mathcal{Q}_m}{\mathcal{B}_k}.$$
 \item[4.] Repeat step $3$.
 \item[5.] Set $b_k=\sqrt{\supbraket{\mathcal{B}_k}{\mathcal{B}_k}}$.
 \item[6.] If $b_k=0$ stop; otherwise set $\supket{\mathcal{Q}_k}=b_k^{-1}\supket{\mathcal{B}_k}$, and go to step $2$.
\end{itemize}
We can find that the values of Lanczos coefficients $b_k$ initially grow as a function of $k$, then drop slowly and become zero at $k=K$, after which the Lanczos algorithm can not generate any more orthogonal operators.
Figure \ref{Lanczos} illustrates the Lanczos sequence for initial observable $\sigma_1^y$ different number of spins of the tilted field Ising model Hamiltonian Eq.(\ref{tising}).

\section{Dimension of Krylov subspace generated from repeated application of a unitary}
\label{krylov_dim}
Here, we delineate the proof of how the repeated application of a single unitary can generate $K\le d^2-d+1$ number of orthogonal operators as demonstrated by Merkel et al. in Ref. \cite{merkel2010random}. The Krylov subspace of operators is obtained as a time series of operators $\mathcal{O}_n=U^{\dagger n}\mathcal{O}U^n$, where $\mathcal{O}$ is a Hermitian operator and $U$ is a fixed unitary. We can determine the dimension of the Krylov subspace of orthogonal operators $\mathcal{A}_A\equiv \mathrm{span}\{\mathcal{O}_n\}$. Let us consider the subspace of operators that are preserved under unitary conjugation by $U$, $\mathcal{G}\equiv \{ \mathrm{g}\in \mathfrak{su}(d)|U\mathrm{g}U^{\dagger}=\mathrm{g}\}$. Let us define $\mathcal{C} \equiv \{ \mathrm{g} \in \mathcal{G}| \tr(\mathrm{g} \mathcal{O})=0\}$. The subspace whose elements are orthogonal to the elements of the Krylov subspace $\mathcal{A}_A$ is $\mathcal{A}_{\perp}$, and operators in this set are not included in the time series of operators. Thus it is clear that $\mathcal{C}\subseteq\mathcal{A}_{\perp}$ since $\forall \mathrm{g} \in \mathcal{C}$
\begin{equation}
    \tr(\mathcal{O}_n \mathrm{g})=\tr [U^{\dagger n}\mathcal{O}U^n \mathrm{g}]=\tr(\mathcal{O}\mathrm{g})=0.
\end{equation}
Now $\mathrm{dim} (\mathcal{A}_A)+\mathrm{dim} (\mathcal{C})\le \mathrm{dim}[\mathfrak{su}(d)]=d^2-1$ as  the two spaces are orthogonal. Thus, for $U$ having non-degenerate eigenvalues $\mathcal{G}$ will be isomorphic to the Cartan subalgebra of $\mathfrak{su}(d)$ which is the largest commuting subalgebra. However, if the eigenspectrum of $U$ has degeneracy, $\mathcal{G}$ will have some additional elements. The dimension of Cartan subalgebra is $d-1$ and hence $\mathrm{dim}(\mathcal{G})\ge d-1$. By definition, $\mathcal{C}$ is obtained from $\mathcal{G}$ by projecting out one direction in operator space and thus $\mathrm{dim}(\mathcal{C})=\mathrm{dim}(\mathcal{G})-1\ge d-2$. Therefore, 
\begin{equation}
    \mathrm{dim}(\mathcal{A}_A)\le \mathrm{dim}[\mathfrak{su}(d)]-\mathrm{dim}(\mathcal{C})\le d^2-d+1.
\end{equation}

%%%%%%%%%%%% figure %%%%%%%%%%%%%%%%%%%%%%%
\begin{figure*}[htbp]
\centering
 \includegraphics[scale=0.55]{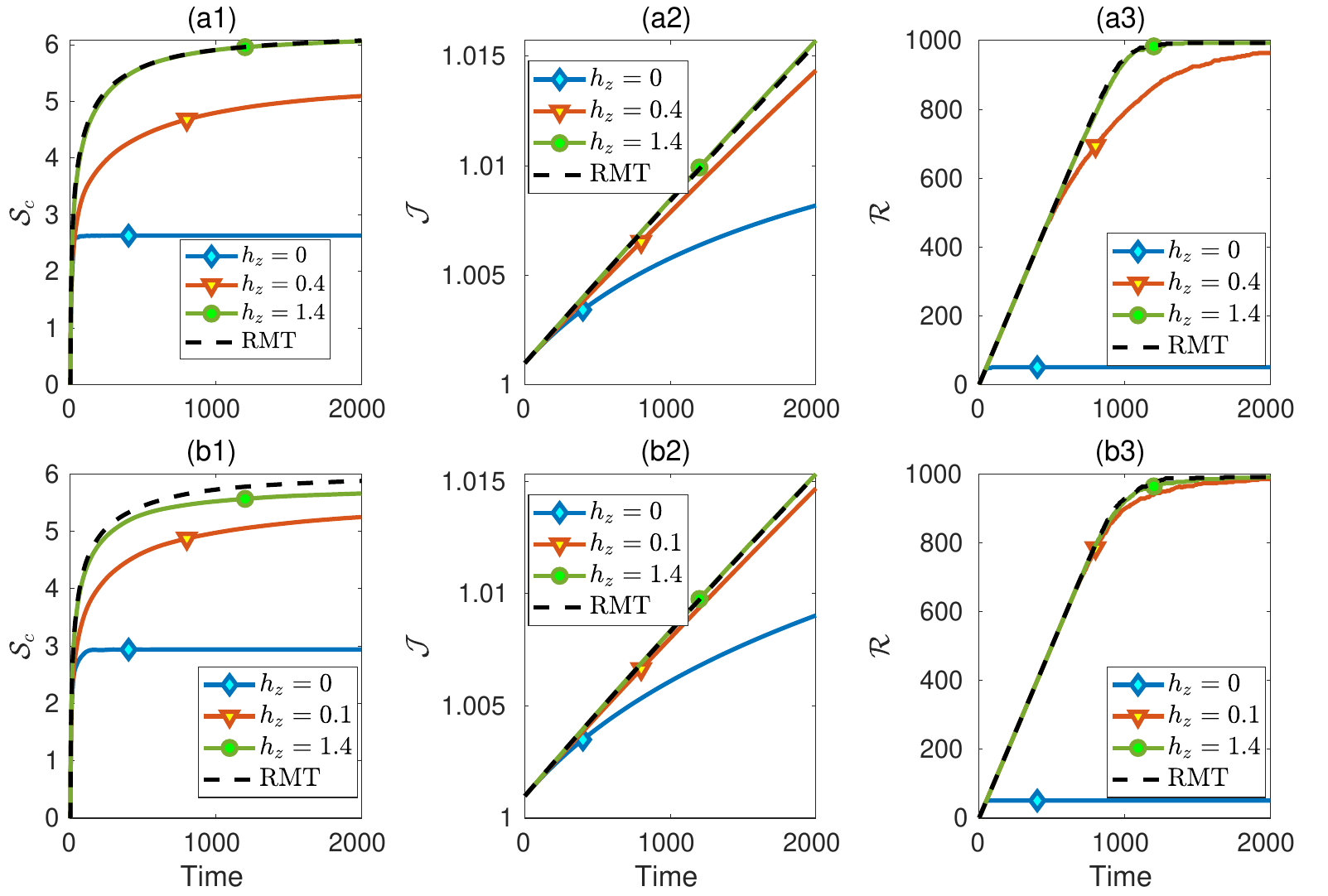}
 \caption{Quantifying operator spreading through various information-theoretic metrics as a function of time with an increase in the extent of chaos. The time series of operators are generated by repeated application of the Floquet map of the time-dependent tilted field kicked Ising model  $U_{TKI}$ as shown in Eq. (\ref{tkising}) for plots (a1)-(a3). The unitary for time-independent tilted field Ising model Eq. (\ref{tising}) generates the time evolved operators for plots (b1)-(b3). All numerical simulations are carried out for the Ising model of $L=5$ spins with $J=1$, $h_x=1.4$, and for an initial random local observable $u_r^{\dagger}s_1^{y}u_r$, where $u_r$ is a single-qubit Haar random unitary. (a1 and b1) The Shannon entropy $\mathcal{S}_c$ of the normalized eigenvalues of the inverse of the covariance matrix of the likelihood function. (a2 and b2) The Fisher information $\mathcal{J}$ for parameter (Bloch vector components) estimation. (a3 and b3) Rank $\mathcal{R}$ of the covariance matrix. In all cases, the values of the quantifiers are higher for higher nonintegrability parameter $h_z$. In the fully chaotic regime, $h_z=1.4$, the results are well predicted by values obtained with a measurement record generated by a Haar-random matrix picked from an appropriate ensemble and are plotted as the dashed line in all plots.}
 \label{plotRMT_ising}
\end{figure*}
%%%%%%%%%%%%%%%%%%%%%% figure %%%%%%%%%%%%%%%%%%%%

{\section{Mutual information and Shannon entropy of the measurement record}
\label{MI_Shannon}
The information obtained about the Bloch vector $\bf r$ from the measurement record $\bf M$ is the mutual information \cite{coverm2006elements}
\begin{equation}
    \mathcal{I}[\textbf{r};\textbf{M}]=\mathcal{S}(\textbf{M})-\mathcal{S}(\textbf{M}|\textbf{r}),
\end{equation}
where $\mathcal{S}$ is the Shannon entropy for the given probability distribution. The entropy of the measurement record, $\mathcal{S}(\textbf{M})$, is entirely due to the shot noise of the probe. Thus, it is a constant irrespective of the state, assuming we have perfect knowledge of the dynamics. One can neglect irrelevant constants to get the mutual information between the Bloch vector and a given measurement record to be specified by the entropy of the conditional probability distribution, Eq. (\ref{likelihood})
\begin{equation}
    \mathcal{I}[\textbf{r};\textbf{M}]=-\mathcal{S}(\textbf{M}|\textbf{r}) = -\frac{1}{2}\log(\det(\textbf{C})) = \log(\frac{1}{V}).
\end{equation}
Here, V is the volume of the error ellipsoid whose semimajor axes are defined by the covariance matrix.
The dynamics that maximizes $1/V=\sqrt{\det(\textbf{C}^{-1})}$ also maximizes the information gain. An important constraint is that after time $t=n$ \cite{madhok2014information},
\begin{equation}
    \tr(\textbf{C}^{-1})=\sum_{i,\alpha}\tilde{\mathcal{O}}_{i\alpha}^2=n\|\mathcal{O}\|^2,
    \label{tr_const}
\end{equation}
where $\|\mathcal{O}\|^2=\sum_{\alpha}\tr(\mathcal{O}E_{\alpha})^2$ is the Euclidean square norm, with initial observable $\mathcal{O}$. The Eq. (\ref{tr_const}) is independent of the choice of dynamics, and the quantity $\tr(\textbf{C}^{-1})$ increases linearly with time. The inequality of arithmetic and geometric means gives us 
\begin{equation}
    \det(\textbf{C}^{-1})\le \left(\frac{1}{d^2-1}\tr(\textbf{C}^{-1}) \right)^{d^2-1}= \left(\frac{n}{d^2-1}\|\mathcal{O}\|^2 \right)^{d^2-1},
\end{equation}
where the rank of the regularized covariance matrix is $d^2-1$. The maximum possible value of mutual information is attained when all eigenvalues are equal, and the above inequality is saturated, making the error ellipsoid a hypersphere. At a given time step, the largest mutual information is achieved when the dynamics mixes the eigenvalues most evenly. We quantify this by Shannon entropy $\mathcal{S}_c$ of the normalized eigenvalue spectrum of $\textbf{C}^{-1}$ as given in Eq. (\ref{shn_ent}). One can extract the maximum information about a random state by measuring all components of the Bloch vector with maximum precision. Given finite time, we can obtain the best estimate by dividing equally between operators in all directions of the operator space.}
{
\section{Random matrix predictions for information-theoretic quantifiers}
\label{app_RMT_Ising}
The tilted field Ising model has a discrete symmetry which makes the spin chain invariant under reflection (also known as "bit-reversal") about the centre of the spin chain \cite{karthik2007entanglement}. Thus, the Floquet map for the kicked Ising model with a tilted magnetic field and the Hamiltonian for the time-independent tilted field Ising model are block diagonal in the eigenbasis of the reflection operator. The model is also invariant under time-reversal operation \cite{zhou2017operator, pal2018entangling}. We generate some random matrices from circular orthogonal ensemble (COE) for the Floquet map and Gaussian orthogonal ensemble (GOE) for time-independent Hamiltonian, which are block diagonal in the basis of the reflection operator.}

%%%%%%%%%%%% figure %%%%%%%%%%%%%%%%%%%%%%%
\begin{figure*}[htbp]
\centering
 \includegraphics[scale=0.58]{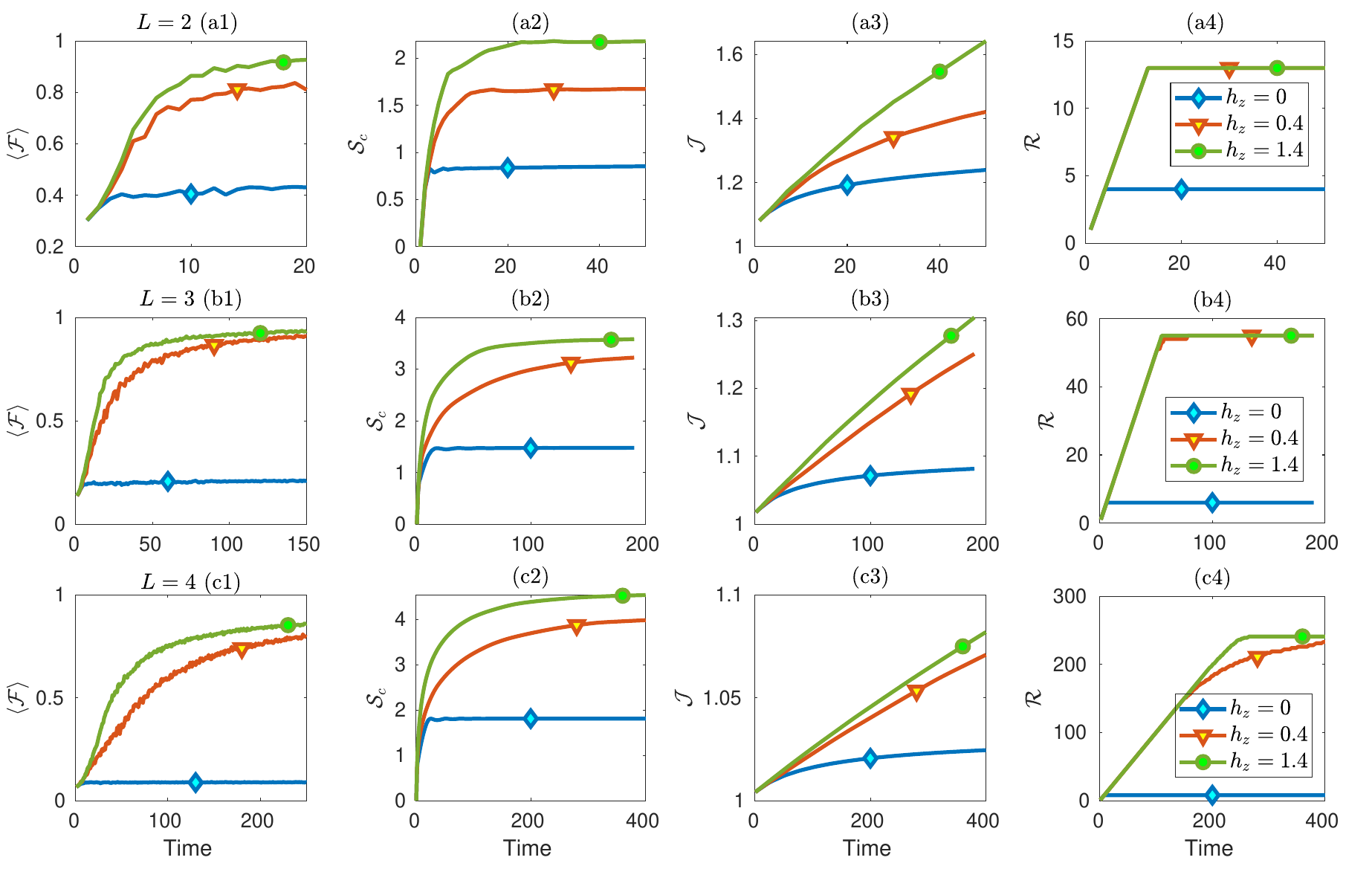}
 \caption{Quantifying operator spreading through various information-theoretic metrics as a function of time with an increase in the extent of chaos [The legends are same for all the plots; $h_z=0$ (diamond), $h_z=0.4$ (triangle) $h_z=1.4$ (circle)]. The time series of operators are generated by repeated application of the Floquet map of the time-dependent tilted field kicked Ising model  $U_{TKI}$ as shown in Eq. (\ref{tkising}) for all plots. The numerical simulations are carried out for the Ising model with $J=1$, $h_x=1.4$, and the initial observable $s_1^{y}$. The number of spins for the plots (a1 - a4) $L=2$, (b1 - b4) $L=3$, and (c1 - c4) $L=4$. (a1, b1, and c1) Average reconstruction fidelity $\langle\mathcal{F}\rangle$ as a function of time. (a2, b2, and c2) The Shannon entropy $\mathcal{S}_c$ of the normalized eigenvalues of the inverse of the covariance matrix of the likelihood function. (a3, b3, and c3) The Fisher information $\mathcal{J}$ for parameter (Bloch vector components) estimation. (a4, b4, and c4) Rank $\mathcal{R}$ of the covariance matrix. In all cases, the values of the quantifiers are more for a higher value of parameter $h_z$. }
 \label{IsingTomo234}
\end{figure*}
%%%%%%%%%%%%%%%%%%%%%% figure %%%%%%%%%%%%%%%%%%%%
{%We show that our information-theoretic quantifiers for fully chaotic dynamics are consistent with random matrix theory. 
We consider a random local observable $\mathcal{O}=u_r^{\dagger}s_1^{y}u_r$, where $u_r$ is a single-qubit random unitary chosen from Haar measure. The measurement record is generated by a time-dependent and time-independent tilted field Ising model to evaluate the Shannon entropy, Fisher information, and rank of the covariance matrix. Also, we generate the time series of observables by the random unitary evolution picked from COE and random Hamiltonian time evolution picked from GOE having the block diagonal structure described above. Figure \ref{plotRMT_ising} shows the behavior of the information-theoretic quantifiers for both the models and the random matrix theory. We see excellent agreement between our predictions from random matrix theory and the calculation for the evolution by both time-dependent and time-independent models in the completely chaotic regime.}

\section{Operator spreading for different numbers of spins in the Ising model with a tilted magnetic field}
\label{appnd_Ising}
We find the initial growth and the saturation value of average fidelity $\mathcal{F}$, Shannon entropy $\mathcal{S}_c$, Fisher information $\mathcal{J}$, and rank $\mathcal{R}$ of the covariance matrix are correlated with the degree of chaos in the dynamics. Thus, all these information-theoretic quantifiers are able to quantify operator spreading. The rank saturates at $\mathcal{R}=d^2-d+1$ for fully chaotic. The rank $\mathcal{R}$ for different numbers of spins is illustrated in Fig. \ref{IsingTomo234} (a4) $L=2$, $\mathcal{R}=13$ (b4) $L=3$, $\mathcal{R}=57$, and (c4) $L=4$, $\mathcal{R}=241$. For all these figures, the initial observable is $\mathcal{O}=s^y_1$, which does not respect the reflection symmetry about the centre of the tilted field Ising spin chain. It is interesting that even in the deep quantum regime, for $L=2$ and $L=3$, we can see the quantum signatures of chaos in the information-theoretic measures.

\bibliography{tmtoc}

\begin{thebibliography}{115}%
\makeatletter
\providecommand \@ifxundefined [1]{%
 \@ifx{#1\undefined}
}%
\providecommand \@ifnum [1]{%
 \ifnum #1\expandafter \@firstoftwo
 \else \expandafter \@secondoftwo
 \fi
}%
\providecommand \@ifx [1]{%
 \ifx #1\expandafter \@firstoftwo
 \else \expandafter \@secondoftwo
 \fi
}%
\providecommand \natexlab [1]{#1}%
\providecommand \enquote  [1]{``#1''}%
\providecommand \bibnamefont  [1]{#1}%
\providecommand \bibfnamefont [1]{#1}%
\providecommand \citenamefont [1]{#1}%
\providecommand \href@noop [0]{\@secondoftwo}%
\providecommand \href [0]{\begingroup \@sanitize@url \@href}%
\providecommand \@href[1]{\@@startlink{#1}\@@href}%
\providecommand \@@href[1]{\endgroup#1\@@endlink}%
\providecommand \@sanitize@url [0]{\catcode `\\12\catcode `\$12\catcode
  `\&12\catcode `\#12\catcode `\^12\catcode `\_12\catcode `\%12\relax}%
\providecommand \@@startlink[1]{}%
\providecommand \@@endlink[0]{}%
\providecommand \url  [0]{\begingroup\@sanitize@url \@url }%
\providecommand \@url [1]{\endgroup\@href {#1}{\urlprefix }}%
\providecommand \urlprefix  [0]{URL }%
\providecommand \Eprint [0]{\href }%
\providecommand \doibase [0]{https://doi.org/}%
\providecommand \selectlanguage [0]{\@gobble}%
\providecommand \bibinfo  [0]{\@secondoftwo}%
\providecommand \bibfield  [0]{\@secondoftwo}%
\providecommand \translation [1]{[#1]}%
\providecommand \BibitemOpen [0]{}%
\providecommand \bibitemStop [0]{}%
\providecommand \bibitemNoStop [0]{.\EOS\space}%
\providecommand \EOS [0]{\spacefactor3000\relax}%
\providecommand \BibitemShut  [1]{\csname bibitem#1\endcsname}%
\let\auto@bib@innerbib\@empty
%</preamble>
\bibitem [{\citenamefont {Von~Keyserlingk}\ \emph {et~al.}(2018)\citenamefont
  {Von~Keyserlingk}, \citenamefont {Rakovszky}, \citenamefont {Pollmann},\ and\
  \citenamefont {Sondhi}}]{von2018operator}%
  \BibitemOpen
  \bibfield  {author} {\bibinfo {author} {\bibfnamefont {C.}~\bibnamefont
  {Von~Keyserlingk}}, \bibinfo {author} {\bibfnamefont {T.}~\bibnamefont
  {Rakovszky}}, \bibinfo {author} {\bibfnamefont {F.}~\bibnamefont
  {Pollmann}},\ and\ \bibinfo {author} {\bibfnamefont {S.~L.}\ \bibnamefont
  {Sondhi}},\ }\bibfield  {title} {\bibinfo {title} {Operator hydrodynamics,
  otocs, and entanglement growth in systems without conservation laws},\
  }\href@noop {} {\bibfield  {journal} {\bibinfo  {journal} {Physical Review
  X}\ }\textbf {\bibinfo {volume} {8}},\ \bibinfo {pages} {021013} (\bibinfo
  {year} {2018})}\BibitemShut {NoStop}%
\bibitem [{\citenamefont {Deutsch}(1991)}]{deutsch1991quantum}%
  \BibitemOpen
  \bibfield  {author} {\bibinfo {author} {\bibfnamefont {J.~M.}\ \bibnamefont
  {Deutsch}},\ }\bibfield  {title} {\bibinfo {title} {Quantum statistical
  mechanics in a closed system},\ }\href@noop {} {\bibfield  {journal}
  {\bibinfo  {journal} {Physical review a}\ }\textbf {\bibinfo {volume} {43}},\
  \bibinfo {pages} {2046} (\bibinfo {year} {1991})}\BibitemShut {NoStop}%
\bibitem [{\citenamefont {Srednicki}(1994)}]{srednicki1994chaos}%
  \BibitemOpen
  \bibfield  {author} {\bibinfo {author} {\bibfnamefont {M.}~\bibnamefont
  {Srednicki}},\ }\bibfield  {title} {\bibinfo {title} {Chaos and quantum
  thermalization},\ }\href@noop {} {\bibfield  {journal} {\bibinfo  {journal}
  {Physical review e}\ }\textbf {\bibinfo {volume} {50}},\ \bibinfo {pages}
  {888} (\bibinfo {year} {1994})}\BibitemShut {NoStop}%
\bibitem [{\citenamefont {Tasaki}(1998)}]{tasaki1998quantum}%
  \BibitemOpen
  \bibfield  {author} {\bibinfo {author} {\bibfnamefont {H.}~\bibnamefont
  {Tasaki}},\ }\bibfield  {title} {\bibinfo {title} {From quantum dynamics to
  the canonical distribution: general picture and a rigorous example},\
  }\href@noop {} {\bibfield  {journal} {\bibinfo  {journal} {Physical review
  letters}\ }\textbf {\bibinfo {volume} {80}},\ \bibinfo {pages} {1373}
  (\bibinfo {year} {1998})}\BibitemShut {NoStop}%
\bibitem [{\citenamefont {Rigol}\ \emph {et~al.}(2008)\citenamefont {Rigol},
  \citenamefont {Dunjko},\ and\ \citenamefont
  {Olshanii}}]{rigol2008thermalization}%
  \BibitemOpen
  \bibfield  {author} {\bibinfo {author} {\bibfnamefont {M.}~\bibnamefont
  {Rigol}}, \bibinfo {author} {\bibfnamefont {V.}~\bibnamefont {Dunjko}},\ and\
  \bibinfo {author} {\bibfnamefont {M.}~\bibnamefont {Olshanii}},\ }\bibfield
  {title} {\bibinfo {title} {Thermalization and its mechanism for generic
  isolated quantum systems},\ }\href@noop {} {\bibfield  {journal} {\bibinfo
  {journal} {Nature}\ }\textbf {\bibinfo {volume} {452}},\ \bibinfo {pages}
  {854} (\bibinfo {year} {2008})}\BibitemShut {NoStop}%
\bibitem [{\citenamefont {Rigol}\ and\ \citenamefont
  {Santos}(2010)}]{rigol2010quantum}%
  \BibitemOpen
  \bibfield  {author} {\bibinfo {author} {\bibfnamefont {M.}~\bibnamefont
  {Rigol}}\ and\ \bibinfo {author} {\bibfnamefont {L.~F.}\ \bibnamefont
  {Santos}},\ }\bibfield  {title} {\bibinfo {title} {Quantum chaos and
  thermalization in gapped systems},\ }\href@noop {} {\bibfield  {journal}
  {\bibinfo  {journal} {Physical Review A}\ }\textbf {\bibinfo {volume} {82}},\
  \bibinfo {pages} {011604} (\bibinfo {year} {2010})}\BibitemShut {NoStop}%
\bibitem [{\citenamefont {Torres-Herrera}\ and\ \citenamefont
  {Santos}(2013)}]{torres2013effects}%
  \BibitemOpen
  \bibfield  {author} {\bibinfo {author} {\bibfnamefont {E.}~\bibnamefont
  {Torres-Herrera}}\ and\ \bibinfo {author} {\bibfnamefont {L.~F.}\
  \bibnamefont {Santos}},\ }\bibfield  {title} {\bibinfo {title} {Effects of
  the interplay between initial state and hamiltonian on the thermalization of
  isolated quantum many-body systems},\ }\href@noop {} {\bibfield  {journal}
  {\bibinfo  {journal} {Physical Review E}\ }\textbf {\bibinfo {volume} {88}},\
  \bibinfo {pages} {042121} (\bibinfo {year} {2013})}\BibitemShut {NoStop}%
\bibitem [{\citenamefont {Hayden}\ and\ \citenamefont
  {Preskill}(2007)}]{hayden2007black}%
  \BibitemOpen
  \bibfield  {author} {\bibinfo {author} {\bibfnamefont {P.}~\bibnamefont
  {Hayden}}\ and\ \bibinfo {author} {\bibfnamefont {J.}~\bibnamefont
  {Preskill}},\ }\bibfield  {title} {\bibinfo {title} {Black holes as mirrors:
  quantum information in random subsystems},\ }\href@noop {} {\bibfield
  {journal} {\bibinfo  {journal} {Journal of high energy physics}\ }\textbf
  {\bibinfo {volume} {2007}},\ \bibinfo {pages} {120} (\bibinfo {year}
  {2007})}\BibitemShut {NoStop}%
\bibitem [{\citenamefont {Sekino}\ and\ \citenamefont
  {Susskind}(2008)}]{sekino2008fast}%
  \BibitemOpen
  \bibfield  {author} {\bibinfo {author} {\bibfnamefont {Y.}~\bibnamefont
  {Sekino}}\ and\ \bibinfo {author} {\bibfnamefont {L.}~\bibnamefont
  {Susskind}},\ }\bibfield  {title} {\bibinfo {title} {Fast scramblers},\
  }\href@noop {} {\bibfield  {journal} {\bibinfo  {journal} {Journal of High
  Energy Physics}\ }\textbf {\bibinfo {volume} {2008}},\ \bibinfo {pages} {065}
  (\bibinfo {year} {2008})}\BibitemShut {NoStop}%
\bibitem [{\citenamefont {Hosur}\ \emph {et~al.}(2016)\citenamefont {Hosur},
  \citenamefont {Qi}, \citenamefont {Roberts},\ and\ \citenamefont
  {Yoshida}}]{hosur2016chaos}%
  \BibitemOpen
  \bibfield  {author} {\bibinfo {author} {\bibfnamefont {P.}~\bibnamefont
  {Hosur}}, \bibinfo {author} {\bibfnamefont {X.-L.}\ \bibnamefont {Qi}},
  \bibinfo {author} {\bibfnamefont {D.~A.}\ \bibnamefont {Roberts}},\ and\
  \bibinfo {author} {\bibfnamefont {B.}~\bibnamefont {Yoshida}},\ }\bibfield
  {title} {\bibinfo {title} {Chaos in quantum channels},\ }\href@noop {}
  {\bibfield  {journal} {\bibinfo  {journal} {Journal of High Energy Physics}\
  }\textbf {\bibinfo {volume} {2016}},\ \bibinfo {pages} {1} (\bibinfo {year}
  {2016})}\BibitemShut {NoStop}%
\bibitem [{\citenamefont {Shenker}\ and\ \citenamefont
  {Stanford}(2014)}]{shenker2014black}%
  \BibitemOpen
  \bibfield  {author} {\bibinfo {author} {\bibfnamefont {S.~H.}\ \bibnamefont
  {Shenker}}\ and\ \bibinfo {author} {\bibfnamefont {D.}~\bibnamefont
  {Stanford}},\ }\bibfield  {title} {\bibinfo {title} {Black holes and the
  butterfly effect},\ }\href@noop {} {\bibfield  {journal} {\bibinfo  {journal}
  {Journal of High Energy Physics}\ }\textbf {\bibinfo {volume} {2014}},\
  \bibinfo {pages} {1} (\bibinfo {year} {2014})}\BibitemShut {NoStop}%
\bibitem [{\citenamefont {McGinley}\ \emph {et~al.}(2022)\citenamefont
  {McGinley}, \citenamefont {Leontica}, \citenamefont {Garratt}, \citenamefont
  {Jovanovic},\ and\ \citenamefont {Simon}}]{mcginley2022quantifying}%
  \BibitemOpen
  \bibfield  {author} {\bibinfo {author} {\bibfnamefont {M.}~\bibnamefont
  {McGinley}}, \bibinfo {author} {\bibfnamefont {S.}~\bibnamefont {Leontica}},
  \bibinfo {author} {\bibfnamefont {S.~J.}\ \bibnamefont {Garratt}}, \bibinfo
  {author} {\bibfnamefont {J.}~\bibnamefont {Jovanovic}},\ and\ \bibinfo
  {author} {\bibfnamefont {S.~H.}\ \bibnamefont {Simon}},\ }\bibfield  {title}
  {\bibinfo {title} {Quantifying information scrambling via classical shadow
  tomography on programmable quantum simulators},\ }\href@noop {} {\bibfield
  {journal} {\bibinfo  {journal} {Physical Review A}\ }\textbf {\bibinfo
  {volume} {106}},\ \bibinfo {pages} {012441} (\bibinfo {year}
  {2022})}\BibitemShut {NoStop}%
\bibitem [{\citenamefont {Bhattacharyya}\ \emph {et~al.}(2022)\citenamefont
  {Bhattacharyya}, \citenamefont {Joshi},\ and\ \citenamefont
  {Sundar}}]{bhattacharyya2022quantum}%
  \BibitemOpen
  \bibfield  {author} {\bibinfo {author} {\bibfnamefont {A.}~\bibnamefont
  {Bhattacharyya}}, \bibinfo {author} {\bibfnamefont {L.~K.}\ \bibnamefont
  {Joshi}},\ and\ \bibinfo {author} {\bibfnamefont {B.}~\bibnamefont
  {Sundar}},\ }\bibfield  {title} {\bibinfo {title} {Quantum information
  scrambling: from holography to quantum simulators},\ }\href@noop {}
  {\bibfield  {journal} {\bibinfo  {journal} {The European Physical Journal C}\
  }\textbf {\bibinfo {volume} {82}},\ \bibinfo {pages} {458} (\bibinfo {year}
  {2022})}\BibitemShut {NoStop}%
\bibitem [{\citenamefont {Xu}\ \emph {et~al.}(2020)\citenamefont {Xu},
  \citenamefont {Scaffidi},\ and\ \citenamefont {Cao}}]{xu2020does}%
  \BibitemOpen
  \bibfield  {author} {\bibinfo {author} {\bibfnamefont {T.}~\bibnamefont
  {Xu}}, \bibinfo {author} {\bibfnamefont {T.}~\bibnamefont {Scaffidi}},\ and\
  \bibinfo {author} {\bibfnamefont {X.}~\bibnamefont {Cao}},\ }\bibfield
  {title} {\bibinfo {title} {Does scrambling equal chaos?},\ }\href@noop {}
  {\bibfield  {journal} {\bibinfo  {journal} {Physical review letters}\
  }\textbf {\bibinfo {volume} {124}},\ \bibinfo {pages} {140602} (\bibinfo
  {year} {2020})}\BibitemShut {NoStop}%
\bibitem [{\citenamefont {Rozenbaum}\ \emph {et~al.}(2020)\citenamefont
  {Rozenbaum}, \citenamefont {Bunimovich},\ and\ \citenamefont
  {Galitski}}]{rozenbaum2020early}%
  \BibitemOpen
  \bibfield  {author} {\bibinfo {author} {\bibfnamefont {E.~B.}\ \bibnamefont
  {Rozenbaum}}, \bibinfo {author} {\bibfnamefont {L.~A.}\ \bibnamefont
  {Bunimovich}},\ and\ \bibinfo {author} {\bibfnamefont {V.}~\bibnamefont
  {Galitski}},\ }\bibfield  {title} {\bibinfo {title} {Early-time exponential
  instabilities in nonchaotic quantum systems},\ }\href@noop {} {\bibfield
  {journal} {\bibinfo  {journal} {Physical Review Letters}\ }\textbf {\bibinfo
  {volume} {125}},\ \bibinfo {pages} {014101} (\bibinfo {year}
  {2020})}\BibitemShut {NoStop}%
\bibitem [{\citenamefont {Pilatowsky-Cameo}\ \emph {et~al.}(2020)\citenamefont
  {Pilatowsky-Cameo}, \citenamefont {Ch{\'a}vez-Carlos}, \citenamefont
  {Bastarrachea-Magnani}, \citenamefont {Str{\'a}nsk{\`y}}, \citenamefont
  {Lerma-Hern{\'a}ndez}, \citenamefont {Santos},\ and\ \citenamefont
  {Hirsch}}]{pilatowsky2020positive}%
  \BibitemOpen
  \bibfield  {author} {\bibinfo {author} {\bibfnamefont {S.}~\bibnamefont
  {Pilatowsky-Cameo}}, \bibinfo {author} {\bibfnamefont {J.}~\bibnamefont
  {Ch{\'a}vez-Carlos}}, \bibinfo {author} {\bibfnamefont {M.~A.}\ \bibnamefont
  {Bastarrachea-Magnani}}, \bibinfo {author} {\bibfnamefont {P.}~\bibnamefont
  {Str{\'a}nsk{\`y}}}, \bibinfo {author} {\bibfnamefont {S.}~\bibnamefont
  {Lerma-Hern{\'a}ndez}}, \bibinfo {author} {\bibfnamefont {L.~F.}\
  \bibnamefont {Santos}},\ and\ \bibinfo {author} {\bibfnamefont {J.~G.}\
  \bibnamefont {Hirsch}},\ }\bibfield  {title} {\bibinfo {title} {Positive
  quantum lyapunov exponents in experimental systems with a regular classical
  limit},\ }\href@noop {} {\bibfield  {journal} {\bibinfo  {journal} {Physical
  Review E}\ }\textbf {\bibinfo {volume} {101}},\ \bibinfo {pages} {010202}
  (\bibinfo {year} {2020})}\BibitemShut {NoStop}%
\bibitem [{\citenamefont {Nahum}\ \emph
  {et~al.}(2018{\natexlab{a}})\citenamefont {Nahum}, \citenamefont {Ruhman},\
  and\ \citenamefont {Huse}}]{nahum2018dynamics}%
  \BibitemOpen
  \bibfield  {author} {\bibinfo {author} {\bibfnamefont {A.}~\bibnamefont
  {Nahum}}, \bibinfo {author} {\bibfnamefont {J.}~\bibnamefont {Ruhman}},\ and\
  \bibinfo {author} {\bibfnamefont {D.~A.}\ \bibnamefont {Huse}},\ }\bibfield
  {title} {\bibinfo {title} {Dynamics of entanglement and transport in
  one-dimensional systems with quenched randomness},\ }\href@noop {} {\bibfield
   {journal} {\bibinfo  {journal} {Physical Review B}\ }\textbf {\bibinfo
  {volume} {98}},\ \bibinfo {pages} {035118} (\bibinfo {year}
  {2018}{\natexlab{a}})}\BibitemShut {NoStop}%
\bibitem [{\citenamefont {Nahum}\ \emph
  {et~al.}(2018{\natexlab{b}})\citenamefont {Nahum}, \citenamefont {Vijay},\
  and\ \citenamefont {Haah}}]{nahum2018operator}%
  \BibitemOpen
  \bibfield  {author} {\bibinfo {author} {\bibfnamefont {A.}~\bibnamefont
  {Nahum}}, \bibinfo {author} {\bibfnamefont {S.}~\bibnamefont {Vijay}},\ and\
  \bibinfo {author} {\bibfnamefont {J.}~\bibnamefont {Haah}},\ }\bibfield
  {title} {\bibinfo {title} {Operator spreading in random unitary circuits},\
  }\href@noop {} {\bibfield  {journal} {\bibinfo  {journal} {Physical Review
  X}\ }\textbf {\bibinfo {volume} {8}},\ \bibinfo {pages} {021014} (\bibinfo
  {year} {2018}{\natexlab{b}})}\BibitemShut {NoStop}%
\bibitem [{\citenamefont {Khemani}\ \emph {et~al.}(2018)\citenamefont
  {Khemani}, \citenamefont {Vishwanath},\ and\ \citenamefont
  {Huse}}]{khemani2018operator}%
  \BibitemOpen
  \bibfield  {author} {\bibinfo {author} {\bibfnamefont {V.}~\bibnamefont
  {Khemani}}, \bibinfo {author} {\bibfnamefont {A.}~\bibnamefont
  {Vishwanath}},\ and\ \bibinfo {author} {\bibfnamefont {D.~A.}\ \bibnamefont
  {Huse}},\ }\bibfield  {title} {\bibinfo {title} {Operator spreading and the
  emergence of dissipative hydrodynamics under unitary evolution with
  conservation laws},\ }\href@noop {} {\bibfield  {journal} {\bibinfo
  {journal} {Physical Review X}\ }\textbf {\bibinfo {volume} {8}},\ \bibinfo
  {pages} {031057} (\bibinfo {year} {2018})}\BibitemShut {NoStop}%
\bibitem [{\citenamefont {Rakovszky}\ \emph {et~al.}(2018)\citenamefont
  {Rakovszky}, \citenamefont {Pollmann},\ and\ \citenamefont
  {Von~Keyserlingk}}]{rakovszky2018diffusive}%
  \BibitemOpen
  \bibfield  {author} {\bibinfo {author} {\bibfnamefont {T.}~\bibnamefont
  {Rakovszky}}, \bibinfo {author} {\bibfnamefont {F.}~\bibnamefont
  {Pollmann}},\ and\ \bibinfo {author} {\bibfnamefont {C.}~\bibnamefont
  {Von~Keyserlingk}},\ }\bibfield  {title} {\bibinfo {title} {Diffusive
  hydrodynamics of out-of-time-ordered correlators with charge conservation},\
  }\href@noop {} {\bibfield  {journal} {\bibinfo  {journal} {Physical Review
  X}\ }\textbf {\bibinfo {volume} {8}},\ \bibinfo {pages} {031058} (\bibinfo
  {year} {2018})}\BibitemShut {NoStop}%
\bibitem [{\citenamefont {Roberts}\ and\ \citenamefont
  {Stanford}(2015)}]{roberts2015diagnosing}%
  \BibitemOpen
  \bibfield  {author} {\bibinfo {author} {\bibfnamefont {D.~A.}\ \bibnamefont
  {Roberts}}\ and\ \bibinfo {author} {\bibfnamefont {D.}~\bibnamefont
  {Stanford}},\ }\bibfield  {title} {\bibinfo {title} {Diagnosing chaos using
  four-point functions in two-dimensional conformal field theory},\ }\href@noop
  {} {\bibfield  {journal} {\bibinfo  {journal} {Physical review letters}\
  }\textbf {\bibinfo {volume} {115}},\ \bibinfo {pages} {131603} (\bibinfo
  {year} {2015})}\BibitemShut {NoStop}%
\bibitem [{\citenamefont {Stanford}(2016)}]{stanford2016many}%
  \BibitemOpen
  \bibfield  {author} {\bibinfo {author} {\bibfnamefont {D.}~\bibnamefont
  {Stanford}},\ }\bibfield  {title} {\bibinfo {title} {Many-body chaos at weak
  coupling},\ }\href@noop {} {\bibfield  {journal} {\bibinfo  {journal}
  {Journal of High Energy Physics}\ }\textbf {\bibinfo {volume} {2016}},\
  \bibinfo {pages} {1} (\bibinfo {year} {2016})}\BibitemShut {NoStop}%
\bibitem [{\citenamefont {Chowdhury}\ and\ \citenamefont
  {Swingle}(2017)}]{chowdhury2017onset}%
  \BibitemOpen
  \bibfield  {author} {\bibinfo {author} {\bibfnamefont {D.}~\bibnamefont
  {Chowdhury}}\ and\ \bibinfo {author} {\bibfnamefont {B.}~\bibnamefont
  {Swingle}},\ }\bibfield  {title} {\bibinfo {title} {Onset of many-body chaos
  in the o (n) model},\ }\href@noop {} {\bibfield  {journal} {\bibinfo
  {journal} {Physical Review D}\ }\textbf {\bibinfo {volume} {96}},\ \bibinfo
  {pages} {065005} (\bibinfo {year} {2017})}\BibitemShut {NoStop}%
\bibitem [{\citenamefont {Patel}\ \emph {et~al.}(2017)\citenamefont {Patel},
  \citenamefont {Chowdhury}, \citenamefont {Sachdev},\ and\ \citenamefont
  {Swingle}}]{patel2017quantum}%
  \BibitemOpen
  \bibfield  {author} {\bibinfo {author} {\bibfnamefont {A.~A.}\ \bibnamefont
  {Patel}}, \bibinfo {author} {\bibfnamefont {D.}~\bibnamefont {Chowdhury}},
  \bibinfo {author} {\bibfnamefont {S.}~\bibnamefont {Sachdev}},\ and\ \bibinfo
  {author} {\bibfnamefont {B.}~\bibnamefont {Swingle}},\ }\bibfield  {title}
  {\bibinfo {title} {Quantum butterfly effect in weakly interacting diffusive
  metals},\ }\href@noop {} {\bibfield  {journal} {\bibinfo  {journal} {Physical
  Review X}\ }\textbf {\bibinfo {volume} {7}},\ \bibinfo {pages} {031047}
  (\bibinfo {year} {2017})}\BibitemShut {NoStop}%
\bibitem [{\citenamefont {Luitz}\ and\ \citenamefont
  {Lev}(2017)}]{luitz2017information}%
  \BibitemOpen
  \bibfield  {author} {\bibinfo {author} {\bibfnamefont {D.~J.}\ \bibnamefont
  {Luitz}}\ and\ \bibinfo {author} {\bibfnamefont {Y.~B.}\ \bibnamefont
  {Lev}},\ }\bibfield  {title} {\bibinfo {title} {Information propagation in
  isolated quantum systems},\ }\href@noop {} {\bibfield  {journal} {\bibinfo
  {journal} {Physical Review B}\ }\textbf {\bibinfo {volume} {96}},\ \bibinfo
  {pages} {020406} (\bibinfo {year} {2017})}\BibitemShut {NoStop}%
\bibitem [{\citenamefont {Heyl}\ \emph {et~al.}(2018)\citenamefont {Heyl},
  \citenamefont {Pollmann},\ and\ \citenamefont
  {D{\'o}ra}}]{heyl2018detecting}%
  \BibitemOpen
  \bibfield  {author} {\bibinfo {author} {\bibfnamefont {M.}~\bibnamefont
  {Heyl}}, \bibinfo {author} {\bibfnamefont {F.}~\bibnamefont {Pollmann}},\
  and\ \bibinfo {author} {\bibfnamefont {B.}~\bibnamefont {D{\'o}ra}},\
  }\bibfield  {title} {\bibinfo {title} {Detecting equilibrium and dynamical
  quantum phase transitions in ising chains via out-of-time-ordered
  correlators},\ }\href@noop {} {\bibfield  {journal} {\bibinfo  {journal}
  {Physical review letters}\ }\textbf {\bibinfo {volume} {121}},\ \bibinfo
  {pages} {016801} (\bibinfo {year} {2018})}\BibitemShut {NoStop}%
\bibitem [{\citenamefont {Lin}\ and\ \citenamefont
  {Motrunich}(2018)}]{lin2018out}%
  \BibitemOpen
  \bibfield  {author} {\bibinfo {author} {\bibfnamefont {C.-J.}\ \bibnamefont
  {Lin}}\ and\ \bibinfo {author} {\bibfnamefont {O.~I.}\ \bibnamefont
  {Motrunich}},\ }\bibfield  {title} {\bibinfo {title} {Out-of-time-ordered
  correlators in a quantum ising chain},\ }\href@noop {} {\bibfield  {journal}
  {\bibinfo  {journal} {Physical Review B}\ }\textbf {\bibinfo {volume} {97}},\
  \bibinfo {pages} {144304} (\bibinfo {year} {2018})}\BibitemShut {NoStop}%
\bibitem [{\citenamefont {Geller}\ \emph {et~al.}(2022)\citenamefont {Geller},
  \citenamefont {Arrasmith}, \citenamefont {Holmes}, \citenamefont {Yan},
  \citenamefont {Coles},\ and\ \citenamefont {Sornborger}}]{geller2022quantum}%
  \BibitemOpen
  \bibfield  {author} {\bibinfo {author} {\bibfnamefont {M.~R.}\ \bibnamefont
  {Geller}}, \bibinfo {author} {\bibfnamefont {A.}~\bibnamefont {Arrasmith}},
  \bibinfo {author} {\bibfnamefont {Z.}~\bibnamefont {Holmes}}, \bibinfo
  {author} {\bibfnamefont {B.}~\bibnamefont {Yan}}, \bibinfo {author}
  {\bibfnamefont {P.~J.}\ \bibnamefont {Coles}},\ and\ \bibinfo {author}
  {\bibfnamefont {A.}~\bibnamefont {Sornborger}},\ }\bibfield  {title}
  {\bibinfo {title} {Quantum simulation of operator spreading in the chaotic
  ising model},\ }\href@noop {} {\bibfield  {journal} {\bibinfo  {journal}
  {Physical Review E}\ }\textbf {\bibinfo {volume} {105}},\ \bibinfo {pages}
  {035302} (\bibinfo {year} {2022})}\BibitemShut {NoStop}%
\bibitem [{\citenamefont {Moudgalya}\ \emph {et~al.}(2019)\citenamefont
  {Moudgalya}, \citenamefont {Devakul}, \citenamefont {Von~Keyserlingk},\ and\
  \citenamefont {Sondhi}}]{moudgalya2019operator}%
  \BibitemOpen
  \bibfield  {author} {\bibinfo {author} {\bibfnamefont {S.}~\bibnamefont
  {Moudgalya}}, \bibinfo {author} {\bibfnamefont {T.}~\bibnamefont {Devakul}},
  \bibinfo {author} {\bibfnamefont {C.}~\bibnamefont {Von~Keyserlingk}},\ and\
  \bibinfo {author} {\bibfnamefont {S.}~\bibnamefont {Sondhi}},\ }\bibfield
  {title} {\bibinfo {title} {Operator spreading in quantum maps},\ }\href@noop
  {} {\bibfield  {journal} {\bibinfo  {journal} {Physical Review B}\ }\textbf
  {\bibinfo {volume} {99}},\ \bibinfo {pages} {094312} (\bibinfo {year}
  {2019})}\BibitemShut {NoStop}%
\bibitem [{\citenamefont {Omanakuttan}\ \emph {et~al.}(2023)\citenamefont
  {Omanakuttan}, \citenamefont {Chinni}, \citenamefont {Blocher},\ and\
  \citenamefont {Poggi}}]{omanakuttan2023scrambling}%
  \BibitemOpen
  \bibfield  {author} {\bibinfo {author} {\bibfnamefont {S.}~\bibnamefont
  {Omanakuttan}}, \bibinfo {author} {\bibfnamefont {K.}~\bibnamefont {Chinni}},
  \bibinfo {author} {\bibfnamefont {P.~D.}\ \bibnamefont {Blocher}},\ and\
  \bibinfo {author} {\bibfnamefont {P.~M.}\ \bibnamefont {Poggi}},\ }\bibfield
  {title} {\bibinfo {title} {Scrambling and quantum chaos indicators from
  long-time properties of operator distributions},\ }\href@noop {} {\bibfield
  {journal} {\bibinfo  {journal} {Physical Review A}\ }\textbf {\bibinfo
  {volume} {107}},\ \bibinfo {pages} {032418} (\bibinfo {year}
  {2023})}\BibitemShut {NoStop}%
\bibitem [{\citenamefont {Maldacena}\ \emph {et~al.}(2016)\citenamefont
  {Maldacena}, \citenamefont {Shenker},\ and\ \citenamefont
  {Stanford}}]{maldacena2016bound}%
  \BibitemOpen
  \bibfield  {author} {\bibinfo {author} {\bibfnamefont {J.}~\bibnamefont
  {Maldacena}}, \bibinfo {author} {\bibfnamefont {S.~H.}\ \bibnamefont
  {Shenker}},\ and\ \bibinfo {author} {\bibfnamefont {D.}~\bibnamefont
  {Stanford}},\ }\bibfield  {title} {\bibinfo {title} {A bound on chaos},\
  }\href@noop {} {\bibfield  {journal} {\bibinfo  {journal} {Journal of High
  Energy Physics}\ }\textbf {\bibinfo {volume} {2016}},\ \bibinfo {pages} {1}
  (\bibinfo {year} {2016})}\BibitemShut {NoStop}%
\bibitem [{\citenamefont {Swingle}(2018)}]{swingle2018unscrambling}%
  \BibitemOpen
  \bibfield  {author} {\bibinfo {author} {\bibfnamefont {B.}~\bibnamefont
  {Swingle}},\ }\bibfield  {title} {\bibinfo {title} {Unscrambling the physics
  of out-of-time-order correlators},\ }\href@noop {} {\bibfield  {journal}
  {\bibinfo  {journal} {Nature Physics}\ }\textbf {\bibinfo {volume} {14}},\
  \bibinfo {pages} {988} (\bibinfo {year} {2018})}\BibitemShut {NoStop}%
\bibitem [{\citenamefont {Seshadri}\ \emph {et~al.}(2018)\citenamefont
  {Seshadri}, \citenamefont {Madhok},\ and\ \citenamefont
  {Lakshminarayan}}]{seshadri2018tripartite}%
  \BibitemOpen
  \bibfield  {author} {\bibinfo {author} {\bibfnamefont {A.}~\bibnamefont
  {Seshadri}}, \bibinfo {author} {\bibfnamefont {V.}~\bibnamefont {Madhok}},\
  and\ \bibinfo {author} {\bibfnamefont {A.}~\bibnamefont {Lakshminarayan}},\
  }\bibfield  {title} {\bibinfo {title} {Tripartite mutual information,
  entanglement, and scrambling in permutation symmetric systems with an
  application to quantum chaos},\ }\href@noop {} {\bibfield  {journal}
  {\bibinfo  {journal} {Physical Review E}\ }\textbf {\bibinfo {volume} {98}},\
  \bibinfo {pages} {052205} (\bibinfo {year} {2018})}\BibitemShut {NoStop}%
\bibitem [{\citenamefont {Prakash}\ and\ \citenamefont
  {Lakshminarayan}(2020)}]{prakash2020scrambling}%
  \BibitemOpen
  \bibfield  {author} {\bibinfo {author} {\bibfnamefont {R.}~\bibnamefont
  {Prakash}}\ and\ \bibinfo {author} {\bibfnamefont {A.}~\bibnamefont
  {Lakshminarayan}},\ }\bibfield  {title} {\bibinfo {title} {Scrambling in
  strongly chaotic weakly coupled bipartite systems: Universality beyond the
  ehrenfest timescale},\ }\href@noop {} {\bibfield  {journal} {\bibinfo
  {journal} {Physical Review B}\ }\textbf {\bibinfo {volume} {101}},\ \bibinfo
  {pages} {121108} (\bibinfo {year} {2020})}\BibitemShut {NoStop}%
\bibitem [{\citenamefont {Xu}\ and\ \citenamefont
  {Swingle}(2020)}]{xu2020accessing}%
  \BibitemOpen
  \bibfield  {author} {\bibinfo {author} {\bibfnamefont {S.}~\bibnamefont
  {Xu}}\ and\ \bibinfo {author} {\bibfnamefont {B.}~\bibnamefont {Swingle}},\
  }\bibfield  {title} {\bibinfo {title} {Accessing scrambling using matrix
  product operators},\ }\href@noop {} {\bibfield  {journal} {\bibinfo
  {journal} {Nature Physics}\ }\textbf {\bibinfo {volume} {16}},\ \bibinfo
  {pages} {199} (\bibinfo {year} {2020})}\BibitemShut {NoStop}%
\bibitem [{\citenamefont {Sreeram}\ \emph {et~al.}(2021)\citenamefont
  {Sreeram}, \citenamefont {Madhok},\ and\ \citenamefont
  {Lakshminarayan}}]{sreeram2021out}%
  \BibitemOpen
  \bibfield  {author} {\bibinfo {author} {\bibfnamefont {P.}~\bibnamefont
  {Sreeram}}, \bibinfo {author} {\bibfnamefont {V.}~\bibnamefont {Madhok}},\
  and\ \bibinfo {author} {\bibfnamefont {A.}~\bibnamefont {Lakshminarayan}},\
  }\bibfield  {title} {\bibinfo {title} {Out-of-time-ordered correlators and
  the loschmidt echo in the quantum kicked top: how low can we go?},\
  }\href@noop {} {\bibfield  {journal} {\bibinfo  {journal} {Journal of Physics
  D: Applied Physics}\ }\textbf {\bibinfo {volume} {54}},\ \bibinfo {pages}
  {274004} (\bibinfo {year} {2021})}\BibitemShut {NoStop}%
\bibitem [{\citenamefont {Varikuti}\ and\ \citenamefont
  {Madhok}(2022)}]{varikuti2022out}%
  \BibitemOpen
  \bibfield  {author} {\bibinfo {author} {\bibfnamefont {N.~D.}\ \bibnamefont
  {Varikuti}}\ and\ \bibinfo {author} {\bibfnamefont {V.}~\bibnamefont
  {Madhok}},\ }\bibfield  {title} {\bibinfo {title} {Out-of-time ordered
  correlators in kicked coupled tops and the role of conserved quantities in
  information scrambling},\ }\href@noop {} {\bibfield  {journal} {\bibinfo
  {journal} {arXiv preprint arXiv:2201.05789}\ } (\bibinfo {year}
  {2022})}\BibitemShut {NoStop}%
\bibitem [{\citenamefont {Nie}\ \emph {et~al.}(2019)\citenamefont {Nie},
  \citenamefont {Nozaki}, \citenamefont {Ryu},\ and\ \citenamefont
  {Tan}}]{nie2019signature}%
  \BibitemOpen
  \bibfield  {author} {\bibinfo {author} {\bibfnamefont {L.}~\bibnamefont
  {Nie}}, \bibinfo {author} {\bibfnamefont {M.}~\bibnamefont {Nozaki}},
  \bibinfo {author} {\bibfnamefont {S.}~\bibnamefont {Ryu}},\ and\ \bibinfo
  {author} {\bibfnamefont {M.~T.}\ \bibnamefont {Tan}},\ }\bibfield  {title}
  {\bibinfo {title} {Signature of quantum chaos in operator entanglement in 2d
  cfts},\ }\href@noop {} {\bibfield  {journal} {\bibinfo  {journal} {Journal of
  Statistical Mechanics: Theory and Experiment}\ }\textbf {\bibinfo {volume}
  {2019}},\ \bibinfo {pages} {093107} (\bibinfo {year} {2019})}\BibitemShut
  {NoStop}%
\bibitem [{\citenamefont {Wang}\ and\ \citenamefont
  {Zhou}(2019)}]{wang2019barrier}%
  \BibitemOpen
  \bibfield  {author} {\bibinfo {author} {\bibfnamefont {H.}~\bibnamefont
  {Wang}}\ and\ \bibinfo {author} {\bibfnamefont {T.}~\bibnamefont {Zhou}},\
  }\bibfield  {title} {\bibinfo {title} {Barrier from chaos: operator
  entanglement dynamics of the reduced density matrix},\ }\href@noop {}
  {\bibfield  {journal} {\bibinfo  {journal} {Journal of High Energy Physics}\
  }\textbf {\bibinfo {volume} {2019}},\ \bibinfo {pages} {1} (\bibinfo {year}
  {2019})}\BibitemShut {NoStop}%
\bibitem [{\citenamefont {Alba}\ \emph {et~al.}(2019)\citenamefont {Alba},
  \citenamefont {Dubail},\ and\ \citenamefont {Medenjak}}]{alba2019operator}%
  \BibitemOpen
  \bibfield  {author} {\bibinfo {author} {\bibfnamefont {V.}~\bibnamefont
  {Alba}}, \bibinfo {author} {\bibfnamefont {J.}~\bibnamefont {Dubail}},\ and\
  \bibinfo {author} {\bibfnamefont {M.}~\bibnamefont {Medenjak}},\ }\bibfield
  {title} {\bibinfo {title} {Operator entanglement in interacting integrable
  quantum systems: the case of the rule 54 chain},\ }\href@noop {} {\bibfield
  {journal} {\bibinfo  {journal} {Physical review letters}\ }\textbf {\bibinfo
  {volume} {122}},\ \bibinfo {pages} {250603} (\bibinfo {year}
  {2019})}\BibitemShut {NoStop}%
\bibitem [{\citenamefont {Styliaris}\ \emph {et~al.}(2021)\citenamefont
  {Styliaris}, \citenamefont {Anand},\ and\ \citenamefont
  {Zanardi}}]{styliaris2021information}%
  \BibitemOpen
  \bibfield  {author} {\bibinfo {author} {\bibfnamefont {G.}~\bibnamefont
  {Styliaris}}, \bibinfo {author} {\bibfnamefont {N.}~\bibnamefont {Anand}},\
  and\ \bibinfo {author} {\bibfnamefont {P.}~\bibnamefont {Zanardi}},\
  }\bibfield  {title} {\bibinfo {title} {Information scrambling over
  bipartitions: Equilibration, entropy production, and typicality},\
  }\href@noop {} {\bibfield  {journal} {\bibinfo  {journal} {Physical Review
  Letters}\ }\textbf {\bibinfo {volume} {126}},\ \bibinfo {pages} {030601}
  (\bibinfo {year} {2021})}\BibitemShut {NoStop}%
\bibitem [{\citenamefont {McCulloch}\ and\ \citenamefont
  {Von~Keyserlingk}(2022)}]{mcculloch2022operator}%
  \BibitemOpen
  \bibfield  {author} {\bibinfo {author} {\bibfnamefont {E.}~\bibnamefont
  {McCulloch}}\ and\ \bibinfo {author} {\bibfnamefont {C.}~\bibnamefont
  {Von~Keyserlingk}},\ }\bibfield  {title} {\bibinfo {title} {Operator
  spreading in the memory matrix formalism},\ }\href@noop {} {\bibfield
  {journal} {\bibinfo  {journal} {Journal of Physics A: Mathematical and
  Theoretical}\ }\textbf {\bibinfo {volume} {55}},\ \bibinfo {pages} {274007}
  (\bibinfo {year} {2022})}\BibitemShut {NoStop}%
\bibitem [{\citenamefont {Parker}\ \emph {et~al.}(2019)\citenamefont {Parker},
  \citenamefont {Cao}, \citenamefont {Avdoshkin}, \citenamefont {Scaffidi},\
  and\ \citenamefont {Altman}}]{parker2019universal}%
  \BibitemOpen
  \bibfield  {author} {\bibinfo {author} {\bibfnamefont {D.~E.}\ \bibnamefont
  {Parker}}, \bibinfo {author} {\bibfnamefont {X.}~\bibnamefont {Cao}},
  \bibinfo {author} {\bibfnamefont {A.}~\bibnamefont {Avdoshkin}}, \bibinfo
  {author} {\bibfnamefont {T.}~\bibnamefont {Scaffidi}},\ and\ \bibinfo
  {author} {\bibfnamefont {E.}~\bibnamefont {Altman}},\ }\bibfield  {title}
  {\bibinfo {title} {A universal operator growth hypothesis},\ }\href@noop {}
  {\bibfield  {journal} {\bibinfo  {journal} {Physical Review X}\ }\textbf
  {\bibinfo {volume} {9}},\ \bibinfo {pages} {041017} (\bibinfo {year}
  {2019})}\BibitemShut {NoStop}%
\bibitem [{\citenamefont {Yates}\ and\ \citenamefont
  {Mitra}(2021)}]{yates2021strong}%
  \BibitemOpen
  \bibfield  {author} {\bibinfo {author} {\bibfnamefont {D.~J.}\ \bibnamefont
  {Yates}}\ and\ \bibinfo {author} {\bibfnamefont {A.}~\bibnamefont {Mitra}},\
  }\bibfield  {title} {\bibinfo {title} {Strong and almost strong modes of
  floquet spin chains in krylov subspaces},\ }\href@noop {} {\bibfield
  {journal} {\bibinfo  {journal} {Physical Review B}\ }\textbf {\bibinfo
  {volume} {104}},\ \bibinfo {pages} {195121} (\bibinfo {year}
  {2021})}\BibitemShut {NoStop}%
\bibitem [{\citenamefont {Rabinovici}\ \emph {et~al.}(2021)\citenamefont
  {Rabinovici}, \citenamefont {S{\'a}nchez-Garrido}, \citenamefont {Shir},\
  and\ \citenamefont {Sonner}}]{rabinovici2021operator}%
  \BibitemOpen
  \bibfield  {author} {\bibinfo {author} {\bibfnamefont {E.}~\bibnamefont
  {Rabinovici}}, \bibinfo {author} {\bibfnamefont {A.}~\bibnamefont
  {S{\'a}nchez-Garrido}}, \bibinfo {author} {\bibfnamefont {R.}~\bibnamefont
  {Shir}},\ and\ \bibinfo {author} {\bibfnamefont {J.}~\bibnamefont {Sonner}},\
  }\bibfield  {title} {\bibinfo {title} {Operator complexity: a journey to the
  edge of krylov space},\ }\href@noop {} {\bibfield  {journal} {\bibinfo
  {journal} {Journal of High Energy Physics}\ }\textbf {\bibinfo {volume}
  {2021}},\ \bibinfo {pages} {1} (\bibinfo {year} {2021})}\BibitemShut
  {NoStop}%
\bibitem [{\citenamefont {Noh}(2021)}]{noh2021operator}%
  \BibitemOpen
  \bibfield  {author} {\bibinfo {author} {\bibfnamefont {J.~D.}\ \bibnamefont
  {Noh}},\ }\bibfield  {title} {\bibinfo {title} {Operator growth in the
  transverse-field ising spin chain with integrability-breaking longitudinal
  field},\ }\href@noop {} {\bibfield  {journal} {\bibinfo  {journal} {Physical
  Review E}\ }\textbf {\bibinfo {volume} {104}},\ \bibinfo {pages} {034112}
  (\bibinfo {year} {2021})}\BibitemShut {NoStop}%
\bibitem [{\citenamefont {Dymarsky}\ and\ \citenamefont
  {Smolkin}(2021)}]{dymarsky2021krylov}%
  \BibitemOpen
  \bibfield  {author} {\bibinfo {author} {\bibfnamefont {A.}~\bibnamefont
  {Dymarsky}}\ and\ \bibinfo {author} {\bibfnamefont {M.}~\bibnamefont
  {Smolkin}},\ }\bibfield  {title} {\bibinfo {title} {Krylov complexity in
  conformal field theory},\ }\href@noop {} {\bibfield  {journal} {\bibinfo
  {journal} {Physical Review D}\ }\textbf {\bibinfo {volume} {104}},\ \bibinfo
  {pages} {L081702} (\bibinfo {year} {2021})}\BibitemShut {NoStop}%
\bibitem [{\citenamefont {Caputa}\ \emph {et~al.}(2022)\citenamefont {Caputa},
  \citenamefont {Magan},\ and\ \citenamefont
  {Patramanis}}]{caputa2022geometry}%
  \BibitemOpen
  \bibfield  {author} {\bibinfo {author} {\bibfnamefont {P.}~\bibnamefont
  {Caputa}}, \bibinfo {author} {\bibfnamefont {J.~M.}\ \bibnamefont {Magan}},\
  and\ \bibinfo {author} {\bibfnamefont {D.}~\bibnamefont {Patramanis}},\
  }\bibfield  {title} {\bibinfo {title} {Geometry of krylov complexity},\
  }\href@noop {} {\bibfield  {journal} {\bibinfo  {journal} {Physical Review
  Research}\ }\textbf {\bibinfo {volume} {4}},\ \bibinfo {pages} {013041}
  (\bibinfo {year} {2022})}\BibitemShut {NoStop}%
\bibitem [{\citenamefont {Rabinovici}\ \emph
  {et~al.}(2022{\natexlab{a}})\citenamefont {Rabinovici}, \citenamefont
  {S{\'a}nchez-Garrido}, \citenamefont {Shir},\ and\ \citenamefont
  {Sonner}}]{rabinovici2022krylov}%
  \BibitemOpen
  \bibfield  {author} {\bibinfo {author} {\bibfnamefont {E.}~\bibnamefont
  {Rabinovici}}, \bibinfo {author} {\bibfnamefont {A.}~\bibnamefont
  {S{\'a}nchez-Garrido}}, \bibinfo {author} {\bibfnamefont {R.}~\bibnamefont
  {Shir}},\ and\ \bibinfo {author} {\bibfnamefont {J.}~\bibnamefont {Sonner}},\
  }\bibfield  {title} {\bibinfo {title} {Krylov localization and suppression of
  complexity},\ }\href@noop {} {\bibfield  {journal} {\bibinfo  {journal}
  {Journal of High Energy Physics}\ }\textbf {\bibinfo {volume} {2022}},\
  \bibinfo {pages} {1} (\bibinfo {year} {2022}{\natexlab{a}})}\BibitemShut
  {NoStop}%
\bibitem [{\citenamefont {Avdoshkin}\ \emph {et~al.}(2022)\citenamefont
  {Avdoshkin}, \citenamefont {Dymarsky},\ and\ \citenamefont
  {Smolkin}}]{avdoshkin2022krylov}%
  \BibitemOpen
  \bibfield  {author} {\bibinfo {author} {\bibfnamefont {A.}~\bibnamefont
  {Avdoshkin}}, \bibinfo {author} {\bibfnamefont {A.}~\bibnamefont
  {Dymarsky}},\ and\ \bibinfo {author} {\bibfnamefont {M.}~\bibnamefont
  {Smolkin}},\ }\bibfield  {title} {\bibinfo {title} {Krylov complexity in
  quantum field theory, and beyond},\ }\href@noop {} {\bibfield  {journal}
  {\bibinfo  {journal} {arXiv preprint arXiv:2212.14429}\ } (\bibinfo {year}
  {2022})}\BibitemShut {NoStop}%
\bibitem [{\citenamefont {Rabinovici}\ \emph
  {et~al.}(2022{\natexlab{b}})\citenamefont {Rabinovici}, \citenamefont
  {S{\'a}nchez-Garrido}, \citenamefont {Shir},\ and\ \citenamefont
  {Sonner}}]{rabinovici2022k}%
  \BibitemOpen
  \bibfield  {author} {\bibinfo {author} {\bibfnamefont {E.}~\bibnamefont
  {Rabinovici}}, \bibinfo {author} {\bibfnamefont {A.}~\bibnamefont
  {S{\'a}nchez-Garrido}}, \bibinfo {author} {\bibfnamefont {R.}~\bibnamefont
  {Shir}},\ and\ \bibinfo {author} {\bibfnamefont {J.}~\bibnamefont {Sonner}},\
  }\bibfield  {title} {\bibinfo {title} {K-complexity from integrability to
  chaos},\ }\href@noop {} {\bibfield  {journal} {\bibinfo  {journal} {arXiv
  preprint arXiv:2207.07701}\ } (\bibinfo {year}
  {2022}{\natexlab{b}})}\BibitemShut {NoStop}%
\bibitem [{\citenamefont {Bhattacharya}\ \emph {et~al.}(2022)\citenamefont
  {Bhattacharya}, \citenamefont {Nandy}, \citenamefont {Nath},\ and\
  \citenamefont {Sahu}}]{bhattacharya2022operator}%
  \BibitemOpen
  \bibfield  {author} {\bibinfo {author} {\bibfnamefont {A.}~\bibnamefont
  {Bhattacharya}}, \bibinfo {author} {\bibfnamefont {P.}~\bibnamefont {Nandy}},
  \bibinfo {author} {\bibfnamefont {P.~P.}\ \bibnamefont {Nath}},\ and\
  \bibinfo {author} {\bibfnamefont {H.}~\bibnamefont {Sahu}},\ }\bibfield
  {title} {\bibinfo {title} {Operator growth and krylov construction in
  dissipative open quantum systems},\ }\href@noop {} {\bibfield  {journal}
  {\bibinfo  {journal} {Journal of High Energy Physics}\ }\textbf {\bibinfo
  {volume} {2022}},\ \bibinfo {pages} {1} (\bibinfo {year} {2022})}\BibitemShut
  {NoStop}%
\bibitem [{\citenamefont {Bhattacharya}\ \emph {et~al.}(2023)\citenamefont
  {Bhattacharya}, \citenamefont {Nandy}, \citenamefont {Nath},\ and\
  \citenamefont {Sahu}}]{bhattacharya2023krylov}%
  \BibitemOpen
  \bibfield  {author} {\bibinfo {author} {\bibfnamefont {A.}~\bibnamefont
  {Bhattacharya}}, \bibinfo {author} {\bibfnamefont {P.}~\bibnamefont {Nandy}},
  \bibinfo {author} {\bibfnamefont {P.~P.}\ \bibnamefont {Nath}},\ and\
  \bibinfo {author} {\bibfnamefont {H.}~\bibnamefont {Sahu}},\ }\bibfield
  {title} {\bibinfo {title} {On krylov complexity in open systems: an approach
  via bi-lanczos algorithm},\ }\href@noop {} {\bibfield  {journal} {\bibinfo
  {journal} {arXiv preprint arXiv:2303.04175}\ } (\bibinfo {year}
  {2023})}\BibitemShut {NoStop}%
\bibitem [{\citenamefont {Suchsland}\ \emph {et~al.}(2023)\citenamefont
  {Suchsland}, \citenamefont {Moessner},\ and\ \citenamefont
  {Claeys}}]{suchsland2023krylov}%
  \BibitemOpen
  \bibfield  {author} {\bibinfo {author} {\bibfnamefont {P.}~\bibnamefont
  {Suchsland}}, \bibinfo {author} {\bibfnamefont {R.}~\bibnamefont
  {Moessner}},\ and\ \bibinfo {author} {\bibfnamefont {P.~W.}\ \bibnamefont
  {Claeys}},\ }\bibfield  {title} {\bibinfo {title} {Krylov complexity and
  trotter transitions in unitary circuit dynamics},\ }\href@noop {} {\bibfield
  {journal} {\bibinfo  {journal} {arXiv preprint arXiv:2308.03851}\ } (\bibinfo
  {year} {2023})}\BibitemShut {NoStop}%
\bibitem [{\citenamefont {Li}\ \emph {et~al.}(2017)\citenamefont {Li},
  \citenamefont {Fan}, \citenamefont {Wang}, \citenamefont {Ye}, \citenamefont
  {Zeng}, \citenamefont {Zhai}, \citenamefont {Peng},\ and\ \citenamefont
  {Du}}]{li2017measuring}%
  \BibitemOpen
  \bibfield  {author} {\bibinfo {author} {\bibfnamefont {J.}~\bibnamefont
  {Li}}, \bibinfo {author} {\bibfnamefont {R.}~\bibnamefont {Fan}}, \bibinfo
  {author} {\bibfnamefont {H.}~\bibnamefont {Wang}}, \bibinfo {author}
  {\bibfnamefont {B.}~\bibnamefont {Ye}}, \bibinfo {author} {\bibfnamefont
  {B.}~\bibnamefont {Zeng}}, \bibinfo {author} {\bibfnamefont {H.}~\bibnamefont
  {Zhai}}, \bibinfo {author} {\bibfnamefont {X.}~\bibnamefont {Peng}},\ and\
  \bibinfo {author} {\bibfnamefont {J.}~\bibnamefont {Du}},\ }\bibfield
  {title} {\bibinfo {title} {Measuring out-of-time-order correlators on a
  nuclear magnetic resonance quantum simulator},\ }\href@noop {} {\bibfield
  {journal} {\bibinfo  {journal} {Physical Review X}\ }\textbf {\bibinfo
  {volume} {7}},\ \bibinfo {pages} {031011} (\bibinfo {year}
  {2017})}\BibitemShut {NoStop}%
\bibitem [{\citenamefont {Green}\ \emph {et~al.}(2022)\citenamefont {Green},
  \citenamefont {Elben}, \citenamefont {Alderete}, \citenamefont {Joshi},
  \citenamefont {Nguyen}, \citenamefont {Zache}, \citenamefont {Zhu},
  \citenamefont {Sundar},\ and\ \citenamefont {Linke}}]{green2022experimental}%
  \BibitemOpen
  \bibfield  {author} {\bibinfo {author} {\bibfnamefont {A.~M.}\ \bibnamefont
  {Green}}, \bibinfo {author} {\bibfnamefont {A.}~\bibnamefont {Elben}},
  \bibinfo {author} {\bibfnamefont {C.~H.}\ \bibnamefont {Alderete}}, \bibinfo
  {author} {\bibfnamefont {L.~K.}\ \bibnamefont {Joshi}}, \bibinfo {author}
  {\bibfnamefont {N.~H.}\ \bibnamefont {Nguyen}}, \bibinfo {author}
  {\bibfnamefont {T.~V.}\ \bibnamefont {Zache}}, \bibinfo {author}
  {\bibfnamefont {Y.}~\bibnamefont {Zhu}}, \bibinfo {author} {\bibfnamefont
  {B.}~\bibnamefont {Sundar}},\ and\ \bibinfo {author} {\bibfnamefont {N.~M.}\
  \bibnamefont {Linke}},\ }\bibfield  {title} {\bibinfo {title} {Experimental
  measurement of out-of-time-ordered correlators at finite temperature},\
  }\href@noop {} {\bibfield  {journal} {\bibinfo  {journal} {Physical Review
  Letters}\ }\textbf {\bibinfo {volume} {128}},\ \bibinfo {pages} {140601}
  (\bibinfo {year} {2022})}\BibitemShut {NoStop}%
\bibitem [{\citenamefont {G{\"a}rttner}\ \emph {et~al.}(2017)\citenamefont
  {G{\"a}rttner}, \citenamefont {Bohnet}, \citenamefont {Safavi-Naini},
  \citenamefont {Wall}, \citenamefont {Bollinger},\ and\ \citenamefont
  {Rey}}]{garttner2017measuring}%
  \BibitemOpen
  \bibfield  {author} {\bibinfo {author} {\bibfnamefont {M.}~\bibnamefont
  {G{\"a}rttner}}, \bibinfo {author} {\bibfnamefont {J.~G.}\ \bibnamefont
  {Bohnet}}, \bibinfo {author} {\bibfnamefont {A.}~\bibnamefont
  {Safavi-Naini}}, \bibinfo {author} {\bibfnamefont {M.~L.}\ \bibnamefont
  {Wall}}, \bibinfo {author} {\bibfnamefont {J.~J.}\ \bibnamefont
  {Bollinger}},\ and\ \bibinfo {author} {\bibfnamefont {A.~M.}\ \bibnamefont
  {Rey}},\ }\bibfield  {title} {\bibinfo {title} {Measuring out-of-time-order
  correlations and multiple quantum spectra in a trapped-ion quantum magnet},\
  }\href@noop {} {\bibfield  {journal} {\bibinfo  {journal} {Nature Physics}\
  }\textbf {\bibinfo {volume} {13}},\ \bibinfo {pages} {781} (\bibinfo {year}
  {2017})}\BibitemShut {NoStop}%
\bibitem [{\citenamefont {Zhu}\ \emph {et~al.}(2016)\citenamefont {Zhu},
  \citenamefont {Hafezi},\ and\ \citenamefont {Grover}}]{zhu2016measurement}%
  \BibitemOpen
  \bibfield  {author} {\bibinfo {author} {\bibfnamefont {G.}~\bibnamefont
  {Zhu}}, \bibinfo {author} {\bibfnamefont {M.}~\bibnamefont {Hafezi}},\ and\
  \bibinfo {author} {\bibfnamefont {T.}~\bibnamefont {Grover}},\ }\bibfield
  {title} {\bibinfo {title} {Measurement of many-body chaos using a quantum
  clock},\ }\href@noop {} {\bibfield  {journal} {\bibinfo  {journal} {Physical
  Review A}\ }\textbf {\bibinfo {volume} {94}},\ \bibinfo {pages} {062329}
  (\bibinfo {year} {2016})}\BibitemShut {NoStop}%
\bibitem [{\citenamefont {Swingle}\ \emph {et~al.}(2016)\citenamefont
  {Swingle}, \citenamefont {Bentsen}, \citenamefont {Schleier-Smith},\ and\
  \citenamefont {Hayden}}]{swingle2016measuring}%
  \BibitemOpen
  \bibfield  {author} {\bibinfo {author} {\bibfnamefont {B.}~\bibnamefont
  {Swingle}}, \bibinfo {author} {\bibfnamefont {G.}~\bibnamefont {Bentsen}},
  \bibinfo {author} {\bibfnamefont {M.}~\bibnamefont {Schleier-Smith}},\ and\
  \bibinfo {author} {\bibfnamefont {P.}~\bibnamefont {Hayden}},\ }\bibfield
  {title} {\bibinfo {title} {Measuring the scrambling of quantum information},\
  }\href@noop {} {\bibfield  {journal} {\bibinfo  {journal} {Physical Review
  A}\ }\textbf {\bibinfo {volume} {94}},\ \bibinfo {pages} {040302} (\bibinfo
  {year} {2016})}\BibitemShut {NoStop}%
\bibitem [{\citenamefont {Yao}\ \emph {et~al.}(2016)\citenamefont {Yao},
  \citenamefont {Grusdt}, \citenamefont {Swingle}, \citenamefont {Lukin},
  \citenamefont {Stamper-Kurn}, \citenamefont {Moore},\ and\ \citenamefont
  {Demler}}]{yao2016interferometric}%
  \BibitemOpen
  \bibfield  {author} {\bibinfo {author} {\bibfnamefont {N.~Y.}\ \bibnamefont
  {Yao}}, \bibinfo {author} {\bibfnamefont {F.}~\bibnamefont {Grusdt}},
  \bibinfo {author} {\bibfnamefont {B.}~\bibnamefont {Swingle}}, \bibinfo
  {author} {\bibfnamefont {M.~D.}\ \bibnamefont {Lukin}}, \bibinfo {author}
  {\bibfnamefont {D.~M.}\ \bibnamefont {Stamper-Kurn}}, \bibinfo {author}
  {\bibfnamefont {J.~E.}\ \bibnamefont {Moore}},\ and\ \bibinfo {author}
  {\bibfnamefont {E.~A.}\ \bibnamefont {Demler}},\ }\bibfield  {title}
  {\bibinfo {title} {Interferometric approach to probing fast scrambling},\
  }\href@noop {} {\bibfield  {journal} {\bibinfo  {journal} {arXiv preprint
  arXiv:1607.01801}\ } (\bibinfo {year} {2016})}\BibitemShut {NoStop}%
\bibitem [{\citenamefont {Halpern}(2017)}]{halpern2017jarzynski}%
  \BibitemOpen
  \bibfield  {author} {\bibinfo {author} {\bibfnamefont {N.~Y.}\ \bibnamefont
  {Halpern}},\ }\bibfield  {title} {\bibinfo {title} {Jarzynski-like equality
  for the out-of-time-ordered correlator},\ }\href@noop {} {\bibfield
  {journal} {\bibinfo  {journal} {Physical Review A}\ }\textbf {\bibinfo
  {volume} {95}},\ \bibinfo {pages} {012120} (\bibinfo {year}
  {2017})}\BibitemShut {NoStop}%
\bibitem [{\citenamefont {Bohrdt}\ \emph {et~al.}(2017)\citenamefont {Bohrdt},
  \citenamefont {Mendl}, \citenamefont {Endres},\ and\ \citenamefont
  {Knap}}]{bohrdt2017scrambling}%
  \BibitemOpen
  \bibfield  {author} {\bibinfo {author} {\bibfnamefont {A.}~\bibnamefont
  {Bohrdt}}, \bibinfo {author} {\bibfnamefont {C.~B.}\ \bibnamefont {Mendl}},
  \bibinfo {author} {\bibfnamefont {M.}~\bibnamefont {Endres}},\ and\ \bibinfo
  {author} {\bibfnamefont {M.}~\bibnamefont {Knap}},\ }\bibfield  {title}
  {\bibinfo {title} {Scrambling and thermalization in a diffusive quantum
  many-body system},\ }\href@noop {} {\bibfield  {journal} {\bibinfo  {journal}
  {New Journal of Physics}\ }\textbf {\bibinfo {volume} {19}},\ \bibinfo
  {pages} {063001} (\bibinfo {year} {2017})}\BibitemShut {NoStop}%
\bibitem [{\citenamefont {Tsuji}\ \emph {et~al.}(2017)\citenamefont {Tsuji},
  \citenamefont {Werner},\ and\ \citenamefont {Ueda}}]{tsuji2017exact}%
  \BibitemOpen
  \bibfield  {author} {\bibinfo {author} {\bibfnamefont {N.}~\bibnamefont
  {Tsuji}}, \bibinfo {author} {\bibfnamefont {P.}~\bibnamefont {Werner}},\ and\
  \bibinfo {author} {\bibfnamefont {M.}~\bibnamefont {Ueda}},\ }\bibfield
  {title} {\bibinfo {title} {Exact out-of-time-ordered correlation functions
  for an interacting lattice fermion model},\ }\href@noop {} {\bibfield
  {journal} {\bibinfo  {journal} {Physical Review A}\ }\textbf {\bibinfo
  {volume} {95}},\ \bibinfo {pages} {011601} (\bibinfo {year}
  {2017})}\BibitemShut {NoStop}%
\bibitem [{\citenamefont {Nie}\ \emph {et~al.}(2020)\citenamefont {Nie},
  \citenamefont {Wei}, \citenamefont {Chen}, \citenamefont {Zhang},
  \citenamefont {Zhao}, \citenamefont {Qiu}, \citenamefont {Tian},
  \citenamefont {Ji}, \citenamefont {Xin}, \citenamefont {Lu} \emph
  {et~al.}}]{nie2020experimental}%
  \BibitemOpen
  \bibfield  {author} {\bibinfo {author} {\bibfnamefont {X.}~\bibnamefont
  {Nie}}, \bibinfo {author} {\bibfnamefont {B.-B.}\ \bibnamefont {Wei}},
  \bibinfo {author} {\bibfnamefont {X.}~\bibnamefont {Chen}}, \bibinfo {author}
  {\bibfnamefont {Z.}~\bibnamefont {Zhang}}, \bibinfo {author} {\bibfnamefont
  {X.}~\bibnamefont {Zhao}}, \bibinfo {author} {\bibfnamefont {C.}~\bibnamefont
  {Qiu}}, \bibinfo {author} {\bibfnamefont {Y.}~\bibnamefont {Tian}}, \bibinfo
  {author} {\bibfnamefont {Y.}~\bibnamefont {Ji}}, \bibinfo {author}
  {\bibfnamefont {T.}~\bibnamefont {Xin}}, \bibinfo {author} {\bibfnamefont
  {D.}~\bibnamefont {Lu}}, \emph {et~al.},\ }\bibfield  {title} {\bibinfo
  {title} {Experimental observation of equilibrium and dynamical quantum phase
  transitions via out-of-time-ordered correlators},\ }\href@noop {} {\bibfield
  {journal} {\bibinfo  {journal} {Physical Review Letters}\ }\textbf {\bibinfo
  {volume} {124}},\ \bibinfo {pages} {250601} (\bibinfo {year}
  {2020})}\BibitemShut {NoStop}%
\bibitem [{\citenamefont {Dressel}\ \emph {et~al.}(2018)\citenamefont
  {Dressel}, \citenamefont {Alonso}, \citenamefont {Waegell},\ and\
  \citenamefont {Halpern}}]{dressel2018strengthening}%
  \BibitemOpen
  \bibfield  {author} {\bibinfo {author} {\bibfnamefont {J.}~\bibnamefont
  {Dressel}}, \bibinfo {author} {\bibfnamefont {J.~R.~G.}\ \bibnamefont
  {Alonso}}, \bibinfo {author} {\bibfnamefont {M.}~\bibnamefont {Waegell}},\
  and\ \bibinfo {author} {\bibfnamefont {N.~Y.}\ \bibnamefont {Halpern}},\
  }\bibfield  {title} {\bibinfo {title} {Strengthening weak measurements of
  qubit out-of-time-order correlators},\ }\href@noop {} {\bibfield  {journal}
  {\bibinfo  {journal} {Physical Review A}\ }\textbf {\bibinfo {volume} {98}},\
  \bibinfo {pages} {012132} (\bibinfo {year} {2018})}\BibitemShut {NoStop}%
\bibitem [{\citenamefont {Joshi}\ \emph {et~al.}(2020)\citenamefont {Joshi},
  \citenamefont {Elben}, \citenamefont {Vermersch}, \citenamefont {Brydges},
  \citenamefont {Maier}, \citenamefont {Zoller}, \citenamefont {Blatt},\ and\
  \citenamefont {Roos}}]{joshi2020quantum}%
  \BibitemOpen
  \bibfield  {author} {\bibinfo {author} {\bibfnamefont {M.~K.}\ \bibnamefont
  {Joshi}}, \bibinfo {author} {\bibfnamefont {A.}~\bibnamefont {Elben}},
  \bibinfo {author} {\bibfnamefont {B.}~\bibnamefont {Vermersch}}, \bibinfo
  {author} {\bibfnamefont {T.}~\bibnamefont {Brydges}}, \bibinfo {author}
  {\bibfnamefont {C.}~\bibnamefont {Maier}}, \bibinfo {author} {\bibfnamefont
  {P.}~\bibnamefont {Zoller}}, \bibinfo {author} {\bibfnamefont
  {R.}~\bibnamefont {Blatt}},\ and\ \bibinfo {author} {\bibfnamefont {C.~F.}\
  \bibnamefont {Roos}},\ }\bibfield  {title} {\bibinfo {title} {Quantum
  information scrambling in a trapped-ion quantum simulator with tunable range
  interactions},\ }\href@noop {} {\bibfield  {journal} {\bibinfo  {journal}
  {Physical Review Letters}\ }\textbf {\bibinfo {volume} {124}},\ \bibinfo
  {pages} {240505} (\bibinfo {year} {2020})}\BibitemShut {NoStop}%
\bibitem [{\citenamefont {Asban}\ \emph {et~al.}(2021)\citenamefont {Asban},
  \citenamefont {Dorfman},\ and\ \citenamefont
  {Mukamel}}]{asban2021interferometric}%
  \BibitemOpen
  \bibfield  {author} {\bibinfo {author} {\bibfnamefont {S.}~\bibnamefont
  {Asban}}, \bibinfo {author} {\bibfnamefont {K.~E.}\ \bibnamefont {Dorfman}},\
  and\ \bibinfo {author} {\bibfnamefont {S.}~\bibnamefont {Mukamel}},\
  }\bibfield  {title} {\bibinfo {title} {Interferometric spectroscopy with
  quantum light: Revealing out-of-time-ordering correlators},\ }\href@noop {}
  {\bibfield  {journal} {\bibinfo  {journal} {The Journal of chemical physics}\
  }\textbf {\bibinfo {volume} {154}} (\bibinfo {year} {2021})}\BibitemShut
  {NoStop}%
\bibitem [{\citenamefont {Vermersch}\ \emph {et~al.}(2019)\citenamefont
  {Vermersch}, \citenamefont {Elben}, \citenamefont {Sieberer}, \citenamefont
  {Yao},\ and\ \citenamefont {Zoller}}]{vermersch2019probing}%
  \BibitemOpen
  \bibfield  {author} {\bibinfo {author} {\bibfnamefont {B.}~\bibnamefont
  {Vermersch}}, \bibinfo {author} {\bibfnamefont {A.}~\bibnamefont {Elben}},
  \bibinfo {author} {\bibfnamefont {L.~M.}\ \bibnamefont {Sieberer}}, \bibinfo
  {author} {\bibfnamefont {N.~Y.}\ \bibnamefont {Yao}},\ and\ \bibinfo {author}
  {\bibfnamefont {P.}~\bibnamefont {Zoller}},\ }\bibfield  {title} {\bibinfo
  {title} {Probing scrambling using statistical correlations between randomized
  measurements},\ }\href@noop {} {\bibfield  {journal} {\bibinfo  {journal}
  {Physical Review X}\ }\textbf {\bibinfo {volume} {9}},\ \bibinfo {pages}
  {021061} (\bibinfo {year} {2019})}\BibitemShut {NoStop}%
\bibitem [{\citenamefont {Blocher}\ \emph {et~al.}(2022)\citenamefont
  {Blocher}, \citenamefont {Asaad}, \citenamefont {Mourik}, \citenamefont
  {Johnson}, \citenamefont {Morello},\ and\ \citenamefont
  {M{\o}lmer}}]{blocher2022measuring}%
  \BibitemOpen
  \bibfield  {author} {\bibinfo {author} {\bibfnamefont {P.~D.}\ \bibnamefont
  {Blocher}}, \bibinfo {author} {\bibfnamefont {S.}~\bibnamefont {Asaad}},
  \bibinfo {author} {\bibfnamefont {V.}~\bibnamefont {Mourik}}, \bibinfo
  {author} {\bibfnamefont {M.~A.}\ \bibnamefont {Johnson}}, \bibinfo {author}
  {\bibfnamefont {A.}~\bibnamefont {Morello}},\ and\ \bibinfo {author}
  {\bibfnamefont {K.}~\bibnamefont {M{\o}lmer}},\ }\bibfield  {title} {\bibinfo
  {title} {Measuring out-of-time-ordered correlation functions without
  reversing time evolution},\ }\href@noop {} {\bibfield  {journal} {\bibinfo
  {journal} {Physical Review A}\ }\textbf {\bibinfo {volume} {106}},\ \bibinfo
  {pages} {042429} (\bibinfo {year} {2022})}\BibitemShut {NoStop}%
\bibitem [{\citenamefont {Sundar}\ \emph {et~al.}(2022)\citenamefont {Sundar},
  \citenamefont {Elben}, \citenamefont {Joshi},\ and\ \citenamefont
  {Zache}}]{sundar2022proposal}%
  \BibitemOpen
  \bibfield  {author} {\bibinfo {author} {\bibfnamefont {B.}~\bibnamefont
  {Sundar}}, \bibinfo {author} {\bibfnamefont {A.}~\bibnamefont {Elben}},
  \bibinfo {author} {\bibfnamefont {L.~K.}\ \bibnamefont {Joshi}},\ and\
  \bibinfo {author} {\bibfnamefont {T.~V.}\ \bibnamefont {Zache}},\ }\bibfield
  {title} {\bibinfo {title} {Proposal for measuring out-of-time-ordered
  correlators at finite temperature with coupled spin chains},\ }\href@noop {}
  {\bibfield  {journal} {\bibinfo  {journal} {New Journal of Physics}\ }\textbf
  {\bibinfo {volume} {24}},\ \bibinfo {pages} {023037} (\bibinfo {year}
  {2022})}\BibitemShut {NoStop}%
\bibitem [{\citenamefont {Merkel}\ \emph {et~al.}(2010)\citenamefont {Merkel},
  \citenamefont {Riofrio}, \citenamefont {Flammia},\ and\ \citenamefont
  {Deutsch}}]{merkel2010random}%
  \BibitemOpen
  \bibfield  {author} {\bibinfo {author} {\bibfnamefont {S.~T.}\ \bibnamefont
  {Merkel}}, \bibinfo {author} {\bibfnamefont {C.~A.}\ \bibnamefont {Riofrio}},
  \bibinfo {author} {\bibfnamefont {S.~T.}\ \bibnamefont {Flammia}},\ and\
  \bibinfo {author} {\bibfnamefont {I.~H.}\ \bibnamefont {Deutsch}},\
  }\bibfield  {title} {\bibinfo {title} {Random unitary maps for quantum state
  reconstruction},\ }\href@noop {} {\bibfield  {journal} {\bibinfo  {journal}
  {Physical Review A}\ }\textbf {\bibinfo {volume} {81}},\ \bibinfo {pages}
  {032126} (\bibinfo {year} {2010})}\BibitemShut {NoStop}%
\bibitem [{\citenamefont {Sreeram}\ and\ \citenamefont
  {Madhok}(2021)}]{sreeram2021quantum}%
  \BibitemOpen
  \bibfield  {author} {\bibinfo {author} {\bibfnamefont {P.}~\bibnamefont
  {Sreeram}}\ and\ \bibinfo {author} {\bibfnamefont {V.}~\bibnamefont
  {Madhok}},\ }\bibfield  {title} {\bibinfo {title} {Quantum tomography with
  random diagonal unitary maps and statistical bounds on information generation
  using random matrix theory},\ }\href@noop {} {\bibfield  {journal} {\bibinfo
  {journal} {Physical Review A}\ }\textbf {\bibinfo {volume} {104}},\ \bibinfo
  {pages} {032404} (\bibinfo {year} {2021})}\BibitemShut {NoStop}%
\bibitem [{\citenamefont {Pesin}(1977)}]{pesin1977characteristic}%
  \BibitemOpen
  \bibfield  {author} {\bibinfo {author} {\bibfnamefont {Y.~B.}\ \bibnamefont
  {Pesin}},\ }\bibfield  {title} {\bibinfo {title} {Characteristic lyapunov
  exponents and smooth ergodic theory},\ }\href@noop {} {\bibfield  {journal}
  {\bibinfo  {journal} {Russian Mathematical Surveys}\ }\textbf {\bibinfo
  {volume} {32}},\ \bibinfo {pages} {55} (\bibinfo {year} {1977})}\BibitemShut
  {NoStop}%
\bibitem [{\citenamefont {Prosen}(2007)}]{prosen2007chaos}%
  \BibitemOpen
  \bibfield  {author} {\bibinfo {author} {\bibfnamefont {T.}~\bibnamefont
  {Prosen}},\ }\bibfield  {title} {\bibinfo {title} {Chaos and complexity of
  quantum motion},\ }\href@noop {} {\bibfield  {journal} {\bibinfo  {journal}
  {Journal of Physics A: Mathematical and Theoretical}\ }\textbf {\bibinfo
  {volume} {40}},\ \bibinfo {pages} {7881} (\bibinfo {year}
  {2007})}\BibitemShut {NoStop}%
\bibitem [{\citenamefont {Pineda}\ and\ \citenamefont
  {Prosen}(2007)}]{pineda2007universal}%
  \BibitemOpen
  \bibfield  {author} {\bibinfo {author} {\bibfnamefont {C.}~\bibnamefont
  {Pineda}}\ and\ \bibinfo {author} {\bibfnamefont {T.}~\bibnamefont
  {Prosen}},\ }\bibfield  {title} {\bibinfo {title} {Universal and nonuniversal
  level statistics in a chaotic quantum spin chain},\ }\href@noop {} {\bibfield
   {journal} {\bibinfo  {journal} {Physical Review E}\ }\textbf {\bibinfo
  {volume} {76}},\ \bibinfo {pages} {061127} (\bibinfo {year}
  {2007})}\BibitemShut {NoStop}%
\bibitem [{\citenamefont {Kukuljan}\ \emph {et~al.}(2017)\citenamefont
  {Kukuljan}, \citenamefont {Grozdanov},\ and\ \citenamefont
  {Prosen}}]{kukuljan2017weak}%
  \BibitemOpen
  \bibfield  {author} {\bibinfo {author} {\bibfnamefont {I.}~\bibnamefont
  {Kukuljan}}, \bibinfo {author} {\bibfnamefont {S.}~\bibnamefont
  {Grozdanov}},\ and\ \bibinfo {author} {\bibfnamefont {T.}~\bibnamefont
  {Prosen}},\ }\bibfield  {title} {\bibinfo {title} {Weak quantum chaos},\
  }\href@noop {} {\bibfield  {journal} {\bibinfo  {journal} {Physical Review
  B}\ }\textbf {\bibinfo {volume} {96}},\ \bibinfo {pages} {060301} (\bibinfo
  {year} {2017})}\BibitemShut {NoStop}%
\bibitem [{\citenamefont {Karthik}\ \emph {et~al.}(2007)\citenamefont
  {Karthik}, \citenamefont {Sharma},\ and\ \citenamefont
  {Lakshminarayan}}]{karthik2007entanglement}%
  \BibitemOpen
  \bibfield  {author} {\bibinfo {author} {\bibfnamefont {J.}~\bibnamefont
  {Karthik}}, \bibinfo {author} {\bibfnamefont {A.}~\bibnamefont {Sharma}},\
  and\ \bibinfo {author} {\bibfnamefont {A.}~\bibnamefont {Lakshminarayan}},\
  }\bibfield  {title} {\bibinfo {title} {Entanglement, avoided crossings, and
  quantum chaos in an ising model with a tilted magnetic field},\ }\href@noop
  {} {\bibfield  {journal} {\bibinfo  {journal} {Physical Review A}\ }\textbf
  {\bibinfo {volume} {75}},\ \bibinfo {pages} {022304} (\bibinfo {year}
  {2007})}\BibitemShut {NoStop}%
\bibitem [{\citenamefont {Prosen}\ and\ \citenamefont
  {{\v{Z}}nidari{\v{c}}}(2007)}]{prosen2007efficiency}%
  \BibitemOpen
  \bibfield  {author} {\bibinfo {author} {\bibfnamefont {T.}~\bibnamefont
  {Prosen}}\ and\ \bibinfo {author} {\bibfnamefont {M.}~\bibnamefont
  {{\v{Z}}nidari{\v{c}}}},\ }\bibfield  {title} {\bibinfo {title} {Is the
  efficiency of classical simulations of quantum dynamics related to
  integrability?},\ }\href@noop {} {\bibfield  {journal} {\bibinfo  {journal}
  {Physical Review E}\ }\textbf {\bibinfo {volume} {75}},\ \bibinfo {pages}
  {015202} (\bibinfo {year} {2007})}\BibitemShut {NoStop}%
\bibitem [{\citenamefont {Santos}(2004)}]{santos2004integrability}%
  \BibitemOpen
  \bibfield  {author} {\bibinfo {author} {\bibfnamefont {L.}~\bibnamefont
  {Santos}},\ }\bibfield  {title} {\bibinfo {title} {Integrability of a
  disordered heisenberg spin-1/2 chain},\ }\href@noop {} {\bibfield  {journal}
  {\bibinfo  {journal} {Journal of Physics A: Mathematical and General}\
  }\textbf {\bibinfo {volume} {37}},\ \bibinfo {pages} {4723} (\bibinfo {year}
  {2004})}\BibitemShut {NoStop}%
\bibitem [{\citenamefont {Santos}\ \emph {et~al.}(2004)\citenamefont {Santos},
  \citenamefont {Rigolin},\ and\ \citenamefont
  {Escobar}}]{santos2004entanglement}%
  \BibitemOpen
  \bibfield  {author} {\bibinfo {author} {\bibfnamefont {L.}~\bibnamefont
  {Santos}}, \bibinfo {author} {\bibfnamefont {G.}~\bibnamefont {Rigolin}},\
  and\ \bibinfo {author} {\bibfnamefont {C.}~\bibnamefont {Escobar}},\
  }\bibfield  {title} {\bibinfo {title} {Entanglement versus chaos in
  disordered spin chains},\ }\href@noop {} {\bibfield  {journal} {\bibinfo
  {journal} {Physical Review A}\ }\textbf {\bibinfo {volume} {69}},\ \bibinfo
  {pages} {042304} (\bibinfo {year} {2004})}\BibitemShut {NoStop}%
\bibitem [{\citenamefont {Bari{\v{s}}i{\'c}}\ \emph {et~al.}(2009)\citenamefont
  {Bari{\v{s}}i{\'c}}, \citenamefont {Prelov{\v{s}}ek}, \citenamefont
  {Metavitsiadis},\ and\ \citenamefont {Zotos}}]{barivsic2009incoherent}%
  \BibitemOpen
  \bibfield  {author} {\bibinfo {author} {\bibfnamefont {O.}~\bibnamefont
  {Bari{\v{s}}i{\'c}}}, \bibinfo {author} {\bibfnamefont {P.}~\bibnamefont
  {Prelov{\v{s}}ek}}, \bibinfo {author} {\bibfnamefont {A.}~\bibnamefont
  {Metavitsiadis}},\ and\ \bibinfo {author} {\bibfnamefont {X.}~\bibnamefont
  {Zotos}},\ }\bibfield  {title} {\bibinfo {title} {Incoherent transport
  induced by a single static impurity in a heisenberg chain},\ }\href@noop {}
  {\bibfield  {journal} {\bibinfo  {journal} {Physical Review B}\ }\textbf
  {\bibinfo {volume} {80}},\ \bibinfo {pages} {125118} (\bibinfo {year}
  {2009})}\BibitemShut {NoStop}%
\bibitem [{\citenamefont {Santos}\ and\ \citenamefont
  {Mitra}(2011)}]{santos2011domain}%
  \BibitemOpen
  \bibfield  {author} {\bibinfo {author} {\bibfnamefont {L.~F.}\ \bibnamefont
  {Santos}}\ and\ \bibinfo {author} {\bibfnamefont {A.}~\bibnamefont {Mitra}},\
  }\bibfield  {title} {\bibinfo {title} {Domain wall dynamics in integrable and
  chaotic spin-1/2 chains},\ }\href@noop {} {\bibfield  {journal} {\bibinfo
  {journal} {Physical Review E}\ }\textbf {\bibinfo {volume} {84}},\ \bibinfo
  {pages} {016206} (\bibinfo {year} {2011})}\BibitemShut {NoStop}%
\bibitem [{\citenamefont {Santos}\ \emph {et~al.}(2012)\citenamefont {Santos},
  \citenamefont {Borgonovi},\ and\ \citenamefont {Izrailev}}]{santos2012chaos}%
  \BibitemOpen
  \bibfield  {author} {\bibinfo {author} {\bibfnamefont {L.~F.}\ \bibnamefont
  {Santos}}, \bibinfo {author} {\bibfnamefont {F.}~\bibnamefont {Borgonovi}},\
  and\ \bibinfo {author} {\bibfnamefont {F.}~\bibnamefont {Izrailev}},\
  }\bibfield  {title} {\bibinfo {title} {Chaos and statistical relaxation in
  quantum systems of interacting particles},\ }\href@noop {} {\bibfield
  {journal} {\bibinfo  {journal} {Physical review letters}\ }\textbf {\bibinfo
  {volume} {108}},\ \bibinfo {pages} {094102} (\bibinfo {year}
  {2012})}\BibitemShut {NoStop}%
\bibitem [{\citenamefont {Gubin}\ and\ \citenamefont
  {F~Santos}(2012)}]{gubin2012quantum}%
  \BibitemOpen
  \bibfield  {author} {\bibinfo {author} {\bibfnamefont {A.}~\bibnamefont
  {Gubin}}\ and\ \bibinfo {author} {\bibfnamefont {L.}~\bibnamefont
  {F~Santos}},\ }\bibfield  {title} {\bibinfo {title} {Quantum chaos: An
  introduction via chains of interacting spins 1/2},\ }\href@noop {} {\bibfield
   {journal} {\bibinfo  {journal} {American Journal of Physics}\ }\textbf
  {\bibinfo {volume} {80}},\ \bibinfo {pages} {246} (\bibinfo {year}
  {2012})}\BibitemShut {NoStop}%
\bibitem [{\citenamefont {Brenes}\ \emph {et~al.}(2018)\citenamefont {Brenes},
  \citenamefont {Mascarenhas}, \citenamefont {Rigol},\ and\ \citenamefont
  {Goold}}]{brenes2018high}%
  \BibitemOpen
  \bibfield  {author} {\bibinfo {author} {\bibfnamefont {M.}~\bibnamefont
  {Brenes}}, \bibinfo {author} {\bibfnamefont {E.}~\bibnamefont {Mascarenhas}},
  \bibinfo {author} {\bibfnamefont {M.}~\bibnamefont {Rigol}},\ and\ \bibinfo
  {author} {\bibfnamefont {J.}~\bibnamefont {Goold}},\ }\bibfield  {title}
  {\bibinfo {title} {High-temperature coherent transport in the xxz chain in
  the presence of an impurity},\ }\href@noop {} {\bibfield  {journal} {\bibinfo
   {journal} {Physical Review B}\ }\textbf {\bibinfo {volume} {98}},\ \bibinfo
  {pages} {235128} (\bibinfo {year} {2018})}\BibitemShut {NoStop}%
\bibitem [{\citenamefont {Brenes}\ \emph {et~al.}(2020)\citenamefont {Brenes},
  \citenamefont {Goold},\ and\ \citenamefont {Rigol}}]{brenes2020low}%
  \BibitemOpen
  \bibfield  {author} {\bibinfo {author} {\bibfnamefont {M.}~\bibnamefont
  {Brenes}}, \bibinfo {author} {\bibfnamefont {J.}~\bibnamefont {Goold}},\ and\
  \bibinfo {author} {\bibfnamefont {M.}~\bibnamefont {Rigol}},\ }\bibfield
  {title} {\bibinfo {title} {Low-frequency behavior of off-diagonal matrix
  elements in the integrable xxz chain and in a locally perturbed
  quantum-chaotic xxz chain},\ }\href@noop {} {\bibfield  {journal} {\bibinfo
  {journal} {Physical Review B}\ }\textbf {\bibinfo {volume} {102}},\ \bibinfo
  {pages} {075127} (\bibinfo {year} {2020})}\BibitemShut {NoStop}%
\bibitem [{\citenamefont {Pandey}\ \emph {et~al.}(2020)\citenamefont {Pandey},
  \citenamefont {Claeys}, \citenamefont {Campbell}, \citenamefont
  {Polkovnikov},\ and\ \citenamefont {Sels}}]{pandey2020adiabatic}%
  \BibitemOpen
  \bibfield  {author} {\bibinfo {author} {\bibfnamefont {M.}~\bibnamefont
  {Pandey}}, \bibinfo {author} {\bibfnamefont {P.~W.}\ \bibnamefont {Claeys}},
  \bibinfo {author} {\bibfnamefont {D.~K.}\ \bibnamefont {Campbell}}, \bibinfo
  {author} {\bibfnamefont {A.}~\bibnamefont {Polkovnikov}},\ and\ \bibinfo
  {author} {\bibfnamefont {D.}~\bibnamefont {Sels}},\ }\bibfield  {title}
  {\bibinfo {title} {Adiabatic eigenstate deformations as a sensitive probe for
  quantum chaos},\ }\href@noop {} {\bibfield  {journal} {\bibinfo  {journal}
  {Physical Review X}\ }\textbf {\bibinfo {volume} {10}},\ \bibinfo {pages}
  {041017} (\bibinfo {year} {2020})}\BibitemShut {NoStop}%
\bibitem [{\citenamefont {Paris}\ and\ \citenamefont
  {Rehacek}(2004)}]{paris2004quantum}%
  \BibitemOpen
  \bibfield  {author} {\bibinfo {author} {\bibfnamefont {M.}~\bibnamefont
  {Paris}}\ and\ \bibinfo {author} {\bibfnamefont {J.}~\bibnamefont
  {Rehacek}},\ }\href@noop {} {\emph {\bibinfo {title} {Quantum state
  estimation}}},\ Vol.\ \bibinfo {volume} {649}\ (\bibinfo  {publisher}
  {Springer Science \& Business Media},\ \bibinfo {year} {2004})\BibitemShut
  {NoStop}%
\bibitem [{\citenamefont {{Mauro D’Ariano}}\ \emph
  {et~al.}(2003)\citenamefont {{Mauro D’Ariano}}, \citenamefont {Paris},\
  and\ \citenamefont {Sacchi}}]{d2003quantum}%
  \BibitemOpen
  \bibfield  {author} {\bibinfo {author} {\bibfnamefont {G.}~\bibnamefont
  {{Mauro D’Ariano}}}, \bibinfo {author} {\bibfnamefont {M.~G.}\ \bibnamefont
  {Paris}},\ and\ \bibinfo {author} {\bibfnamefont {M.~F.}\ \bibnamefont
  {Sacchi}},\ }\bibfield  {title} {\bibinfo {title} {Quantum tomography}\
  }(\bibinfo  {publisher} {Elsevier},\ \bibinfo {year} {2003})\ pp.\ \bibinfo
  {pages} {205--308}\BibitemShut {NoStop}%
\bibitem [{\citenamefont {Smith}\ \emph {et~al.}(2006)\citenamefont {Smith},
  \citenamefont {Silberfarb}, \citenamefont {Deutsch},\ and\ \citenamefont
  {Jessen}}]{smith2006efficient}%
  \BibitemOpen
  \bibfield  {author} {\bibinfo {author} {\bibfnamefont {G.~A.}\ \bibnamefont
  {Smith}}, \bibinfo {author} {\bibfnamefont {A.}~\bibnamefont {Silberfarb}},
  \bibinfo {author} {\bibfnamefont {I.~H.}\ \bibnamefont {Deutsch}},\ and\
  \bibinfo {author} {\bibfnamefont {P.~S.}\ \bibnamefont {Jessen}},\ }\bibfield
   {title} {\bibinfo {title} {Efficient quantum-state estimation by continuous
  weak measurement and dynamical control},\ }\href@noop {} {\bibfield
  {journal} {\bibinfo  {journal} {Physical review letters}\ }\textbf {\bibinfo
  {volume} {97}},\ \bibinfo {pages} {180403} (\bibinfo {year}
  {2006})}\BibitemShut {NoStop}%
\bibitem [{\citenamefont {Silberfarb}\ \emph {et~al.}(2005)\citenamefont
  {Silberfarb}, \citenamefont {Jessen},\ and\ \citenamefont
  {Deutsch}}]{silberfarb2005quantum}%
  \BibitemOpen
  \bibfield  {author} {\bibinfo {author} {\bibfnamefont {A.}~\bibnamefont
  {Silberfarb}}, \bibinfo {author} {\bibfnamefont {P.~S.}\ \bibnamefont
  {Jessen}},\ and\ \bibinfo {author} {\bibfnamefont {I.~H.}\ \bibnamefont
  {Deutsch}},\ }\bibfield  {title} {\bibinfo {title} {Quantum state
  reconstruction via continuous measurement},\ }\href@noop {} {\bibfield
  {journal} {\bibinfo  {journal} {Physical review letters}\ }\textbf {\bibinfo
  {volume} {95}},\ \bibinfo {pages} {030402} (\bibinfo {year}
  {2005})}\BibitemShut {NoStop}%
\bibitem [{\citenamefont {Riofr{\'\i}o}\ \emph {et~al.}(2011)\citenamefont
  {Riofr{\'\i}o}, \citenamefont {Jessen},\ and\ \citenamefont
  {Deutsch}}]{riofrio2011quantum}%
  \BibitemOpen
  \bibfield  {author} {\bibinfo {author} {\bibfnamefont {C.~A.}\ \bibnamefont
  {Riofr{\'\i}o}}, \bibinfo {author} {\bibfnamefont {P.~S.}\ \bibnamefont
  {Jessen}},\ and\ \bibinfo {author} {\bibfnamefont {I.~H.}\ \bibnamefont
  {Deutsch}},\ }\bibfield  {title} {\bibinfo {title} {Quantum tomography of the
  full hyperfine manifold of atomic spins via continuous measurement on an
  ensemble},\ }\href@noop {} {\bibfield  {journal} {\bibinfo  {journal}
  {Journal of Physics B: Atomic, Molecular and Optical Physics}\ }\textbf
  {\bibinfo {volume} {44}},\ \bibinfo {pages} {154007} (\bibinfo {year}
  {2011})}\BibitemShut {NoStop}%
\bibitem [{\citenamefont {Smith}\ \emph {et~al.}(2013)\citenamefont {Smith},
  \citenamefont {Riofr{\'\i}o}, \citenamefont {Anderson}, \citenamefont
  {Sosa-Martinez}, \citenamefont {Deutsch},\ and\ \citenamefont
  {Jessen}}]{smith2013quantum}%
  \BibitemOpen
  \bibfield  {author} {\bibinfo {author} {\bibfnamefont {A.}~\bibnamefont
  {Smith}}, \bibinfo {author} {\bibfnamefont {C.}~\bibnamefont {Riofr{\'\i}o}},
  \bibinfo {author} {\bibfnamefont {B.}~\bibnamefont {Anderson}}, \bibinfo
  {author} {\bibfnamefont {H.}~\bibnamefont {Sosa-Martinez}}, \bibinfo {author}
  {\bibfnamefont {I.}~\bibnamefont {Deutsch}},\ and\ \bibinfo {author}
  {\bibfnamefont {P.}~\bibnamefont {Jessen}},\ }\bibfield  {title} {\bibinfo
  {title} {Quantum state tomography by continuous measurement and compressed
  sensing},\ }\href@noop {} {\bibfield  {journal} {\bibinfo  {journal}
  {Physical Review A}\ }\textbf {\bibinfo {volume} {87}},\ \bibinfo {pages}
  {030102} (\bibinfo {year} {2013})}\BibitemShut {NoStop}%
\bibitem [{\citenamefont {Madhok}\ \emph {et~al.}(2014)\citenamefont {Madhok},
  \citenamefont {Riofr{\'\i}o}, \citenamefont {Ghose},\ and\ \citenamefont
  {Deutsch}}]{madhok2014information}%
  \BibitemOpen
  \bibfield  {author} {\bibinfo {author} {\bibfnamefont {V.}~\bibnamefont
  {Madhok}}, \bibinfo {author} {\bibfnamefont {C.~A.}\ \bibnamefont
  {Riofr{\'\i}o}}, \bibinfo {author} {\bibfnamefont {S.}~\bibnamefont
  {Ghose}},\ and\ \bibinfo {author} {\bibfnamefont {I.~H.}\ \bibnamefont
  {Deutsch}},\ }\bibfield  {title} {\bibinfo {title} {Information gain in
  tomography--a quantum signature of chaos},\ }\href@noop {} {\bibfield
  {journal} {\bibinfo  {journal} {Physical review letters}\ }\textbf {\bibinfo
  {volume} {112}},\ \bibinfo {pages} {014102} (\bibinfo {year}
  {2014})}\BibitemShut {NoStop}%
\bibitem [{\citenamefont {Madhok}\ \emph {et~al.}(2016)\citenamefont {Madhok},
  \citenamefont {Riofr{\'\i}o},\ and\ \citenamefont
  {Deutsch}}]{madhok2016characterizing}%
  \BibitemOpen
  \bibfield  {author} {\bibinfo {author} {\bibfnamefont {V.}~\bibnamefont
  {Madhok}}, \bibinfo {author} {\bibfnamefont {C.~A.}\ \bibnamefont
  {Riofr{\'\i}o}},\ and\ \bibinfo {author} {\bibfnamefont {I.~H.}\ \bibnamefont
  {Deutsch}},\ }\bibfield  {title} {\bibinfo {title} {Characterizing and
  quantifying quantum chaos with quantum tomography},\ }\href@noop {}
  {\bibfield  {journal} {\bibinfo  {journal} {Pramana}\ }\textbf {\bibinfo
  {volume} {87}},\ \bibinfo {pages} {1} (\bibinfo {year} {2016})}\BibitemShut
  {NoStop}%
\bibitem [{\citenamefont {Sahu}\ \emph
  {et~al.}(2022{\natexlab{a}})\citenamefont {Sahu}, \citenamefont {Sreeram},\
  and\ \citenamefont {Madhok}}]{sahu2022effect}%
  \BibitemOpen
  \bibfield  {author} {\bibinfo {author} {\bibfnamefont {A.}~\bibnamefont
  {Sahu}}, \bibinfo {author} {\bibfnamefont {P.}~\bibnamefont {Sreeram}},\ and\
  \bibinfo {author} {\bibfnamefont {V.}~\bibnamefont {Madhok}},\ }\bibfield
  {title} {\bibinfo {title} {Effect of chaos on information gain in quantum
  tomography},\ }\href@noop {} {\bibfield  {journal} {\bibinfo  {journal}
  {Physical Review E}\ }\textbf {\bibinfo {volume} {106}},\ \bibinfo {pages}
  {024209} (\bibinfo {year} {2022}{\natexlab{a}})}\BibitemShut {NoStop}%
\bibitem [{\citenamefont {Sahu}\ \emph
  {et~al.}(2022{\natexlab{b}})\citenamefont {Sahu}, \citenamefont {Varikuti},\
  and\ \citenamefont {Madhok}}]{sahu2022quantum}%
  \BibitemOpen
  \bibfield  {author} {\bibinfo {author} {\bibfnamefont {A.}~\bibnamefont
  {Sahu}}, \bibinfo {author} {\bibfnamefont {N.~D.}\ \bibnamefont {Varikuti}},\
  and\ \bibinfo {author} {\bibfnamefont {V.}~\bibnamefont {Madhok}},\
  }\bibfield  {title} {\bibinfo {title} {Quantum tomography under perturbed
  hamiltonian evolution and scrambling of errors--a quantum signature of
  chaos},\ }\href@noop {} {\bibfield  {journal} {\bibinfo  {journal} {arXiv
  preprint arXiv:2211.11221}\ } (\bibinfo {year}
  {2022}{\natexlab{b}})}\BibitemShut {NoStop}%
\bibitem [{\citenamefont {Ben-Israel}\ and\ \citenamefont
  {Greville}(2003)}]{ben2003generalized}%
  \BibitemOpen
  \bibfield  {author} {\bibinfo {author} {\bibfnamefont {A.}~\bibnamefont
  {Ben-Israel}}\ and\ \bibinfo {author} {\bibfnamefont {T.~N.}\ \bibnamefont
  {Greville}},\ }\href@noop {} {\emph {\bibinfo {title} {Generalized inverses:
  theory and applications}}},\ Vol.~\bibinfo {volume} {15}\ (\bibinfo
  {publisher} {Springer Science \& Business Media},\ \bibinfo {year}
  {2003})\BibitemShut {NoStop}%
\bibitem [{\citenamefont {Vandenberghe}\ and\ \citenamefont
  {Boyd}(1996)}]{vandenberghe1996semidefinite}%
  \BibitemOpen
  \bibfield  {author} {\bibinfo {author} {\bibfnamefont {L.}~\bibnamefont
  {Vandenberghe}}\ and\ \bibinfo {author} {\bibfnamefont {S.}~\bibnamefont
  {Boyd}},\ }\bibfield  {title} {\bibinfo {title} {Semidefinite programming},\
  }\href@noop {} {\bibfield  {journal} {\bibinfo  {journal} {SIAM review}\
  }\textbf {\bibinfo {volume} {38}},\ \bibinfo {pages} {49} (\bibinfo {year}
  {1996})}\BibitemShut {NoStop}%
\bibitem [{\citenamefont {Shastry}\ and\ \citenamefont
  {Sutherland}(1990)}]{shastry1990twisted}%
  \BibitemOpen
  \bibfield  {author} {\bibinfo {author} {\bibfnamefont {B.~S.}\ \bibnamefont
  {Shastry}}\ and\ \bibinfo {author} {\bibfnamefont {B.}~\bibnamefont
  {Sutherland}},\ }\bibfield  {title} {\bibinfo {title} {Twisted boundary
  conditions and effective mass in heisenberg-ising and hubbard rings},\
  }\href@noop {} {\bibfield  {journal} {\bibinfo  {journal} {Physical review
  letters}\ }\textbf {\bibinfo {volume} {65}},\ \bibinfo {pages} {243}
  (\bibinfo {year} {1990})}\BibitemShut {NoStop}%
\bibitem [{\citenamefont {Cazalilla}\ \emph {et~al.}(2011)\citenamefont
  {Cazalilla}, \citenamefont {Citro}, \citenamefont {Giamarchi}, \citenamefont
  {Orignac},\ and\ \citenamefont {Rigol}}]{cazalilla2011one}%
  \BibitemOpen
  \bibfield  {author} {\bibinfo {author} {\bibfnamefont {M.~A.}\ \bibnamefont
  {Cazalilla}}, \bibinfo {author} {\bibfnamefont {R.}~\bibnamefont {Citro}},
  \bibinfo {author} {\bibfnamefont {T.}~\bibnamefont {Giamarchi}}, \bibinfo
  {author} {\bibfnamefont {E.}~\bibnamefont {Orignac}},\ and\ \bibinfo {author}
  {\bibfnamefont {M.}~\bibnamefont {Rigol}},\ }\bibfield  {title} {\bibinfo
  {title} {One dimensional bosons: From condensed matter systems to ultracold
  gases},\ }\href@noop {} {\bibfield  {journal} {\bibinfo  {journal} {Reviews
  of Modern Physics}\ }\textbf {\bibinfo {volume} {83}},\ \bibinfo {pages}
  {1405} (\bibinfo {year} {2011})}\BibitemShut {NoStop}%
\bibitem [{\citenamefont {Joel}\ \emph {et~al.}(2013)\citenamefont {Joel},
  \citenamefont {Kollmar},\ and\ \citenamefont
  {Santos}}]{joel2013introduction}%
  \BibitemOpen
  \bibfield  {author} {\bibinfo {author} {\bibfnamefont {K.}~\bibnamefont
  {Joel}}, \bibinfo {author} {\bibfnamefont {D.}~\bibnamefont {Kollmar}},\ and\
  \bibinfo {author} {\bibfnamefont {L.~F.}\ \bibnamefont {Santos}},\ }\bibfield
   {title} {\bibinfo {title} {An introduction to the spectrum, symmetries, and
  dynamics of spin-1/2 heisenberg chains},\ }\href@noop {} {\bibfield
  {journal} {\bibinfo  {journal} {American Journal of Physics}\ }\textbf
  {\bibinfo {volume} {81}},\ \bibinfo {pages} {450} (\bibinfo {year}
  {2013})}\BibitemShut {NoStop}%
\bibitem [{\citenamefont {\ifmmode \check{R}\else
  \v{R}\fi{}eh\'a\ifmmode~\check{c}\else \v{c}\fi{}ek}\ and\ \citenamefont
  {Hradil}(2002)}]{hradil02}%
  \BibitemOpen
  \bibfield  {author} {\bibinfo {author} {\bibfnamefont {J.}~\bibnamefont
  {\ifmmode \check{R}\else \v{R}\fi{}eh\'a\ifmmode~\check{c}\else
  \v{c}\fi{}ek}}\ and\ \bibinfo {author} {\bibfnamefont {Z.}~\bibnamefont
  {Hradil}},\ }\bibfield  {title} {\bibinfo {title} {Invariant information and
  quantum state estimation},\ }\href
  {https://doi.org/10.1103/PhysRevLett.88.130401} {\bibfield  {journal}
  {\bibinfo  {journal} {Phys. Rev. Lett.}\ }\textbf {\bibinfo {volume} {88}},\
  \bibinfo {pages} {130401} (\bibinfo {year} {2002})}\BibitemShut {NoStop}%
\bibitem [{\citenamefont {Boyd}\ \emph {et~al.}(2004)\citenamefont {Boyd},
  \citenamefont {Boyd},\ and\ \citenamefont {Vandenberghe}}]{boyd2004convex}%
  \BibitemOpen
  \bibfield  {author} {\bibinfo {author} {\bibfnamefont {S.}~\bibnamefont
  {Boyd}}, \bibinfo {author} {\bibfnamefont {S.~P.}\ \bibnamefont {Boyd}},\
  and\ \bibinfo {author} {\bibfnamefont {L.}~\bibnamefont {Vandenberghe}},\
  }\href@noop {} {\emph {\bibinfo {title} {Convex optimization}}}\ (\bibinfo
  {publisher} {Cambridge university press},\ \bibinfo {year}
  {2004})\BibitemShut {NoStop}%
\bibitem [{\citenamefont {Espa{\~n}ol}\ and\ \citenamefont
  {Wisniacki}(2023)}]{espanol2023assessing}%
  \BibitemOpen
  \bibfield  {author} {\bibinfo {author} {\bibfnamefont {B.~L.}\ \bibnamefont
  {Espa{\~n}ol}}\ and\ \bibinfo {author} {\bibfnamefont {D.~A.}\ \bibnamefont
  {Wisniacki}},\ }\bibfield  {title} {\bibinfo {title} {Assessing the
  saturation of krylov complexity as a measure of chaos},\ }\href@noop {}
  {\bibfield  {journal} {\bibinfo  {journal} {Physical Review E}\ }\textbf
  {\bibinfo {volume} {107}},\ \bibinfo {pages} {024217} (\bibinfo {year}
  {2023})}\BibitemShut {NoStop}%
\bibitem [{\citenamefont {Lysne}\ \emph {et~al.}(2020)\citenamefont {Lysne},
  \citenamefont {Kuper}, \citenamefont {Poggi}, \citenamefont {Deutsch},\ and\
  \citenamefont {Jessen}}]{lysne2020small}%
  \BibitemOpen
  \bibfield  {author} {\bibinfo {author} {\bibfnamefont {N.~K.}\ \bibnamefont
  {Lysne}}, \bibinfo {author} {\bibfnamefont {K.~W.}\ \bibnamefont {Kuper}},
  \bibinfo {author} {\bibfnamefont {P.~M.}\ \bibnamefont {Poggi}}, \bibinfo
  {author} {\bibfnamefont {I.~H.}\ \bibnamefont {Deutsch}},\ and\ \bibinfo
  {author} {\bibfnamefont {P.~S.}\ \bibnamefont {Jessen}},\ }\bibfield  {title}
  {\bibinfo {title} {Small, highly accurate quantum processor for
  intermediate-depth quantum simulations},\ }\href@noop {} {\bibfield
  {journal} {\bibinfo  {journal} {Physical review letters}\ }\textbf {\bibinfo
  {volume} {124}},\ \bibinfo {pages} {230501} (\bibinfo {year}
  {2020})}\BibitemShut {NoStop}%
\bibitem [{\citenamefont {Mu{\~n}oz-Arias}\ \emph {et~al.}(2020)\citenamefont
  {Mu{\~n}oz-Arias}, \citenamefont {Poggi}, \citenamefont {Jessen},\ and\
  \citenamefont {Deutsch}}]{munoz2020simulating}%
  \BibitemOpen
  \bibfield  {author} {\bibinfo {author} {\bibfnamefont {M.~H.}\ \bibnamefont
  {Mu{\~n}oz-Arias}}, \bibinfo {author} {\bibfnamefont {P.~M.}\ \bibnamefont
  {Poggi}}, \bibinfo {author} {\bibfnamefont {P.~S.}\ \bibnamefont {Jessen}},\
  and\ \bibinfo {author} {\bibfnamefont {I.~H.}\ \bibnamefont {Deutsch}},\
  }\bibfield  {title} {\bibinfo {title} {Simulating nonlinear dynamics of
  collective spins via quantum measurement and feedback},\ }\href@noop {}
  {\bibfield  {journal} {\bibinfo  {journal} {Physical review letters}\
  }\textbf {\bibinfo {volume} {124}},\ \bibinfo {pages} {110503} (\bibinfo
  {year} {2020})}\BibitemShut {NoStop}%
\bibitem [{\citenamefont {Krithika}\ \emph {et~al.}(2023)\citenamefont
  {Krithika}, \citenamefont {Santhanam},\ and\ \citenamefont
  {Mahesh}}]{krithika2023nmr}%
  \BibitemOpen
  \bibfield  {author} {\bibinfo {author} {\bibfnamefont {V.}~\bibnamefont
  {Krithika}}, \bibinfo {author} {\bibfnamefont {M.}~\bibnamefont
  {Santhanam}},\ and\ \bibinfo {author} {\bibfnamefont {T.}~\bibnamefont
  {Mahesh}},\ }\bibfield  {title} {\bibinfo {title} {Nmr investigations of
  dynamical tunneling in spin systems},\ }\href@noop {} {\bibfield  {journal}
  {\bibinfo  {journal} {Physical Review A}\ }\textbf {\bibinfo {volume}
  {108}},\ \bibinfo {pages} {032207} (\bibinfo {year} {2023})}\BibitemShut
  {NoStop}%
\bibitem [{\citenamefont {Maurya}\ \emph {et~al.}(2022)\citenamefont {Maurya},
  \citenamefont {Kannan}, \citenamefont {Patel}, \citenamefont {Dutta},
  \citenamefont {Biswas}, \citenamefont {Mangaonkar}, \citenamefont
  {Santhanam},\ and\ \citenamefont {Rapol}}]{maurya2022control}%
  \BibitemOpen
  \bibfield  {author} {\bibinfo {author} {\bibfnamefont {S.~S.}\ \bibnamefont
  {Maurya}}, \bibinfo {author} {\bibfnamefont {S.~B.}\ \bibnamefont {Kannan}},
  \bibinfo {author} {\bibfnamefont {K.}~\bibnamefont {Patel}}, \bibinfo
  {author} {\bibfnamefont {P.}~\bibnamefont {Dutta}}, \bibinfo {author}
  {\bibfnamefont {K.}~\bibnamefont {Biswas}}, \bibinfo {author} {\bibfnamefont
  {J.}~\bibnamefont {Mangaonkar}}, \bibinfo {author} {\bibfnamefont
  {M.}~\bibnamefont {Santhanam}},\ and\ \bibinfo {author} {\bibfnamefont
  {U.~D.}\ \bibnamefont {Rapol}},\ }\bibfield  {title} {\bibinfo {title}
  {Control of dynamical localization in atom-optics kicked rotor},\ }\href@noop
  {} {\bibfield  {journal} {\bibinfo  {journal} {arXiv preprint
  arXiv:2202.02820}\ } (\bibinfo {year} {2022})}\BibitemShut {NoStop}%
\bibitem [{\citenamefont {Chaudhury}\ \emph {et~al.}(2009)\citenamefont
  {Chaudhury}, \citenamefont {Smith}, \citenamefont {Anderson}, \citenamefont
  {Ghose},\ and\ \citenamefont {Jessen}}]{chaudhury2009quantum}%
  \BibitemOpen
  \bibfield  {author} {\bibinfo {author} {\bibfnamefont {S.}~\bibnamefont
  {Chaudhury}}, \bibinfo {author} {\bibfnamefont {A.}~\bibnamefont {Smith}},
  \bibinfo {author} {\bibfnamefont {B.}~\bibnamefont {Anderson}}, \bibinfo
  {author} {\bibfnamefont {S.}~\bibnamefont {Ghose}},\ and\ \bibinfo {author}
  {\bibfnamefont {P.~S.}\ \bibnamefont {Jessen}},\ }\bibfield  {title}
  {\bibinfo {title} {Quantum signatures of chaos in a kicked top},\ }\href@noop
  {} {\bibfield  {journal} {\bibinfo  {journal} {Nature}\ }\textbf {\bibinfo
  {volume} {461}},\ \bibinfo {pages} {768} (\bibinfo {year}
  {2009})}\BibitemShut {NoStop}%
\bibitem [{\citenamefont {Viswanath}\ and\ \citenamefont
  {M{\"u}ller}(2008)}]{viswanath2008recursion}%
  \BibitemOpen
  \bibfield  {author} {\bibinfo {author} {\bibfnamefont {V.}~\bibnamefont
  {Viswanath}}\ and\ \bibinfo {author} {\bibfnamefont {G.}~\bibnamefont
  {M{\"u}ller}},\ }\href@noop {} {\emph {\bibinfo {title} {The recursion
  method: application to many-body dynamics}}},\ Vol.~\bibinfo {volume} {23}\
  (\bibinfo  {publisher} {Springer Science \& Business Media},\ \bibinfo {year}
  {2008})\BibitemShut {NoStop}%
\bibitem [{\citenamefont {Parlett}(1998)}]{parlett1998thesym}%
  \BibitemOpen
  \bibfield  {author} {\bibinfo {author} {\bibfnamefont {B.~N.}\ \bibnamefont
  {Parlett}},\ }\href {https://doi.org/10.1137/1.9781611971163} {\emph
  {\bibinfo {title} {The Symmetric Eigenvalue Problem}}}\ (\bibinfo
  {publisher} {Society for Industrial and Applied Mathematics},\ \bibinfo
  {year} {1998})\BibitemShut {NoStop}%
\bibitem [{\citenamefont {Thomas M.~Cover}(2006)}]{coverm2006elements}%
  \BibitemOpen
  \bibfield  {author} {\bibinfo {author} {\bibfnamefont {J.~A.~T.}\
  \bibnamefont {Thomas M.~Cover}},\ }\href@noop {} {\emph {\bibinfo {title}
  {Elements of Information Theory}}}\ (\bibinfo  {publisher} {John Wiley \&
  Sons, Ltd},\ \bibinfo {year} {2006})\BibitemShut {NoStop}%
\bibitem [{\citenamefont {Zhou}\ and\ \citenamefont
  {Luitz}(2017)}]{zhou2017operator}%
  \BibitemOpen
  \bibfield  {author} {\bibinfo {author} {\bibfnamefont {T.}~\bibnamefont
  {Zhou}}\ and\ \bibinfo {author} {\bibfnamefont {D.~J.}\ \bibnamefont
  {Luitz}},\ }\bibfield  {title} {\bibinfo {title} {Operator entanglement
  entropy of the time evolution operator in chaotic systems},\ }\href@noop {}
  {\bibfield  {journal} {\bibinfo  {journal} {Physical Review B}\ }\textbf
  {\bibinfo {volume} {95}},\ \bibinfo {pages} {094206} (\bibinfo {year}
  {2017})}\BibitemShut {NoStop}%
\bibitem [{\citenamefont {Pal}\ and\ \citenamefont
  {Lakshminarayan}(2018)}]{pal2018entangling}%
  \BibitemOpen
  \bibfield  {author} {\bibinfo {author} {\bibfnamefont {R.}~\bibnamefont
  {Pal}}\ and\ \bibinfo {author} {\bibfnamefont {A.}~\bibnamefont
  {Lakshminarayan}},\ }\bibfield  {title} {\bibinfo {title} {Entangling power
  of time-evolution operators in integrable and nonintegrable many-body
  systems},\ }\href@noop {} {\bibfield  {journal} {\bibinfo  {journal}
  {Physical Review B}\ }\textbf {\bibinfo {volume} {98}},\ \bibinfo {pages}
  {174304} (\bibinfo {year} {2018})}\BibitemShut {NoStop}%
\end{thebibliography}%
\onecolumngrid

\end{document}